\documentclass[%
 aip,
 amsmath,amssymb,
 reprint,%
]{revtex4-2}

\usepackage{graphicx}
\usepackage{dcolumn}
\usepackage{bm}

\usepackage[utf8]{inputenc}
\usepackage[T1]{fontenc}
\usepackage{mathptmx}
\usepackage{etoolbox}

\usepackage[english]{babel}
\usepackage{longtable}
\usepackage{multirow}
\usepackage{xcolor}

\makeatletter
\def\@email#1#2{%
 \endgroup
 \patchcmd{\titleblock@produce}
  {\frontmatter@RRAPformat}
  {\frontmatter@RRAPformat{\produce@RRAP{*#1\href{mailto:#2}{#2}}}\frontmatter@RRAPformat}
  {}{}
}%
\makeatother
\begin{document}

\preprint{AIP/123-QED}

\title{Neural operator transformers capture bifurcating drift wave turbulence in fusion plasma simulations}
\author{J. J. van de Wetering}
\thanks{vandeweterin1@llnl.gov}
\affiliation{Lawrence Livermore National Laboratory, Livermore, CA 94551, USA}

\author{B. Zhu}%
\thanks{ben.zhu@columbia.edu}
\affiliation{Department of Applied Physics and Applied Mathematics, Columbia University, New York, NY 10027, USA}

\date{\today}

\begin{abstract}
Self-consistent modeling of turbulence-driven transport is critical for optimizing confinement in magnetically confined fusion plasmas, such as in tokamaks and stellarators. In particular, capturing the long-term co-evolution of turbulence, flow, and background plasma profiles remains computationally challenging. Direct numerical simulation of these multiscale, highly nonlinear processes is often demanding and impractical for real-time control or design optimization. To address this bottleneck, we investigate transformer-based neural operator PDE surrogates for emulating the dynamics of drift-wave turbulence bifurcation mediated by zonal flows, using the modified Hasegawa-Wakatani (MHW) model as a prototypical system. We find that the finetuned neural operator model has excellent performance in capturing the multi-spatiotemporal-scales of MHW turbulence bifurcation and is robust to testing on rare and out-of-distribution dynamics. Specifically, we demonstrate that a single unified model accurately predicts both quasi-steady-state turbulence and a wide range of dynamical transition processes, such as nonlinear saturation, spontaneous suppression of turbulence and the emergence of macroscopic zonal flows, over time horizons vastly exceeding the local turbulence correlation time. This computationally efficient approach establishes a strong foundation for fast, AI-based modeling of complex, multiscale phenomena in magnetized fusion plasmas.
\end{abstract}

\maketitle

\section{\label{sec:level1}Introduction}

Turbulence-driven transport is the primary loss mechanism in magnetically confined fusion plasmas. Consequently, understanding and predicting this process has been pursued extensively within the fusion community. However, predictive modeling remains exceptionally challenging because it is an inherently multi-spatiotemporal problem, characterized by the nonlinear interaction between fast, small-scale turbulent fluctuations and slower, large-scale background profiles and flows. A prime example is the spontaneous bifurcation between distinct confinement regimes, such as the transition from low- to high-confinement modes (L--H transition) in tokamaks, which can reduce deleterious edge turbulence by over an order of magnitude. The dynamics that trigger and sustain these bifurcations are computationally prohibitive to fully resolve over long time horizons using high-fidelity Direct Numerical Simulation (DNS).  This gap in fast, predictive modeling limits our confidence in extrapolating to future devices, planning discharge scenarios, optimizing magnetic and divertor configurations, and developing real-time control strategies.

The extreme computational expense of running DNS for these highly nonlinear processes has motivated the development of AI-based surrogate models designed to emulate turbulent transport at a fraction of the cost. To date, much of this work has focused on emulating long-term dynamically steady-state turbulence behavior. For instance, reduced-order models \cite{gahr_scientific_2024} and generative AI frameworks \cite{clavier_generative_2025} have been developed to capture the statistics of isotropic turbulence using the original Hasegawa-Wakatani system \cite{hasegawa_plasma_1983, wakatani_collisional_1984}, which does not exhibit turbulence bifurcations. Other approaches, such as StyleGAN \cite{castagna2024stylegan}, have been applied to the modified Hasegawa-Wakatani (MHW) model \cite{numata2007bifurcation} to reconstruct high-resolution fields from lower-resolution Large Eddy Simulations (LES), capturing the co-evolution of turbulence and zonal flows. Despite these advances, constructing AI-based surrogate models capable of capturing transient shifts and dynamical bifurcations remains a distinct challenge. Our study aims to go beyond the emulation of steady-state turbulence to directly model dynamical transitions. 

In magnetized plasmas, phenomena like the L--H transition represent a complex self-organization process from a fully turbulent state to a turbulence-suppressed regime. The dynamics of this transition, beyond established scaling laws \cite{martin_power_2008}, is known to be sensitive to many factors such as divertor geometry and plasma shape \cite{schmitz_role_2017,gohil_lh_2011,ryter_survey_2013}. Hence, for an AI-based surrogate model to be useful in this regard, it must not only accurately integrate these stiff temporal dynamics but also be capable of generalizing across arbitrary geometries, meshes, and spatial domains.

This dual requirement naturally motivates us to investigate neural operators\cite{kovachki_neural_2023}, a class of AI surrogate models that learn the solution operator of parametric partial differential equations (PDEs) from data, as a candidate to emulate the dynamics of such a system. Current transformer\cite{vaswani_attention_2017}-based implementations \cite{wu_transolver_2024,wen_geometry_2025,alkin_ab-upt_2025} have demonstrated PDE emulation on large, variable 2D/3D meshes and geometries in the context of computational fluid dynamics. Furthermore, these architectures are also being adapted to handle emulation of full 5D gyrokinetics\cite{paischer_gyroswin_2025}, making them even more pertinent to the modeling of magnetic fusion plasmas.

In this work, we apply a transformer-based neural operator framework \cite{wen_geometry_2025} to the MHW model to emulate the dynamics of bifurcating drift-wave turbulence. The MHW system serves as an ideal prototypical testbed; it is a well-established reduced model that retains the critical nonlinear feedback mechanisms between turbulent fluctuations and zonal flows -- arguably the minimal model capable of reproducing L--H transition-like phenomena. We present results of the predictive capability of these emulators for both short timescale turbulent dynamics and long timescale generation/destruction of zonal flows, laying the groundwork for future studies involving more complete plasma turbulence simulations in realistic geometries. 

The remainder of this paper is organized as follows. Section~\ref{sec:mhw} introduces the Modified Hasegawa-Wakatani (MHW) equations, while Section~\ref{sec:nos} describes the transformer-based neural operator architecture used in this study. Section~\ref{sec:data} details the generation of the datasets, and Section~\ref{sec:training} outlines the specific pretraining and finetuning strategies employed for short- and long-horizon prediction tasks. In Section~\ref{sec:results}, we present the performance evaluation of the surrogate models, systematically analyzing short-term trajectory predictions, long-horizon statistical fidelity, and the model's capacity to generalize to both gradual and abrupt dynamical transitions. Finally, Section~\ref{sec:con} summarizes our main findings and discusses the computational speed-up over DNS. Additional details regarding the spatial inhomogeneity of the MHW turbulence and the model's performance (supplementary figures) are provided in the Appendices.

\section{Modified Hasegawa-Wakatani Equations}\label{sec:mhw}

The nonlinear dynamical system investigated in this study is the Modified Hasegawa-Wakatani (MHW) model~\cite{numata2007bifurcation} which describes the electrostatic drift-wave instability in an isothermal plasma, driven by an inhomogeneous density background in a slab geometry with a uniform magnetic field. Despite being a minimal model, it contains key ingredients in plasma turbulence such as zonal-flow generation and shear flow stabilization.
The model consists of two coupled partial differential equations governing the evolution of the perturbed plasma density $n$ and the electrostatic potential $\phi$.

Adopting standard drift-scale normalization where lengths are scaled by the ion sound Larmor radius $\rho_\text{s}$, time by the inverse ion cyclotron frequency $\omega_\text{ci}^{-1}$, potential by $T_e/e$, and density by the background density $n_0$, the dimensionless governing equations are
\begin{align}
    \frac{\partial \varpi}{\partial t} &= -[\phi, \varpi] + \alpha(\tilde{\phi} - \tilde{n}) - \nu_\varpi \nabla_\perp^{2k} \varpi \\
    \frac{\partial n}{\partial t} &= -[\phi, n] + \alpha(\tilde{\phi} - \tilde{n}) - \kappa \frac{\partial \phi}{\partial y} - \nu_n \nabla_\perp^{2k} n.
\end{align}
Here $\varpi=\nabla_\perp^2\phi$ denotes vorticity. For any field $f$, the zonal (or, $y$-averaged) and non-zonal (fluctuating) components are defined as
\begin{equation}
    \langle f\rangle = \frac{1}{L_y}\int f dy, \quad \tilde{f} = f - \langle f\rangle.
\end{equation}

The nonlinear terms are expressed via the Poisson bracket, $[\phi, f] = \hat{\boldsymbol{z}} \cdot (\nabla \phi \times \nabla f)$, which represents the $E \times B$ advection of the respective fields. The parameter $\alpha$, often referred to as the adiabaticity, denotes the resistive coupling between $n$ and $\phi$ due to parallel electron dynamics. In two-dimensional simulations, a characteristic parallel wavevector $k_z$ is assumed, rendering $\alpha = T_{e}k_z^2 / (\eta_\parallel n_0 \omega_\text{ci} e^2)$ a constant. The drive for the instability is provided by the background density gradient $\kappa = n_0/L_n = - \partial (\ln n_0) / \partial x$. To ensure numerical stability and facilitate energy dissipation at the grid scale, hyper-diffusion terms of the form $\nu \nabla_\perp^{2k}$ with $k=2,3$ are also included in simulations.

\begin{figure}[!htbp]
    \centering
    \includegraphics[width=\linewidth]{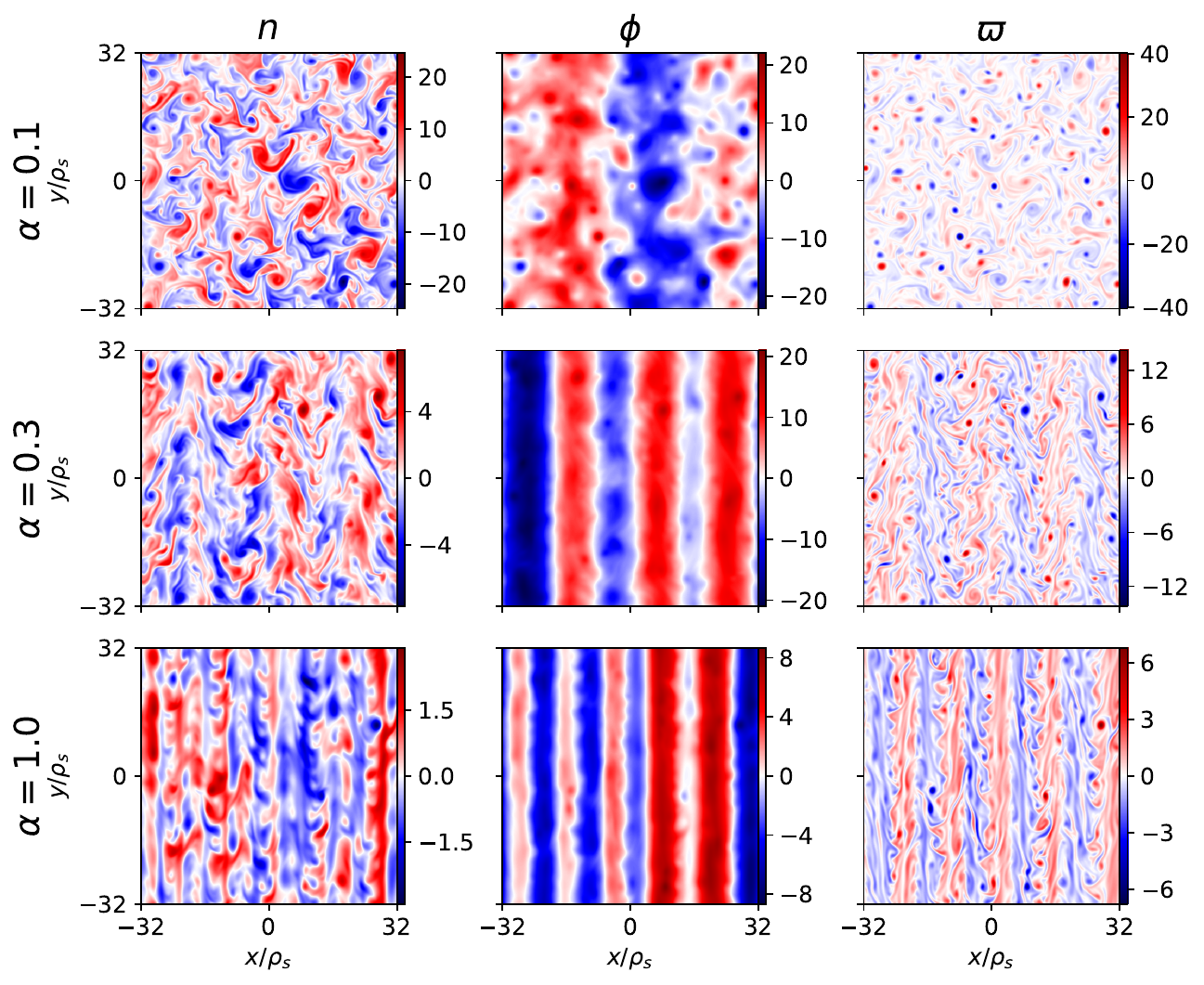}
    \caption{Snapshots of the perturbed density $n$, potential $\phi$ and vorticity $\varpi$ for the MHW model in the turbulence saturation stage with background density gradient $\kappa=1.0$ and adiabaticity $\alpha \in \{0.1, 0.3, 1.0\}$.}
    \label{fig:Fig1}
\end{figure}

The MHW model differs from the original Hasegawa-Wakatani formulation~\cite{hasegawa1987self} by the subtraction of the zonal mean from the resistive coupling terms. This modification accounts for the implication that zonal flow could not be resistively dissipated in this model. Consequently, the MHW model permits the spontaneous generation of zonal flows, which play a critical role in the self-regulation of turbulent transport, and the system's behavior is primarily governed by $\alpha$, which delineates two distinct physical regimes. In the hydrodynamic limit ($\alpha \to 0$) where parallel resistivity dominates, the system decouples and tends toward isotropic, neutral-fluid-like turbulence; while in the adiabatic limit ($\alpha \to \infty$) where electrons respond nearly instantaneously to the potential, the system recovers the Hasegawa-Mima model~\cite{hasegawa1977stationary}, characterized by the emergence of large-scale coherent (e.g.,  zonal) structures and suppressed turbulent transport. Figure \ref{fig:Fig1} shows the density and potential snapshots for different $\alpha$ values in the turbulence saturation stage.

The macroscopic state of the MHW system can be characterized by its total energy and generalized enstrophy,
\begin{align}
E = \frac{1}{2}\int(n^2+|\nabla\phi|^2)\text{d}x\text{d}y, \,\,\,\, \,\,\,\, W = \frac{1}{2}\int(n-\varpi)^2\text{d}x\text{d}y.
\end{align}
The total energy $E$ comprises both internal and kinetic energy, each of which can be split into zonal and non-zonal (turbulent) components. Another important pair of metrics used in our study to quantify the system's transport behavior are the turbulent flux and resistive dissipation,
\begin{align}
    \Gamma_n = -\kappa \langle \tilde{n} \frac{\partial \tilde{\phi}}{\partial y}\rangle, \,\,\,\, \,\,\,\, D_\alpha = \alpha \langle(\tilde{n}-\tilde{\phi})^2\rangle. 
\end{align}
Here $\langle \dots \rangle$ represents the spatial domain average. The turbulent particle flux $\Gamma_n$ captures the cross-field transport driven by the correlation between density fluctuations and the radial $E \times B$ velocity, serving as the energy source for the system. Conversely, the resistive dissipation $D_\alpha$ represents the energy loss due to finite parallel electron resistivity (non-adiabaticity). It acts as the primary physical energy sink of the system, with the remaining small-scale energy dissipated numerically via the hyper-diffusion terms.

\section{Neural Operators}\label{sec:nos}

Neural operators are a recently popularized class of PDE surrogate models designed to learn a representation of the solution operator
\begin{align}
    \mathcal{S} : a \mapsto u(a)
\end{align}
in a data-driven manner, where $\mathcal{S}$ maps a parameter or input function $a$ to the corresponding solution $u(a)$ of a parametric PDE, for example
\begin{align}
    \mathcal{L}(u(a); a) = f(a) \quad \text{in domain } D,
\end{align}
with suitable boundary conditions. Their main advantages over other types of surrogates are that they output functions and can be queried at arbitrary points in space. Some more recent implementations are also approximately invariant to the choice of mesh discretization and can accept inputs specified at arbitrary points, which some argue are essential properties of ``true'' neural operators \cite{kovachki_neural_2023}.

Neural operators are constructed by defining a single layer update of the hidden representations $v$ to $u$ with a learnable non-local kernel integration 
\begin{align}
    u(x) = \sigma\left( Wv(x) + \int_DK(x, y)v(y)\,\mathrm{d}\mu(y) + b(x) \right) \quad \forall x \in D
\end{align}
for weights $W$, biases $b$, activations $\sigma$, measure $\mu$ and learnable kernel function $K$. The challenge here is to parameterize and approximate the kernel integral in a way that is both expressive and computationally efficient, for example via spectral convolutions, low rank decompositions, or graph-based message passing. Some established operator learning frameworks include DeepONets \cite{lu_learning_2021}, Graph Neural Operators (GNO) \cite{li_neural_2020}, Fourier Neural Operators (FNO) \cite{li_fourier_2021}, Convolutional Neural Operators (CNO) \cite{raonic_convolutional_2023}, Laplace Neural Operators (LNO) \cite{cao_laplace_2024} and, more recently, transformer \cite{vaswani_attention_2017}-based architectures adapted to operator learning.

In this study, we adapt the Geometry--Aware Operator Transformer (GAOT) \cite{wen_geometry_2025}, an open source neural operator framework for both time-independent and time-dependent parametric PDEs. It is comprised of a standard U-shaped vision transformer (ViT \cite{50650,ovadia_vito_2024}) as its processor, with additional features available for its GNO encoder and decoder to enable efficient operator learning on arbitrary domains. Although the geometry-generalization features were ablated in this study due to the 2D fully periodic square domain of the current turbulence model, such capability is necessary for subsequent studies in the more complex and variable geometries of realistic tokamak edge plasma turbulence where the background magnetic field topology includes both closed and open field lines and often features singularities (e.g., X-points).

For time-dependent PDEs, there is a choice of time-stepping strategies. As described in \cite{mousavi_rigno_2025,wen_geometry_2025}, we have opted for the ``time-derivative'' stepping strategy. This is where the model $\hat{\mathcal{S}}_\theta$ (neglecting normalizations) learns a representation of the forward difference approximation of the time derivative
\begin{align}
    \hat{\mathcal{S}}_\theta(x,t,\tau,a(t)) = \frac{u(t+\tau)-u(t)}{\tau},
\end{align}
where $u(t)$ is the solution at time $t$ and $\tau>0$ is the lead-time. There is a design choice to make for the range of lead-times to train on. In this framework, the default is the ``all2all'' training strategy \cite{herde_poseidon_2024}, which provides all $N(N+1)/2$ pairs up to a specified maximum lead-time $T = N\Delta\tau$ with sampling timestep $\Delta\tau$. This, depending on the dynamics, can greatly enhance the available data for training on trajectories. The time derivative timestepping is also subject to Z-score data normalization, see \cite{mousavi_rigno_2025} for further details. Testing within and extrapolation beyond the maximum lead time $T$ are performed with autoregressive rollouts \cite{lippe_pde-refiner_2023}.

\section{Simulation Dataset}\label{sec:data}

The simulation data used in this study were generated using a reduced version of the Global Drift-Ballooning (GDB) code~\cite{zhu2018gdb}, employing a numerical implementation consistent with the full model. The governing equations are solved in a two-dimensional, fully periodic domain of size $L_x = L_y = 64\rho_\text{s}$ with a spatial resolution of $512 \times 512$ grid points.
The Poisson brackets are evaluated with the Arakawa scheme~\cite{arakawa1966computational} to ensure the conservation of kinetic energy and enstrophy, which is critical for maintaining the physical fidelity of spectral energy transfers and the long-term accuracy of the coupled turbulence-zonal flow dynamics.
To maintain numerical stability and suppress grid-scale oscillations without affecting large-scale dynamics, a small ($\nu_\varpi = \nu_n = 1 \times 10^{-5}$) sixth-order hyper-diffusion term ($k=3$) is applied to both the vorticity and density~\cite{greif2023physics}. The hyper-diffusion and the inversion of the Poisson equation for the potential ($\phi = \nabla_\perp^{-2} \varpi$) are performed with the spectral method.
A second-order accurate leapfrog-trapezoidal method is used for time-stepping. 
All simulations presented employ a fixed timestep of $\Delta t = 5 \times 10^{-3}$, which satisfies the Courant-Friedrichs-Lewy (CFL) condition for the chosen parameters and spatial resolution. In terms of computational cost, this configuration takes approximately 2.8 seconds of wall-clock time per unit of simulation time when executed on a single CPU node of the NERSC Perlmutter system.

In this study, the background density gradient is fixed at $\kappa=1$, while the adiabaticity parameter $\alpha$ is varied to explore different turbulence regimes. Two distinct types of numerical experiments were conducted to generate the database.

\begin{figure*}[t]
    \centering
    \includegraphics[width=\linewidth]{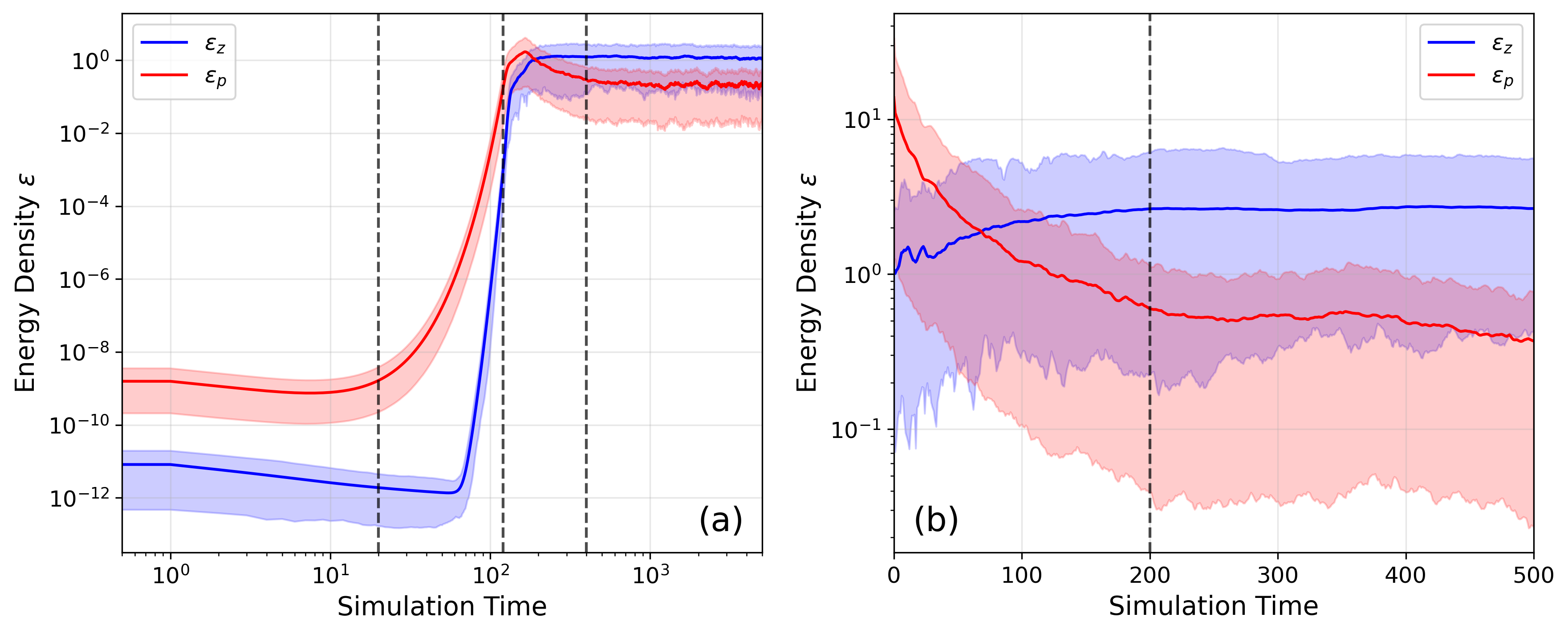}
    \caption{Time evolution of the zonal $\epsilon_z$ and perturbed $\epsilon_p$ energy densities for (a) a steady-state simulation with $(\alpha, \kappa) = (1.0, 1.0)$ and (b) a dynamical transition simulation where $\alpha$ is increased instantaneously from 0.1 to 1.0 while $\kappa = 1.0$ remains fixed. Shaded regions represent the 10th to 90th percentiles of spatial variation. Vertical dashed lines roughly delineate different phases of temporal evolution.}
    \label{fig:energy_evo}
\end{figure*}

The first dataset comprises of \textit{steady-state} simulations initialized with small-amplitude ($10^{-4}$) random noise in the density and potential fields. The system is allowed to evolve freely until it reaches a fully saturated turbulent state, as indicated by the quasi-stationary total energy and generalized enstrophy. While saturation typically occurs within a few hundred normalized time units, the simulations are extended to a total duration of $t = 5,000$ to provide a statistically significant ensemble for turbulence analysis and model training. Data snapshots are recorded every 200 timesteps, corresponding to one normalized time unit, yielding a total of 5,000 frames per run. To capture a broad spectrum of turbulent dynamics, fourteen independent simulations were performed varying the adiabaticity parameter $\alpha$. Ten of these runs ($\alpha \in \{0.1, 0.2, 0.3, 0.4, 0.5, 0.6, 0.7, 0.8, 0.9, 1.0\}$) comprise the standard training and testing sets. The remaining four runs are strictly reserved for evaluating the model's generalization, with two assessing interpolation ($\alpha \in \{0.25, 0.75\}$) and two assessing extrapolation ($\alpha \in \{0.08, 1.1\}$).
As illustrated in the energy evolution plot in Figure~\ref{fig:energy_evo}(a), a typical steady-state simulation exhibits four distinct phases: (1) relaxation, where the initial random noise reorganizes into the most unstable eigenmode structure (often $t \leq 20$), (2) linear growth, characterized by the exponential growth of the instability ($t \sim 20 - 100$), (3) nonlinear interaction, here dominant nonlinearities drive the transition to turbulence, potentially accompanied by profound zonal flow generation and turbulence suppression ($t \sim 100 - 400$), and (4) saturation, where the system reaches a statistically stationary state ($t > 500$).

Additionally, \textit{dynamical transition} simulations were also performed to investigate the plasma response to parametric variations. These runs are initialized by restarting a fully saturated steady-state simulation and impulsively shifting the adiabaticity parameter $\alpha$. Capturing these transients between distinct plasma regimes (e.g., the transition from the hydrodynamic to the adiabatic limit) allows us to further evaluate the model's capability to generalize towards transient dynamics. Consistent with the steady-state runs, snapshots are recorded at intervals of one normalized time unit for a total duration of 500 frames per transition. To explore the effects of transition direction and magnitude, twenty simulations were conducted, including twelve forward and backward transitions among $\alpha \in \{0.1, 0.4, 0.7, 1.0\}$ for the standard training and testing sets and eight additional runs to assess the model's generalization to out-of-distribution parameter shifts, specifically: $\{0.1 \to 0.08, 0.1 \to 0.25, 0.1 \to 0.75, 0.1 \to 1.1\}$ and $\{1.0 \to 0.08, 1.0 \to 0.25, 1.0 \to 0.75, 1.0 \to 1.1\}$.
Unsurprisingly, the \textit{dynamical transition} simulations exhibit a transient phase, as illustrated in Figure~\ref{fig:energy_evo}(b). Zonal flow reorganization, whether generation or collapse typically occurs within $t\leq 200$, allowing the system to reach a new quasi-steady-state by approximately $t\sim 300$. Notably, this evolution is relatively slow compared to the accompanying turbulence dynamics (often $\tau\sim 1-10$), highlighting the multi-temporal scale nature of the model.

Despite only slight modifications to the original governing equations, the dynamics within the MHW model are significantly richer and more complex. For example, as the adiabaticity increases from the hydrodynamic limit (e.g., $\alpha=0.1$) to the near-adiabatic regime (e.g. $\alpha=1.0$), the system accesses a broader range of energies and fluxes due to enhanced flow shearing. Furthermore, it exhibits a profound shift in energy partitioning, transitioning from a state dominated by non-zonal density perturbations (i.e., internal energy) to one dominated by zonal flows (i.e., kinetic energy). In contrast, these key transport quantities in the original HW model vary slowly and smoothly with $\alpha$, exhibiting no qualitative changes (see Figure~\ref{fig:energy_flux}). 
More importantly, the turbulent behavior becomes spatially inhomogeneous. Even in the low-adiabaticity limit, the presence of small but finite zonal flows breaks spatial symmetry, modifying the spatial distribution of the autocorrelation time. As the adiabaticity increases, zonal flows become stronger and more localized, resulting in striated patterns in both the autocorrelation times and the characteristic length scales of the perturbed quantities that spans a wider dynamic range. Physically, this corresponds to the co-existence of the short-lived, elongated eddies in regions of strong shear and the long-lived, highly coherent structures in regions of weak shear. Consequently, capturing these dynamical transitions across the parameter space is non-trivial. A detailed description of the analysis methodology and results is provided in Appendix A.

\section{Training Strategy}\label{sec:training}

The data were uniformly subsampled down to a resolution of $128\times128$ for computational efficiency. All three physical channels $[n, \phi, \varpi]$ and conditional channel $\alpha$ were presented to the model and were normalized independently. While the system formally only requires the density along with either the potential or vorticity with initial conditions to propagate forwards in time, we decided to provide both the potential $\phi$ and vorticity $\varpi=\nabla_\perp^2\phi$ to mitigate potential errors associated with spatial subsampling, and let the model learn the relationship between them from data. The adiabaticity paramater $\alpha$ was also spatially embedded using a simple 2D Gaussian profile in an effort to promote interpolation of $\alpha$-dependent dynamics by the model. 

Aside from ablating the additional geometry-aware features in the GNO encoder and decoder, we implemented mostly the same default model architecture and optimizer settings reported in the GAOT paper\cite{wen_geometry_2025} (note that some of these architectural choices differ from the code available on GitHub\cite{wen_geometry_2025} as of the writing of this paper). We also increased the number of transformer blocks from 5 to 10 in the ViT processor, which we found was necessary to capture the fine-grained turbulent dynamics in the density and vorticity fields, resulting in a $\sim$11M parameter model.

To handle the multi-timescale nature of this data, we employed a curriculum learning\cite{bengio_curriculum_2009} strategy, where we separated the model training into a pretraining stage on short trajectories from the \textit{steady-state} dataset and a finetuning stage on a mix of both the \textit{steady-state} and \textit{dynamical transition} datasets. 

In pretraining, we only used the \textit{steady-state} dataset comprising of the long 5,000 frame trajectories. We held out the simulations with $\alpha \in \{0.08, 0.25, 0.75, 1.1\}$, leaving 10 independent trajectories, each corresponding with a different value of $\alpha$. This allowed us to test the model's ability to interpolate and extrapolate in $\alpha$. The remaining simulations were segmented into short trajectories comprising of 19 frames each, resulting in 263 segmented trajectories for each $\alpha$. Using a train/val/test split of 2374/128/128, the model was trained up to a maximum lead time of $T = 12$ with a minimum timestep of $\Delta\tau = 2$, holding out the last 6 frames of each trajectory for time extrapolation testing. As this framework uses the ``all2all'' training strategy\cite{herde_poseidon_2024}, this corresponds with a data enhancement factor of 21. The pretraining process takes about 38 hours on one NVIDIA A100 (80 GB) GPU.

For finetuning, we used a mix of both the \textit{steady-state} and \textit{dynamical transition} datasets. Just like in pretraining, all simulations with $\alpha \in \{0.08, 0.25, 0.75, 1.1\}$ were held out. This left trajectories with $\alpha\in \{0.1,0.2,0.3,0.4,0.5,0.6,0.7,0.8,0.9,1.0\}$ for \textit{steady-state} and $\alpha\in\{0.1\leftrightarrow0.4,0.1\leftrightarrow0.7,0.1\leftrightarrow1.0,0.4\leftrightarrow0.7,0.4\leftrightarrow1.0,0.7\leftrightarrow1.0\}$ for \textit{dynamical transition}. Only the first 500 frames of each of the \textit{steady-state} simulations were included in the mix, as this is the full duration of the \textit{dynamical transition} simulations. This yields a roughly even mix of the original \textit{steady-state} and new \textit{dynamical transition} data, which is a simple way to prevent catastrophic forgetting \cite{french_catastrophic_1999}.

We split the finetuning into two stages; a short-trajectory stage with segmentation identical to the pretraining procedure, and a long-trajectory stage.
For long-trajectory training, we segmented the data into trajectories of 96 frames each, producing 5 trajectories for each case. We used a 86/16/8 split, with a maximum lead time of $T = 72$ and a minimum timestep of $\Delta\tau = 6$, holding out the last 24 frames for time extrapolation tests. As before, this results in an ``all2all'' data enhancement factor of 21. Learning rates were reduced by a factor of 100 for finetuning. The total finetuning process takes about 21 hours on one NVIDIA A100 (80 GB) GPU (stage one 1 hour, stage two 20 hours).

\section{Results}\label{sec:results}

In this section, we first focus on short-term trajectory prediction for durations up to $12\omega_\text{ci}^{-1}$. Depending on the plasma parameters and spatial region, this prediction window is comparable to, or several times longer than, the local perturbation correlation time. Model performance is evaluated for two distinct scenarios: within the saturated, statistically stationary turbulence regime and during the transient evolutionary phase. Subsequently, we push the models to perform long-time horizon forecasting up to $72\omega_\text{ci}^{-1}$, a substantially more challenging task for chaotic turbulent systems. To maintain a concise results section, only several essential figures are included in the main content, while supplementary visualizations are provided in Appendix B.

\subsection{Short trajectories}

Here we present results from the pretrained and finetuned models evaluated on short trajectory $\tau\leq12$ prediction tasks. All results here were produced using autoregressive rollouts with $\Delta\tau=2$. The model was tested on new initial conditions and held-out adiabaticities to examine its ability to interpolate ($\alpha\in\{0.25,0.75\}$) and extrapolate ($\alpha\in\{0.08,1.1\}$) adiabaticity-dependent \textit{steady-state} turbulent dynamics, as well as on some held-out $\alpha$'s from the \textit{dynamical transition} dataset. 

Because the autocorrelation time depends strongly on the adiabaticity and the initial turbulent/zonal energy partition, \textit{any} model's point-wise prediction performance is expected to degrade at lower $\alpha$ values and smaller zonal energy fractions due to stronger chaotic mixing. Quantitatively, a lead time of $\tau=12$ in the \textit{steady-state} phase corresponds to several autocorrelation times for $\alpha=0.1$, but only about one autocorrelation time for $\alpha=1.0$ in regions of weak shear (see Figures~\ref{fig:scale_mhw_alpha01} and \ref{fig:scale_mhw_alpha10} in Appendix A). 

\subsubsection{Steady-state turbulence}
\begin{figure}[!htbp]
    \centering
    \includegraphics[width=\linewidth]{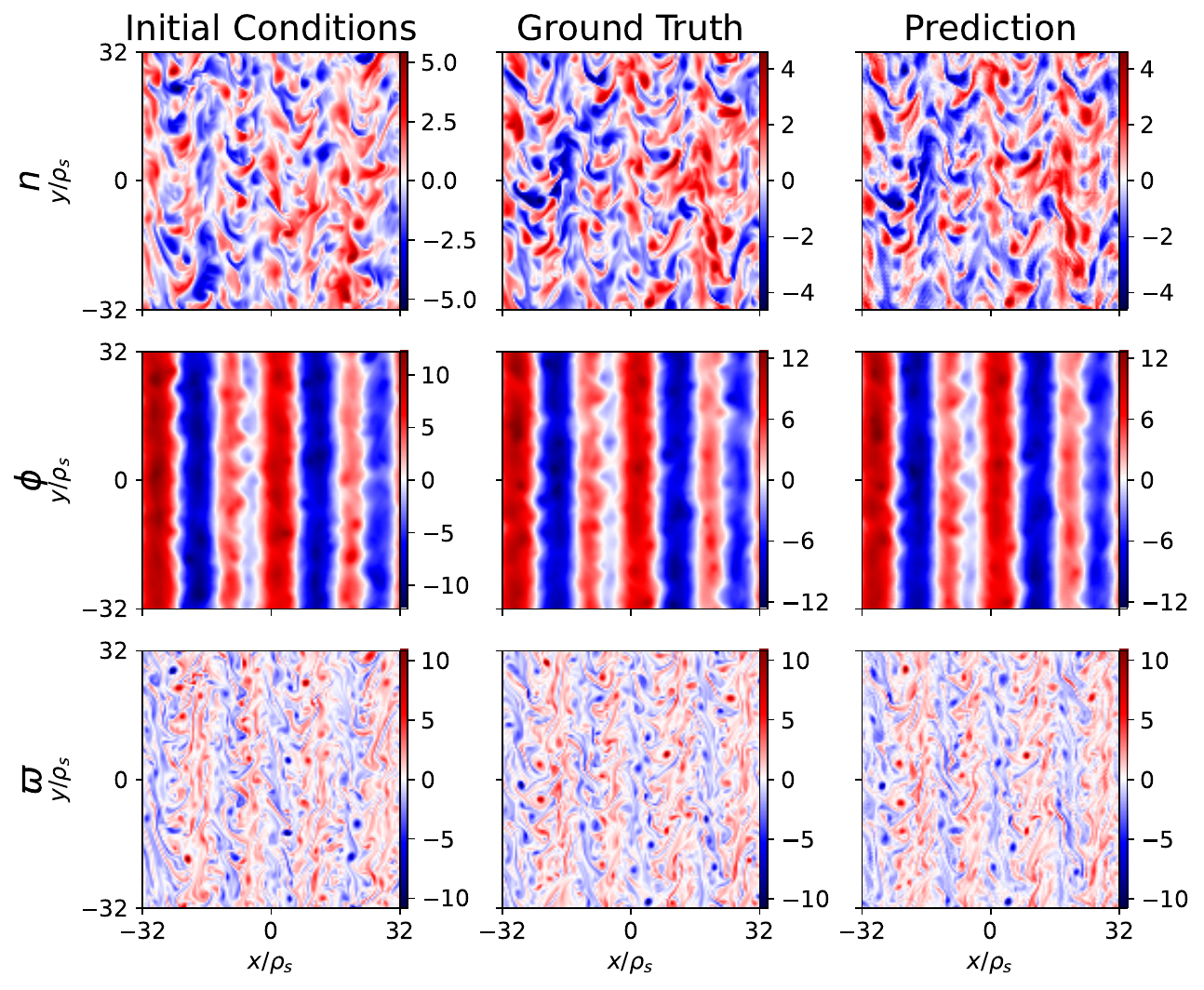}
    \caption{Initial conditions, ground truth and pretrained model prediction after $12\omega_\text{ci}^{-1}$ for an $\alpha=0.4$ \textit{steady state} trajectory.}
    \label{fig:Fig2_04}
\end{figure}

\begin{figure}[!htbp]
    \centering
    \includegraphics[width=\linewidth]{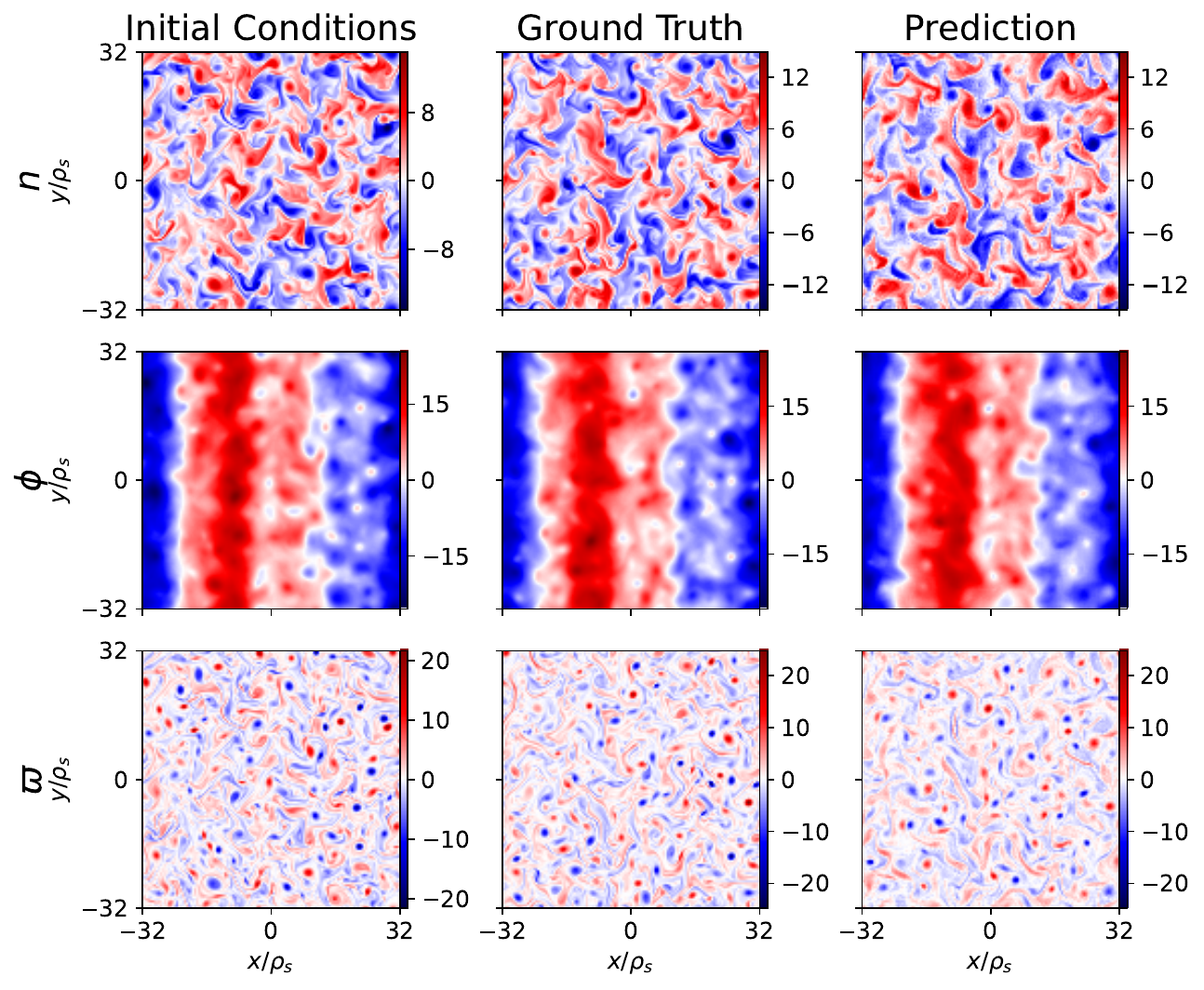}
    \caption{Initial conditions, ground truth and pretrained model prediction after $12\omega_\text{ci}^{-1}$ for an $\alpha=0.2$ \textit{steady state} trajectory.}
    \label{fig:Fig2_02}
\end{figure}

Model performance is first assessed using the reserved test set from pretraining (i.e., the held-out \textit{steady-state} turbulence data). In Figure \ref{fig:Fig2_04}, we see that the pretrained model can provide good point-wise predictions for the $\alpha=0.4$ trajectory at $\tau=12$, as the system is dominated by zonal flows and is therefore sufficiently predictable. Conversely, Figure \ref{fig:Fig2_02} shows that the $\alpha=0.2$ trajectory (a turbulence dominated system) is too chaotic for direct prediction up to this time. Despite this chaotic nature, the energy partition of each component does not undergo any quantitative changes, and the predicted turbulent flux distribution, for instance, agrees well with the DNS even at $\tau=12$. This is likely because as a neural operator autoregressively rolled out with a sufficiently small step $\Delta\tau=2$, the model still manages to produce a physically reasonable prediction that is strongly correlated with the simulation result. Therefore, the pretrained model is able to produce a statistically valid solution in both cases, demonstrating a robust capability for emulating steady-state turbulence within these short time trajectories.

Building on the model's success with the standard test set, we next evaluate its capacity to generalize to unseen adiabaticities. For the held-out interpolation cases ($\alpha \in \{0.25, 0.75\}$), both the pretrained and finetuned models maintain strong predictive performance (Figures \ref{fig:Fig2_025long} and \ref{fig:Fig2_075long} in Appendix B). This demonstrates that a single, unified model successfully learns a continuous representation of the $\alpha$-dependent, non-trivial, turbulent dynamics. It is particularly noteworthy that the model succeeds at $\alpha=0.25$; as illustrated in Figure~\ref{fig:energy_flux}(a), this represents the system transitioning from a turbulence-dominated to a zonal-flow-dominated state (marked by the intersection of the zonal flow and density fluctuation energies). Somewhat surprisingly, even when predicting the out-of-distribution extrapolation cases ($\alpha \in \{0.08, 1.1\}$, or Figures~\ref{fig:Fig2_008} and~\ref{fig:Fig2_11} in Appendix B), we observe similarly strong performance, especially for the more predictable $\alpha=1.1$ case.

Importantly, although the finetuning procedure is primarily designed to optimize performance for dynamical transitions and extended rollout horizons, the finetuned model performs as well as the pretrained model in these out-of-distribution regimes. These results confirm that careful finetuning introduces no performance penalty for short-trajectory \textit{steady-state} predictions.

\begin{figure}[!htbp]
    \centering
    \includegraphics[width=0.49\linewidth]{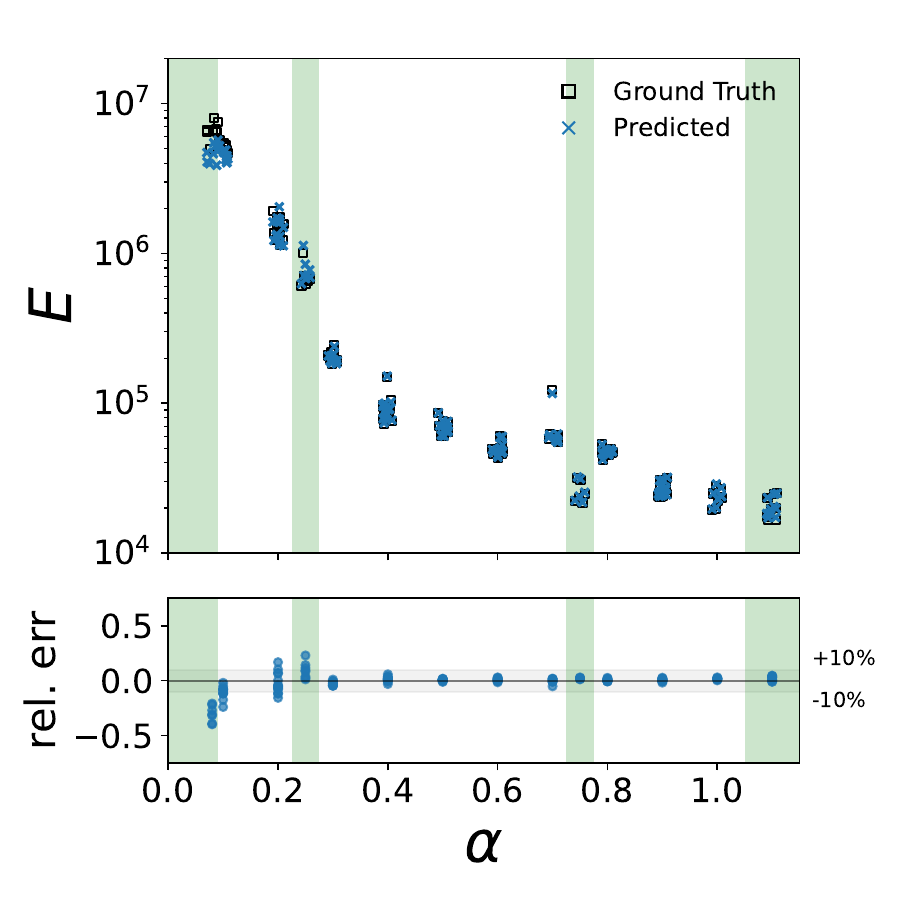}
    \includegraphics[width=0.49\linewidth]{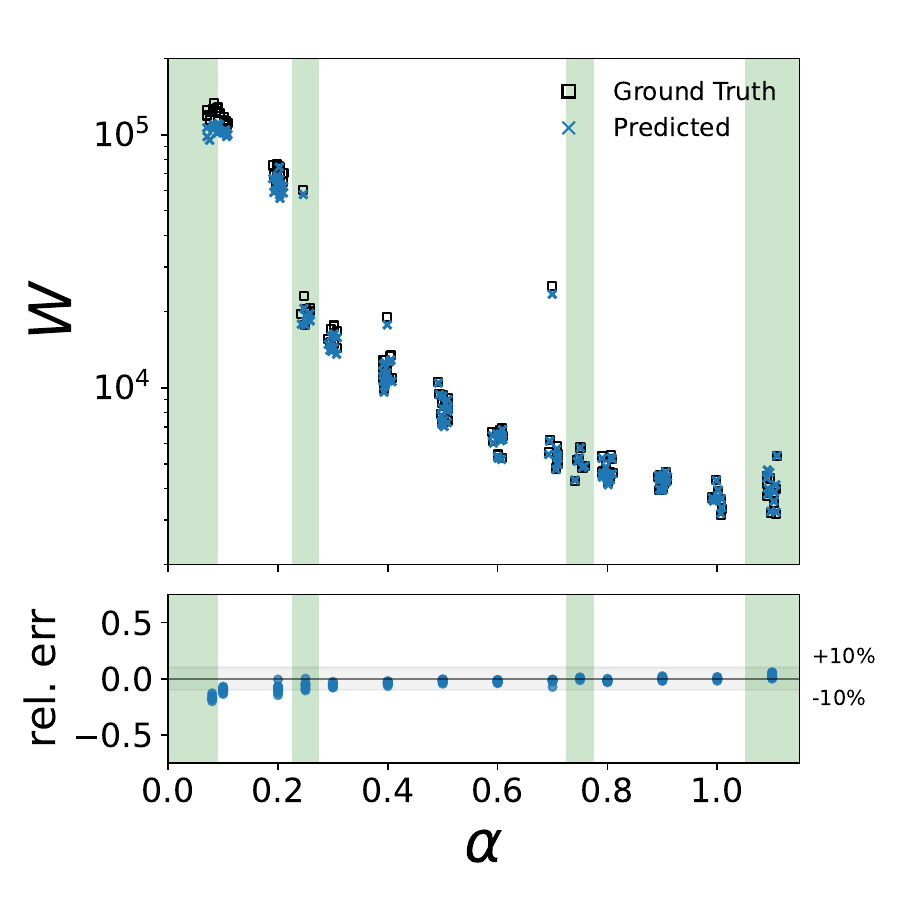}
    \includegraphics[width=0.49\linewidth]{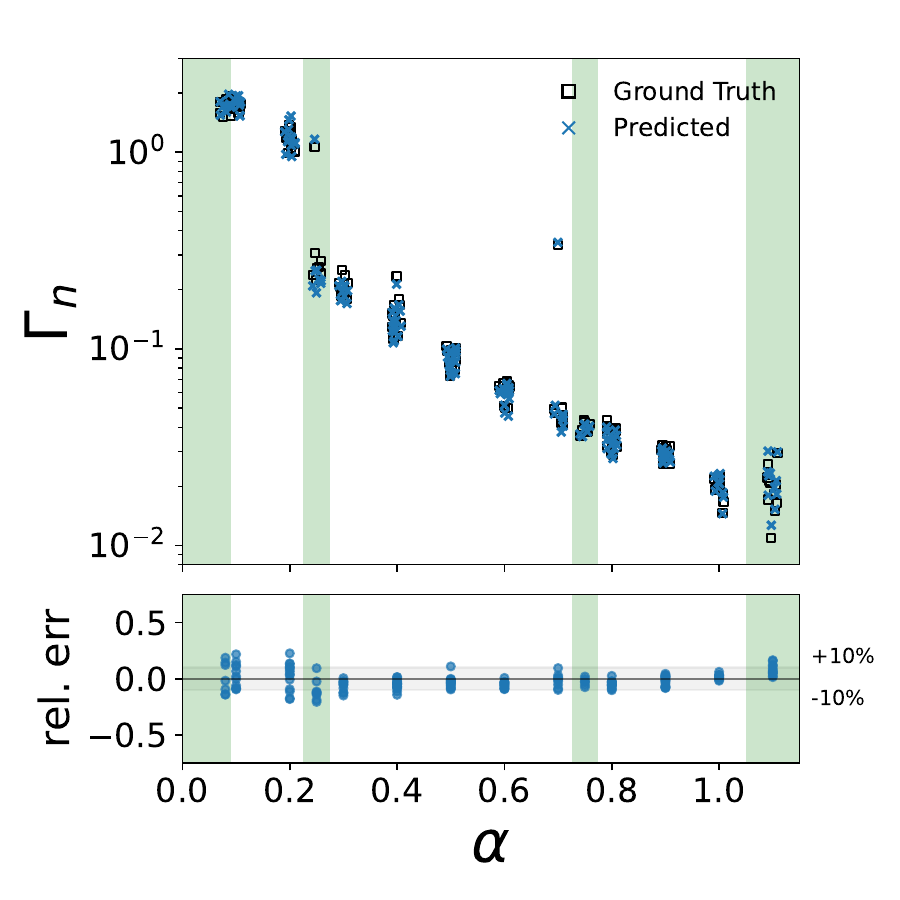}
    \includegraphics[width=0.49\linewidth]{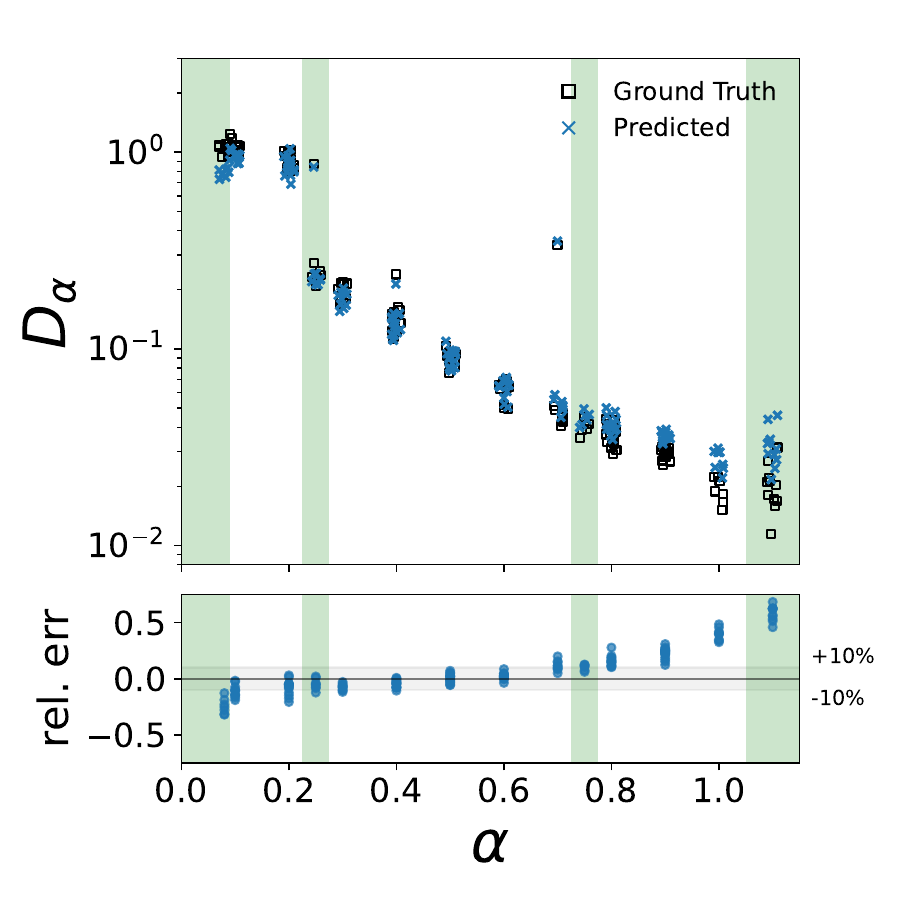}
    \caption{Pairwise comparisons between the ground truth and pretrained model predictions of 96 random test trajectories, including held-out $\alpha\in\{0.08,0.25,0.75,1.1\}$ (green shaded regions), after $12\omega_\text{ci}^{-1}$ of the total energy $E$, enstrophy $W$, turbulent flux $\Gamma_n$ and resistive dissipation $D_\alpha$.}
    \label{fig:Fig2_stats}
\end{figure}

The results for the full range of $\alpha$ is summarized in Figure \ref{fig:Fig2_stats}, where we evaluate the pretrained model performance based on the metrics $E$, $W$, $\Gamma_n$ and $D_\alpha$ across all randomly selected test cases. Note that due to this random sampling, a few test cases, such as one instance at $\alpha=0.7$, fall outside the saturated turbulent steady-state phase and instead within the nonlinear interaction phase; therefore, they do not cluster with the rest of the data for that $\alpha$. Nevertheless, we keep them here for completeness. Overall, the pretrained model performs exceptionally well, with the predictions (crosses) lying close to the DNS ground truth (squares) at $12\omega_\text{ci}^{-1}$ (often within 10\% relative errors). The small $\alpha<0.3$ cases generally have a larger relative error, which is expected due to their more turbulent behavior. Otherwise, model performance is largely insensitive to $\alpha$ for all of the metrics except for the apparent resistive dissipation $D_\alpha$, which tends to be overestimated with larger $\alpha$. Despite this, the total energy $E$ remains excellent. We will revisit this in the next section as it pertains to long trajectories. We also observe that the held-out interpolation $\alpha\in\{0.25,0.75\}$ and extrapolation $\alpha\in\{0.08,1.1\}$ also do pretty well on this short trajectory prediction task. 

\subsubsection{Dynamical evolution}

\begin{figure}[!htbp]
    \centering
    \includegraphics[width=\linewidth]{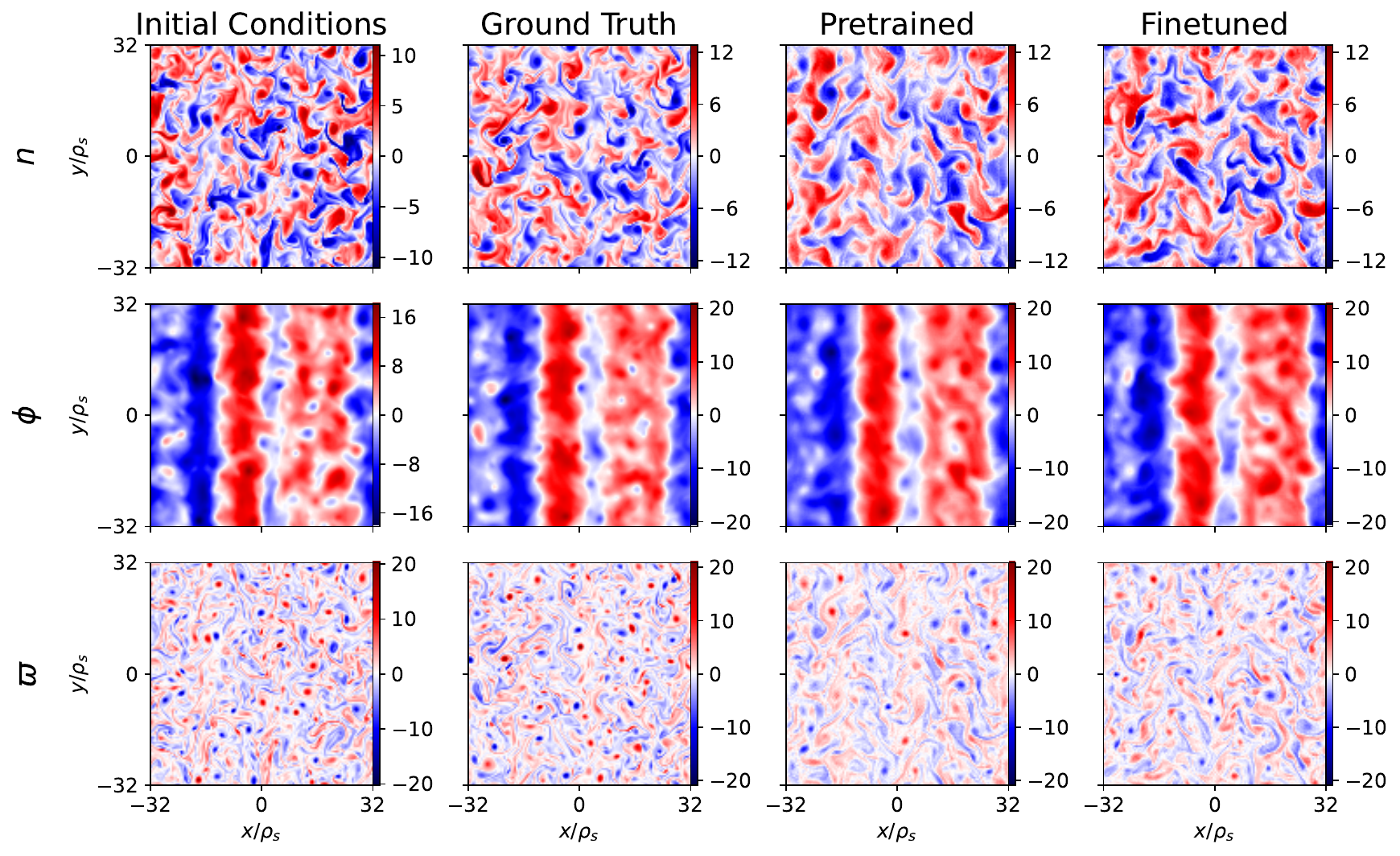}
    \caption{Initial conditions, ground truth, pretrained and finetuned model predictions after $12\omega_\text{ci}^{-1}$ for a held-out $\alpha=0.25$, $\omega_\text{ci}t_0=266$ trajectory in the nonlinear interaction phase of the \textit{steady-state} dataset.}
    \label{fig:Fig2_025}
\end{figure}

\begin{figure}[!htbp]
    \centering
    \includegraphics[width=\linewidth]{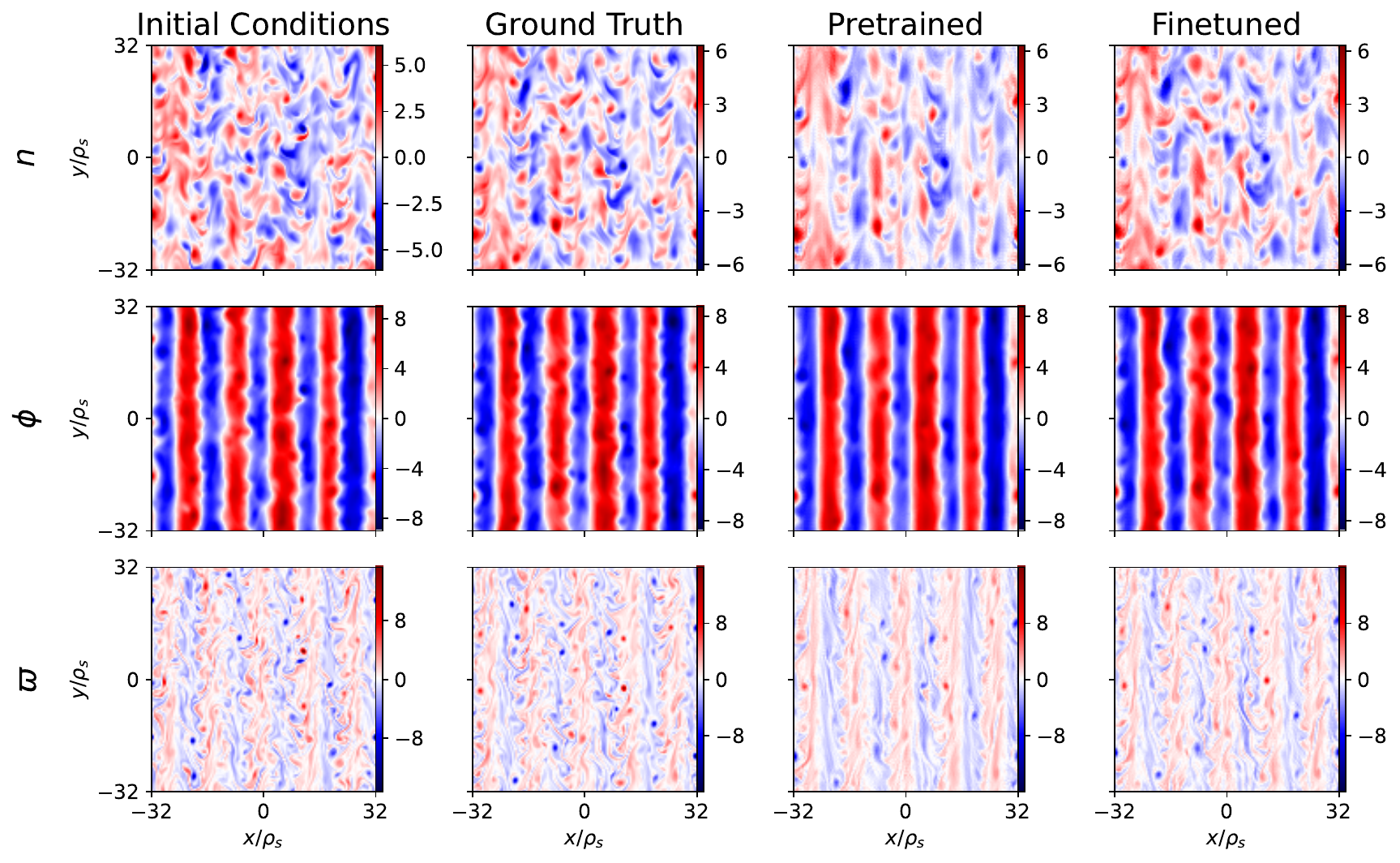}
    \caption{Initial conditions, ground truth, pretrained and finetuned model predictions after $12\omega_\text{ci}^{-1}$ for a held-out $\alpha=0.75$, $\omega_\text{ci}t_0=228$ trajectory in the nonlinear interaction phase of the \textit{steady-state} dataset.}
    \label{fig:Fig2_075}
\end{figure}

Expanding our evaluation to non-stationary dynamics, we test the models on the pre-saturation evolutionary phase of the \textit{steady-state} and the \textit{dynamical transition} test sets. For instance, Figures \ref{fig:Fig2_025} and \ref{fig:Fig2_075} demonstrate that for the unseen parameters $\alpha=0.25$ and $\alpha=0.75$, both the pre-trained and finetuned models qualitatively capture the evolving fields with high visual fidelity. Crucially, by $\tau=12\omega_{\text{ci}}^{-1}$, the ground truth has evolved substantially from the initial conditions, e.g. the shifts of maximum fluctuation amplitude. This confirms that the models are actively predicting the transient physics rather than merely generating a state from the same statistical distribution as the initial condition. While both models perform comparably in these scenarios, the finetuned model exhibits noticeably better accuracy, particularly in resolving sharp eddy structures.

Similarly, both pretrained and finetuned models successfully capture the gradual \textit{dynamical transition}s. For example, they accurately emulate the enhancement of zonal flows during an $\alpha=0.1 \to 0.25$ shift, and the enhancement of turbulence during an $\alpha=1.0 \to 0.75$ shift (Figures~\ref{fig:Fig2_01_025} and \ref{fig:Fig2_10_075} in Appendix B). This result is somewhat surprising for the pretrained model, given that it was never exposed to such transient scenarios during training. However, it could be argued that these transitions are relatively mild, as the system does not undergo a qualitative regime change. For instance, the $\alpha=1.0 \to 0.75$ test remains zonal-flow dominated throughout the simulation, and vice versa. Therefore, once the model has learned a continuous representation of the $\alpha$-dependent \textit{steady-state} dynamics, it can readily handle these unseen scenarios to a certain extent. In other words, the model treats the given initial condition as an out-of-distribution sample and naturally evolves the system toward the corresponding $\alpha$-dependent steady state.

An more intriguing scenario would involve a transition across two fundamentally distinct states. To address this, a more comprehensive analysis of dynamical transitions, expanding to more rapid parameter shifts and evaluated over extended prediction horizons (which inherently encompass these short-term dynamics and are the primary interests of this study), is presented in the following subsection.

\subsection{Long trajectories}

We now turn our attention to the prediction of long trajectories with lead times of $\tau\sim72\omega_\text{ci}^{-1}$, well beyond the autocorrelation times of any perturbations in the simulations considered in this study. Unless stated otherwise, the results presented here were produced using autoregressive rollouts with $\Delta\tau=2$.

The chaotic nature of this turbulent system over this long time horizon guarantees that a direct point-wise prediction is infeasible. However, having a surrogate model in this context is still useful not only for producing ensembles for statistical analysis, but also for capturing the long term \textit{dynamical transition} physics in response to changes in $\alpha$.

\subsubsection{Steady-state turbulence}

\begin{figure}[!htbp]
    \centering
    \includegraphics[width=\linewidth]{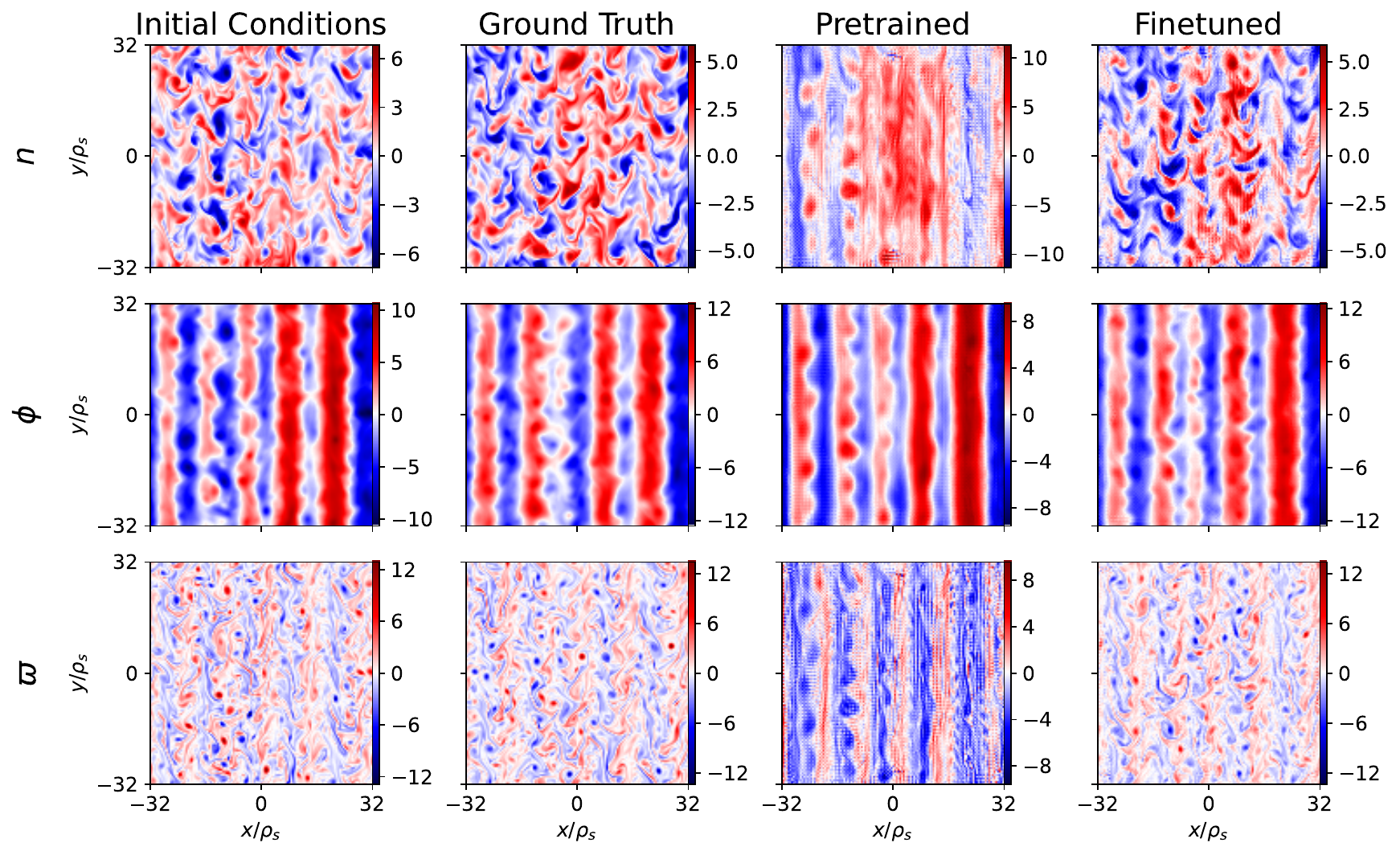}
    \includegraphics[width=\linewidth]{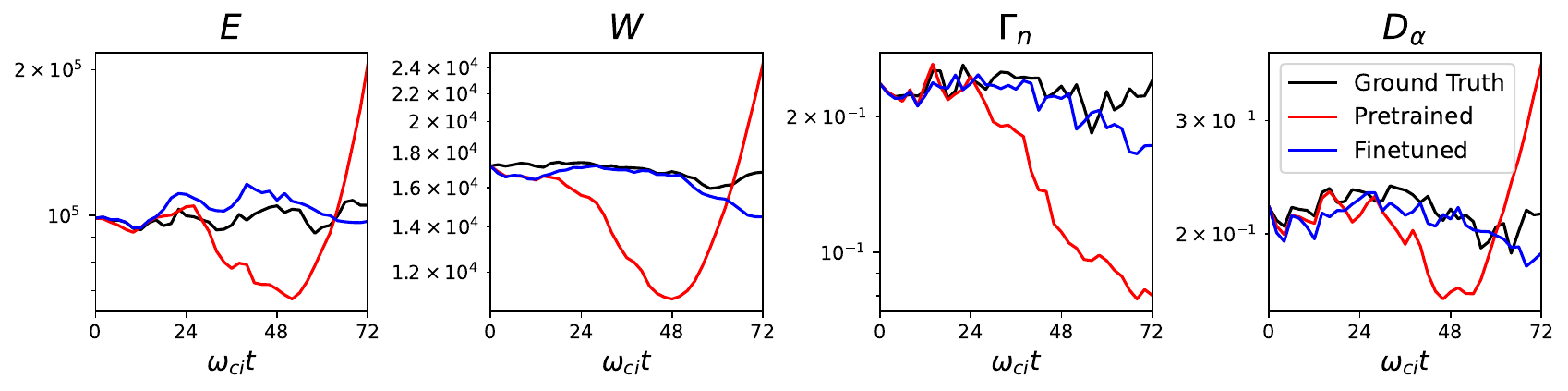}
    \caption{(Top) initial conditions, ground truth, pretrained and finetuned model predictions after $72\omega_\text{ci}^{-1}$ for \textit{steady-state} $\alpha=0.4$. (Bottom) time evolution of the energy $E$, enstrophy $W$, turbulent flux $\Gamma_n$ and resistive dissipation $D_\alpha$ from the ground truth, pretrained and finetuned models.}
    \label{fig:Fig3_04}
\end{figure}

We first establish the finetuned model's performance for long \textit{steady-state} prediction tasks. As shown in Figure \ref{fig:Fig3_04} where $\alpha=0.4$, the pretrained model produces unphysical results when extrapolated to $6\times$ the lead time it was originally trained on. Although the electrostatic potential $\phi$ appears visually reasonable, the corresponding density $n$ and vorticity $\varpi$ fields exhibit unrealistic zonal structures and grid-scale oscillations which are similar to numerical instability. This breakdown is also apparent when tracing its full trajectory prediction of the total energy $E$, generalized enstrophy $W$, turbulent flux $\Gamma_n$ and resistive dissipation $D_\alpha$. These key statistical metrics begin to drift from the DNS ground truth after $t \sim 24\omega_\text{ci}^{-1}$ and ultimately diverge after $t \sim 48\omega_\text{ci}^{-1}$. Meanwhile, the finetuned model is able to preserve the turbulent statistics and maintains coherent zonal flow structures. Its predicted evolution of the enstrophy and fluxes also remains close to the ground truth over this long time window.

\begin{figure}[!htbp]
    \centering
    \includegraphics[width=0.49\linewidth]{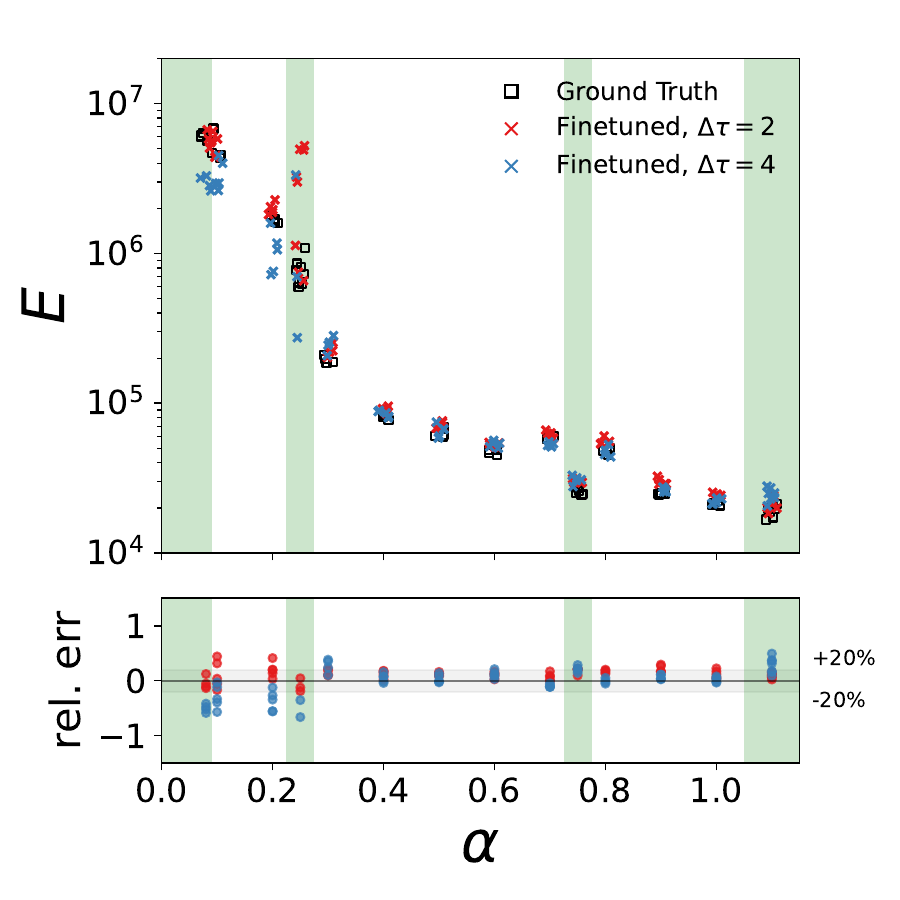}
    \includegraphics[width=0.49\linewidth]{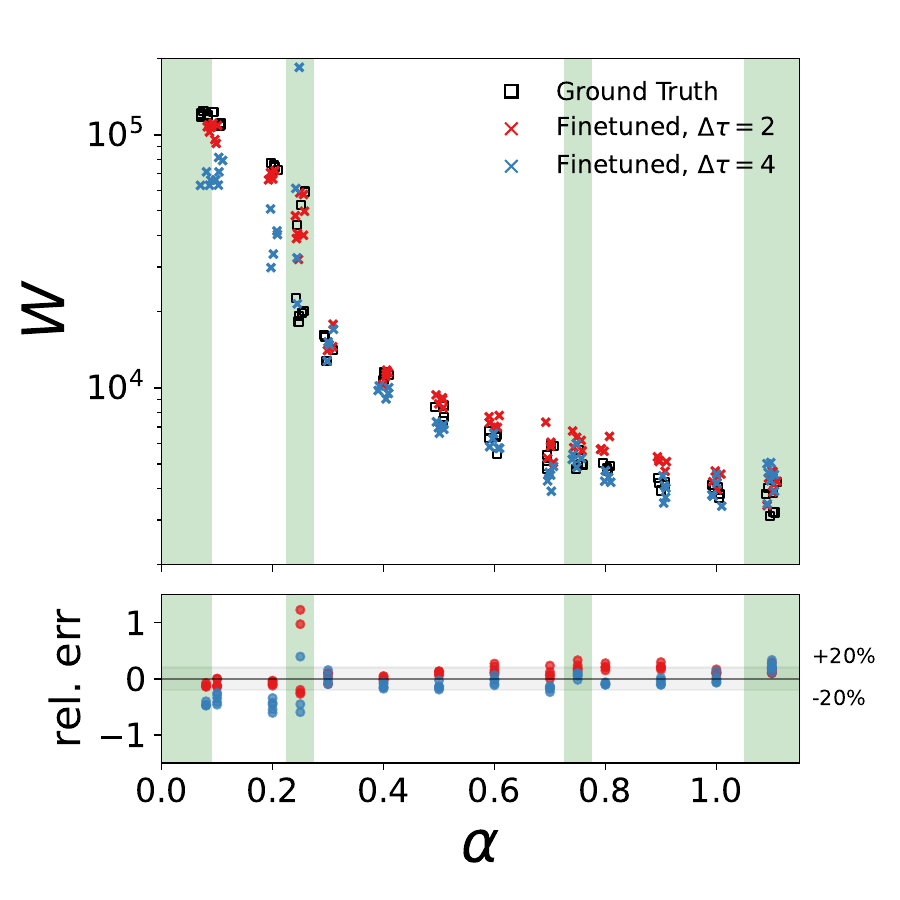}
    \includegraphics[width=0.49\linewidth]{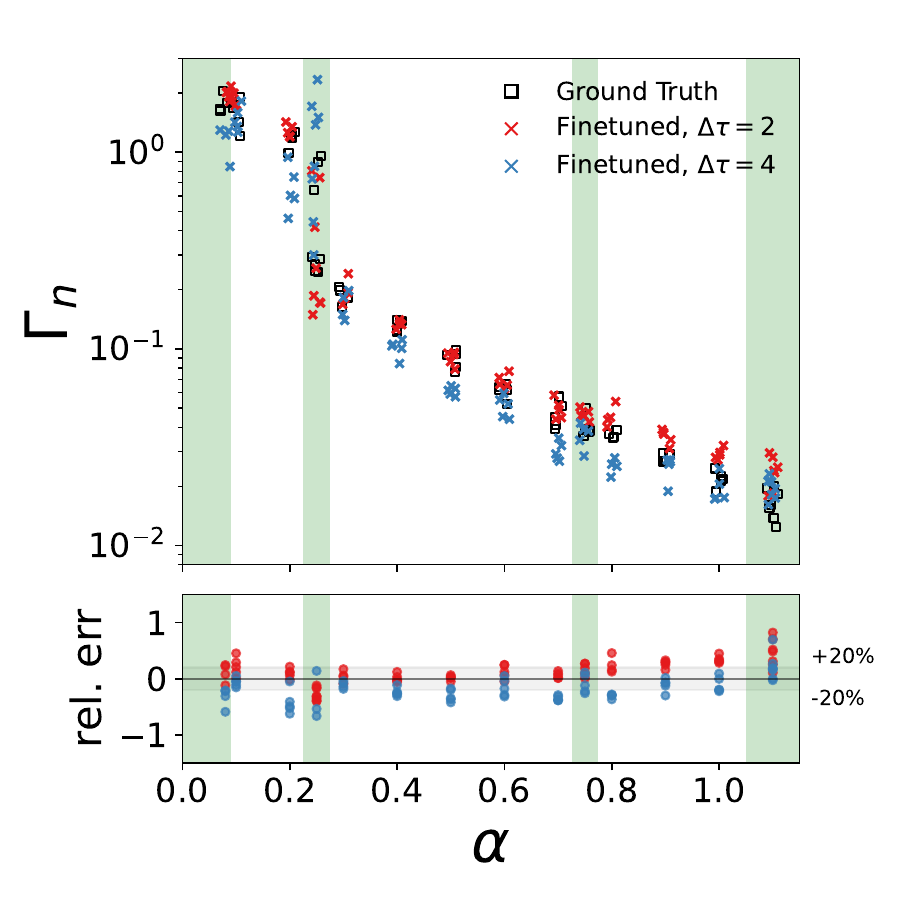}
    \includegraphics[width=0.49\linewidth]{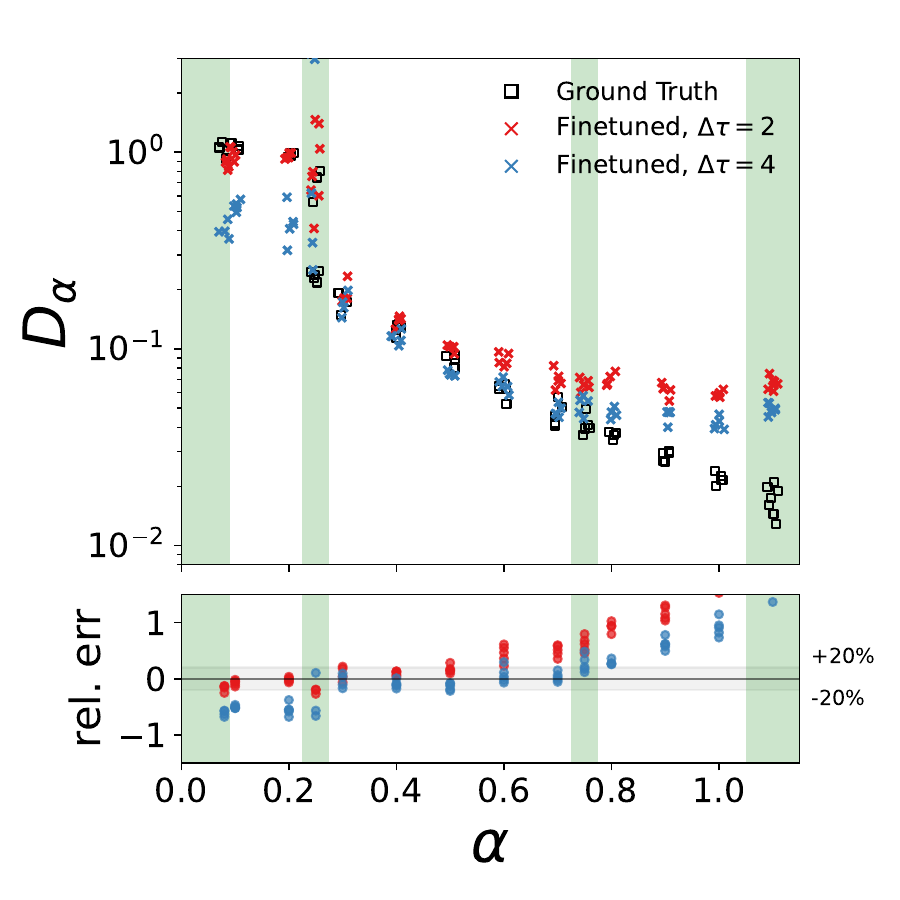}
    \caption{Pairwise comparisons between the ground truth and finetuned model predictions of 72 random held-out \textit{steady-state} test trajectories, including held-out $\alpha\in\{0.08,0.25,0.75,1.1\}$ (green shaded regions), after $48\omega_\text{ci}^{-1}$ of the total energy $E$, enstrophy $W$, turbulent flux $\Gamma_n$ and resistive dissipation $D_\alpha$, using autoregressive rollouts of timesteps $\Delta\tau=2$ (red) and $\Delta\tau=4$ (blue).}
    \label{fig:Fig2_stats_mid}
\end{figure}

During pretraining and finetuning, the original \textit{steady-state} database ($\alpha \in \{0.1, 0.2, \dots, 1.0\}$) was randomly sampled as short trajectories. To prevent potential data leakage and rigorously evaluate the finetuned model's performance across entire $\alpha$ range and long-horizon trajectories, we generated a completely independent test dataset by extending the steady-state simulations for an additional 500 simulation time units (i.e., from $t = 5000\omega_\text{ci}^{-1}$ to $5500\omega_\text{ci}^{-1}$). This data was entirely unseen during both pretraining and finetuning. The results are summarized in Figure~\ref{fig:Fig2_stats_mid}, which evaluates key statistical quantities at $t=48\omega_\text{ci}^{-1}$, a time horizon four times longer than what was originally seen in pretraining. We also compare the performance between autoregressive rollouts with timesteps $\Delta\tau\in\{2,4\}$. 

Focusing on the directly trained $\alpha$ values, at least one of the finetuned model variants performs well in predicting the total energy $E$ and generalized enstrophy $W$. Due to the nature of autoregressive rollouts, we expect a larger accumulated error over this long time horizon compared with our short-trajectory benchmarks. This introduces another design choice; we should always use the largest timestep $\Delta\tau$ possible to minimize both the accumulated error and reduce the required computational resources. For the highly turbulent scenarios ($\alpha=0.1$ and $0.2$), we cannot get around using the original $\Delta\tau=2$ timestep, as the model must be able to capture the short turbulent correlation times. In fact, the $\Delta\tau=2$ model outperforms its counterpart up to $\alpha=0.4$. However, as the system becomes more dominated by macroscopic zonal flows, employing the larger $\Delta\tau=4$ time step becomes highly advantageous, and in fact outperforms $\Delta\tau=2$ for $\alpha\gtrsim0.6$ due to the reduction in accumulated autoregressive errors. This performance discrepancy between the two time steps becomes even more pronounced when the predictions are extended to a full $t=72\omega_\text{ci}^{-1}$ horizon (Figure~\ref{fig:Fig2_stats_long} in Appendix B).

Perhaps unsurprisingly, the model struggles with the $\alpha=0.25$ interpolation task, as this happens to be right at the transition between the system tending to a turbulent or zonal state (i.e., the marginal scenario). The ground truth in this regime produces large and persistent fluctuations in potential $\phi$ with more than double the amplitude of the most extreme cases seen in the training set (see Figure \ref{fig:Fig2_025long} in Appendix B). This ends up being far too out-of-distribution with the data normalization procedure we used for the model to handle, leading to it producing unphysical results. This simply stresses the fact that this model, like any other machine learning model, requires careful assessment in parameter ranges where the physical solution is non-smooth or even discontinuous. Additional techniques such as active sampling could be employed to avoid or mitigate locally enhanced interpolation errors like this. 

In contrast, the performance of the other interpolated ($\alpha=0.75$) and two extrapolated scenarios ($\alpha \in \{0.08, 1.1\}$) remains fairly consistent. This is most likely because the macroscopic system behavior varies smoothly across these parameter regimes, allowing the model's learned continuous representation to extrapolate effectively to a certain extent.

We do note that the systematic overestimation of $D_\alpha$ at large $\alpha$ values accumulates further, though this is partially mitigated by performing rollouts with $\Delta\tau=4$ instead.
Interestingly, while the apparent resistive dissipation $D_\alpha$ is overestimated, the turbulent flux $\Gamma_n$ and, more importantly, the total energy $E$ remain accurate and stable, suggesting that the neural operator has learned a robust sub-grid internal representation of $D_\alpha$ during extended autoregressive rollouts.

\subsubsection{Nonlinear interaction}

Before moving on to the fully nonlinear \textit{dynamical transition} results, we first evaluate a prediction task during the nonlinear interaction stage. This phase is arguably even more challenging to predict than the dynamical transitions themselves, as the model must accurately capture both the linear instability growth rates and the nonlinear saturation mechanisms to correctly predict the linear to nonlinear transition timeline and fluctuation amplitude. In addition, this type of transient saturation phase is an exceptionally \textit{rare} event in our datasets (comprising $\sim 3\%$ of the pretraining data and $\sim 10\%$ of the finetuning data), making it a rigorous test of the model's ability to generalize rather than simply overfit to steady-state behaviors.

\begin{figure}[!htbp]
    \centering
    \includegraphics[width=\linewidth]{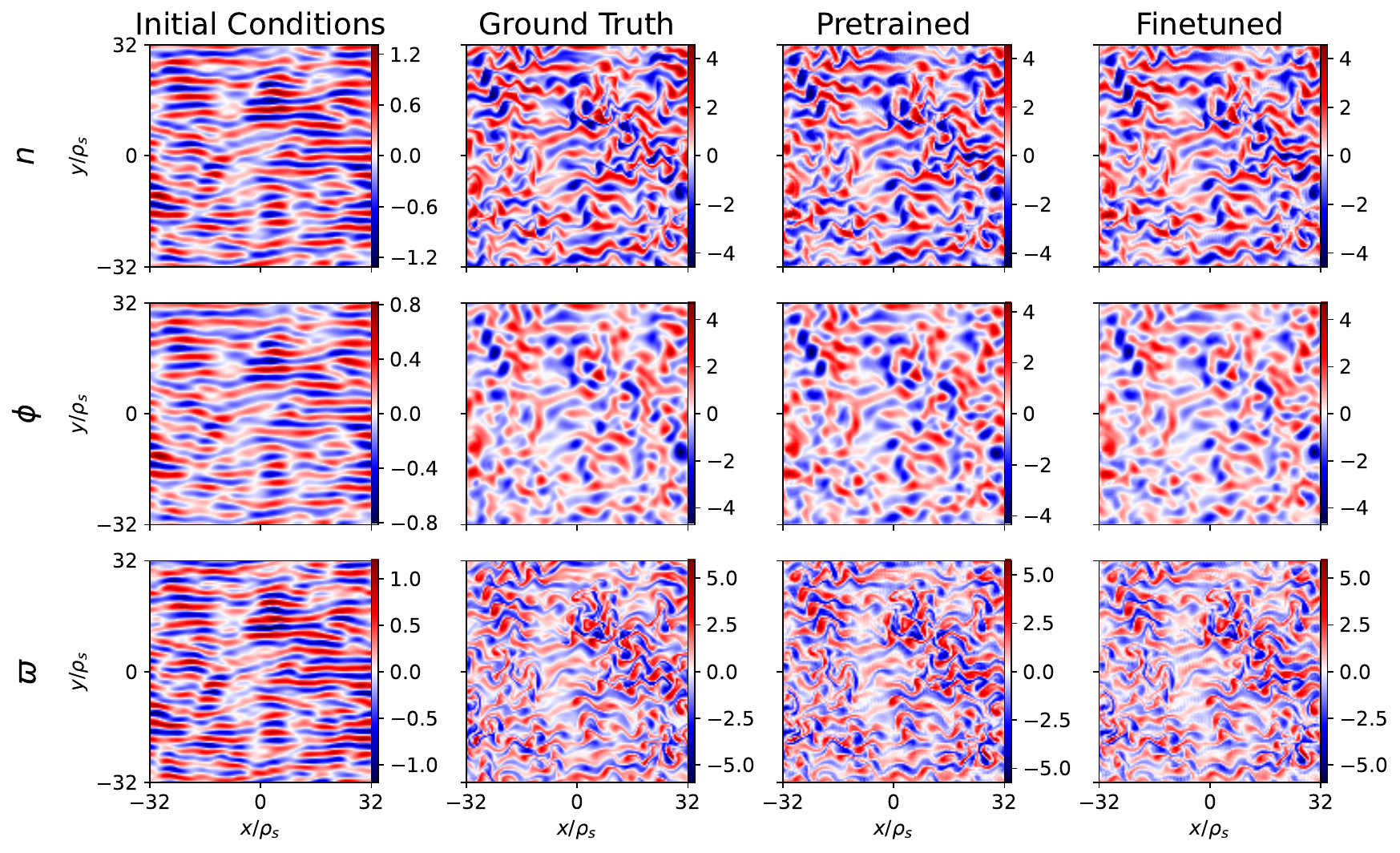}
    \includegraphics[width=\linewidth]{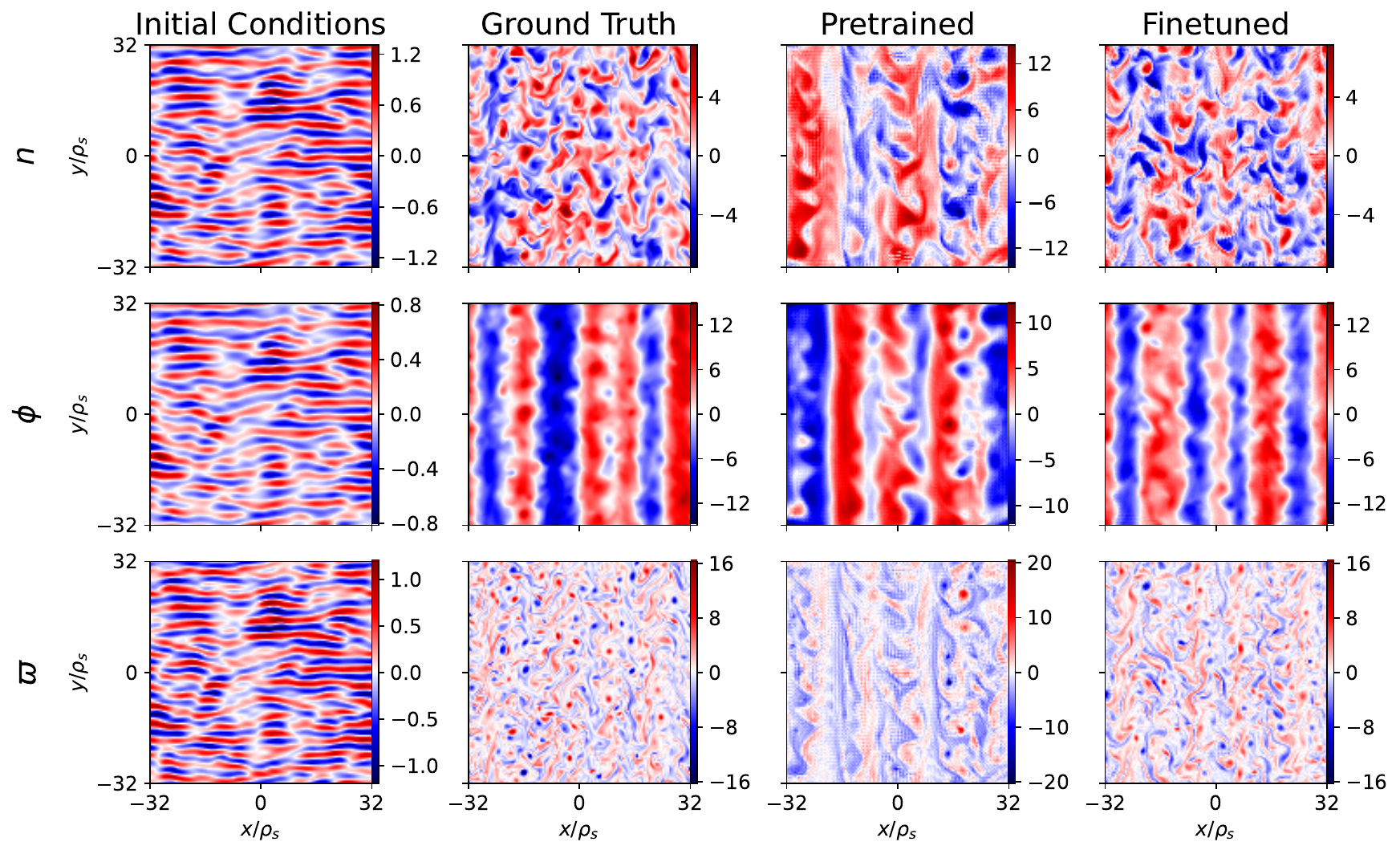}
    \includegraphics[width=\linewidth]{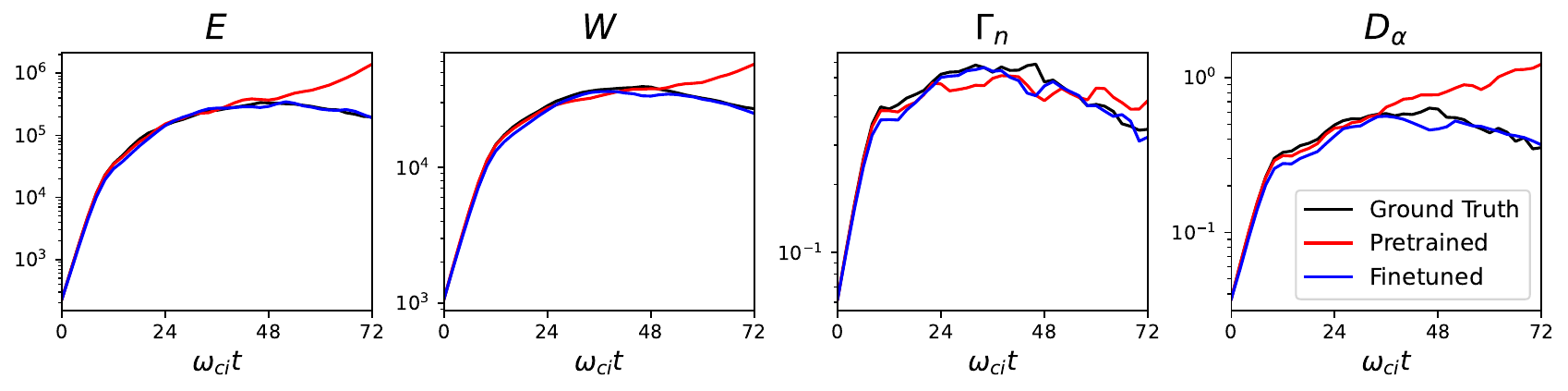}
    \includegraphics[width=\linewidth]{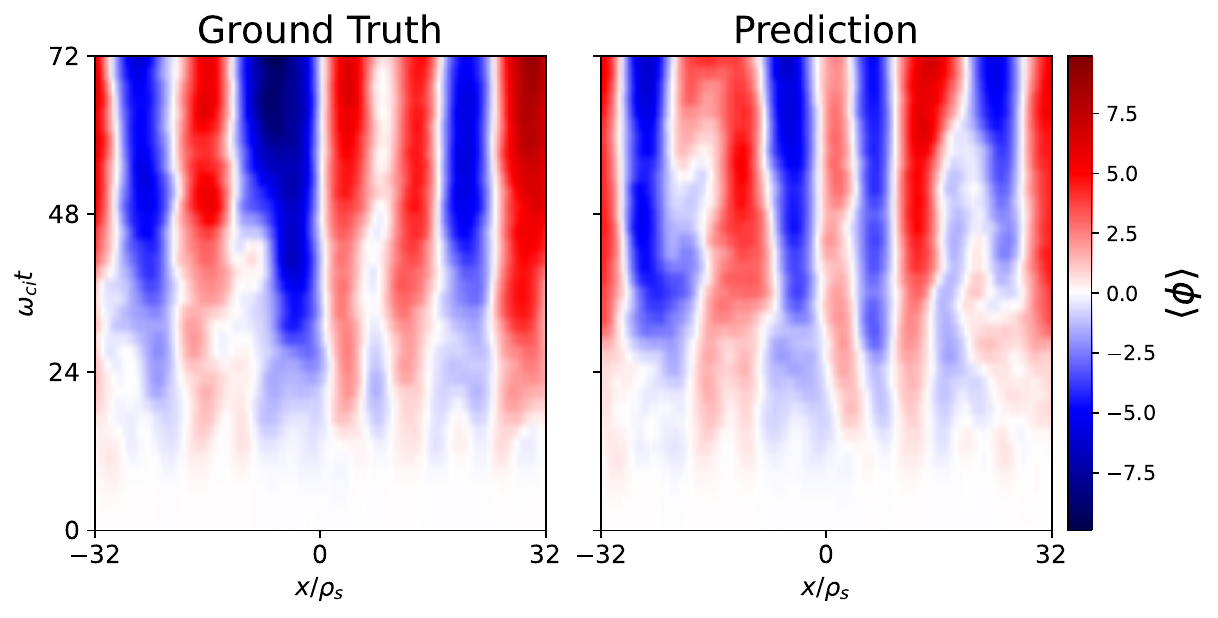}
    \caption{(Top) initial conditions, ground truth, pretrained and finetuned model predictions after $12\omega_\text{ci}^{-1}$ and $72\omega_\text{ci}^{-1}$ respectively for $\alpha=0.5$, $\omega_\text{ci}t_0=89$. (Middle) time evolution of the energy $E$, enstrophy $W$, turbulent flux $\Gamma_n$ and resistive dissipation $D_\alpha$ from the ground truth, pretrained and finetuned models. (Bottom) ground truth and finetuned model trajectories of the zonal component of the potential $\langle\phi\rangle$. }
    \label{fig:Fig3_05_weak}
\end{figure}

We acknowledge that the current model typically fails in the early linear stage due to internal model noise superseding or heavily influencing the small physical perturbations (e.g., amplitudes of $\sim10^{-4}-10^{-2}$). However, the model produces excellent predictions when initialized in the late-linear or weakly nonlinear phase, as shown in Figure \ref{fig:Fig3_05_weak} for an independently generated $\alpha=0.5$ test trajectory.

For shorter timescale predictions ($12\omega_\text{ci}^{-1}$), both the pretrained and finetuned models exhibit excellent point-wise accuracy, successfully tracking the rapid growth from the late-linear stage to the onset of weak nonlinearity. For the extended $72\omega_\text{ci}^{-1}$ predictions, the analysis of the evolving statistical properties reveals that both models align with the DNS well into the moderately nonlinear stage, successfully reproducing the initial turnover point. However, beyond $t \sim 36\omega_\text{ci}^{-1}$, the pretrained model begins to diverge and produces unphysical results. In contrast, the finetuned model maintains physically reasonable integrated quantities ($E$, $W$, $\Gamma_n$, and $D_\alpha$). Most notably, the finetuned model successfully captures the complete ``rollover'' in all these curves. This rollover is a critical indicator as it signifies the transient nonlinear saturation process: the initial generation of zonal flows, the subsequent suppression of turbulence, and the system's ultimate self-organization into a saturated state.  Capturing this complex sequence, despite its overall sparsity in the training data, demonstrates the architecture's powerful capacity to learn the physics of relatively rare transient events. While the exact point-wise structure of the emergent zonal flows is eventually lost due to the chaotic nature of the extended time horizon, the macroscopic zonal saturation structure (i.e., the dominant mode $k_x\simeq 4$) and its overall amplitude level are well captured.

\subsubsection{Dynamical transition}

Beyond the saturated turbulence and nonlinear interaction phases of the \textit{steady-state} simulations, the finetuned model is specifically trained to handle long-horizon \textit{dynamical transition} tasks. In these scenarios, the adiabaticity parameter $\alpha$ is abruptly shifted, forcing the system to transition between zonal-flow-dominated and turbulence-dominated states. For these problems, we are particularly interested in assessing whether the model can accurately capture the timescales over which zonal flows collapse and turbulence emerges (or vice versa), the transient dynamics of the transition process itself, and the final macroscopic zonal structure and saturation level.

As both pretrained and finetuned models have shown good performance on the short-term, gradual transition predictions, here we focus on significantly more challenging tasks where the system undergoes an abrupt bifurcation between two distinctly separated states.

\begin{figure}[!htbp]
    \centering
    \includegraphics[width=\linewidth]{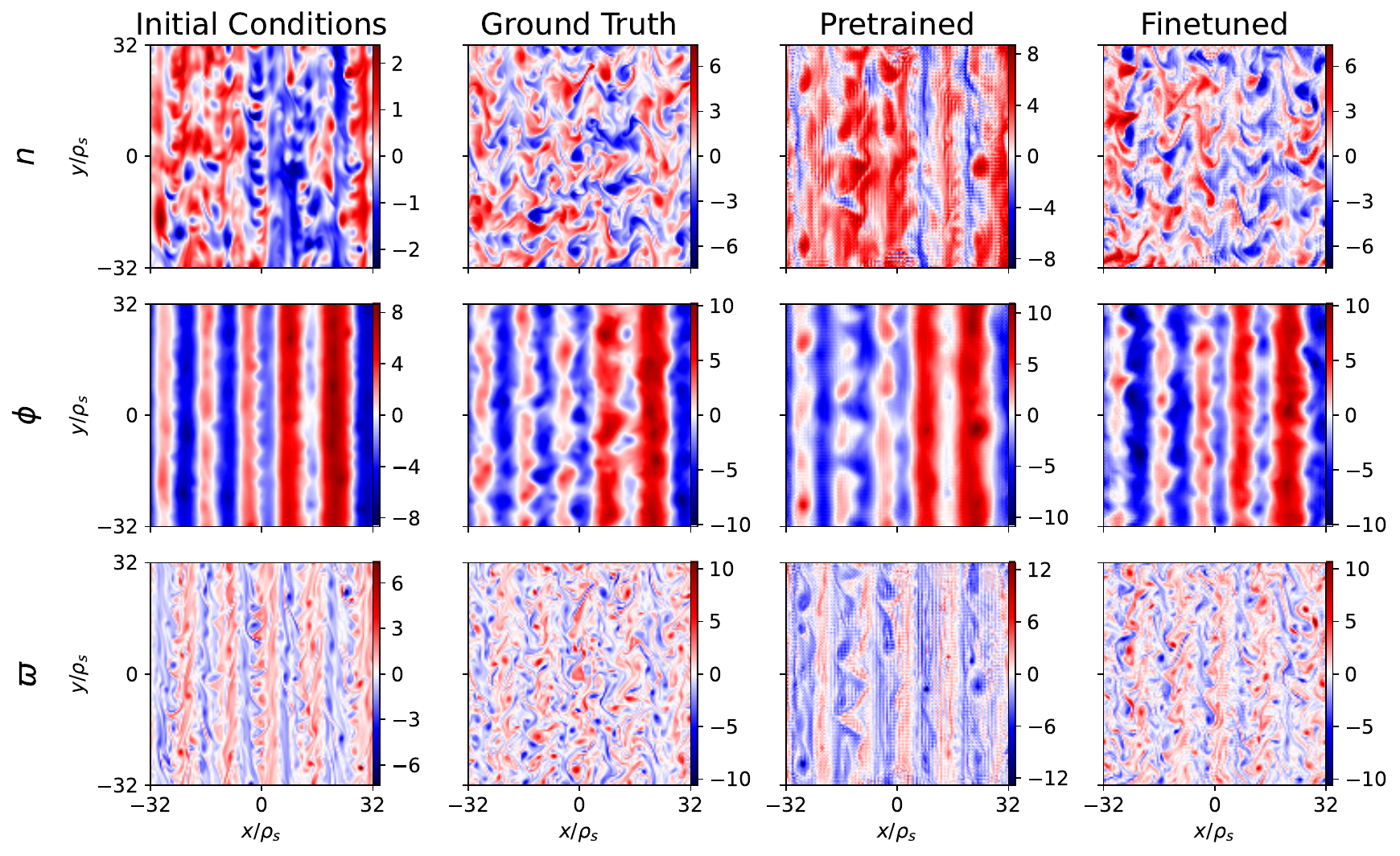}
    \includegraphics[width=\linewidth]{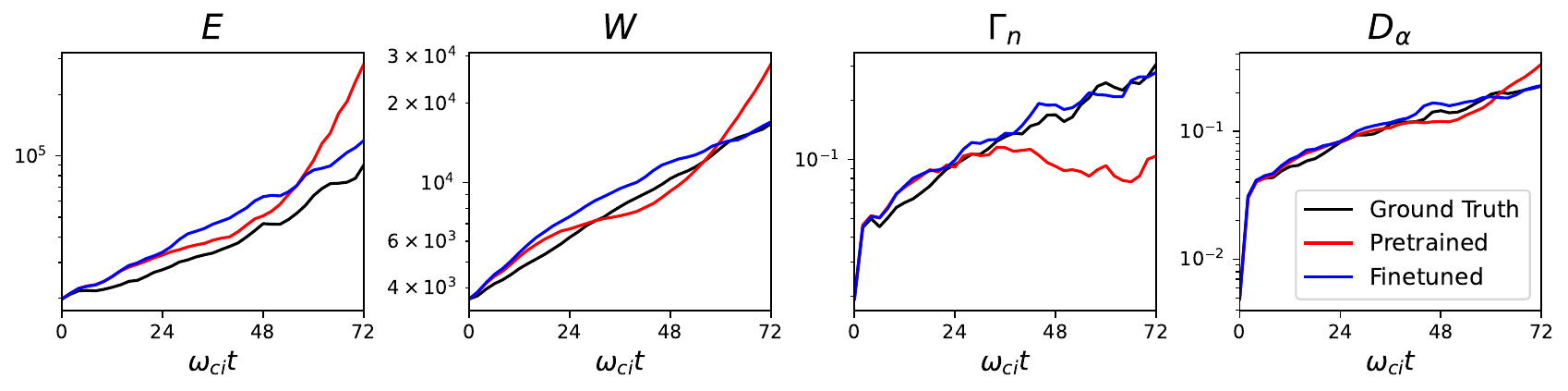}
    \includegraphics[width=\linewidth]{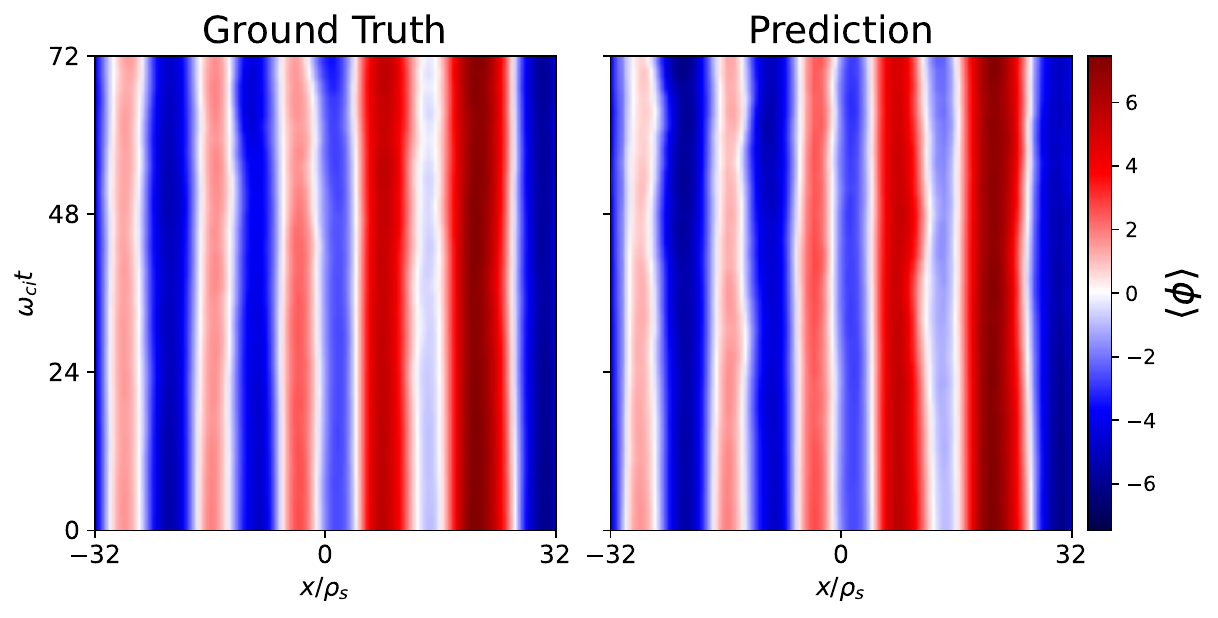}
    \caption{(Top) initial conditions, ground truth, pretrained and finetuned model predictions after $72\omega_\text{ci}^{-1}$ for held-out \textit{dynamical transition} $\alpha=1.0\rightarrow0.25$, $\omega_\text{ci}t_0=0$. (Middle) time evolution of the energy $E$, enstrophy $W$, turbulent flux $\Gamma_n$ and resistive dissipation $D_\alpha$ from the ground truth, pretrained and finetuned models. (Bottom) ground truth and finetuned model trajectories of the zonal component of the potential $\langle\phi\rangle$. }
    \label{fig:Fig3_10_025}
\end{figure}

We first examine a case of zonal flow suppression and turbulence enhancement (Figure \ref{fig:Fig3_10_025}), representing a held-out interpolation task from $\alpha=1.0 \to 0.25$ starting at $t=0$. Notably, $\alpha=0.25$ represents a marginal state located near the critical bifurcation point separating the zonal-dominated and turbulence-dominated regimes. Both the pretrained and finetuned models correctly capture the overall direction of the system's evolution, despite an initial discrepancy where both total energy $E$ and generalized enstrophy $W$ are slightly overestimated at the very first prediction step. Interestingly, the highly sensitive derived quantities, turbulent flux $\Gamma_n$ and resistive dissipation $D_\alpha$, experience a massive initial jump (by a factor of $4\times$ or more) due to the abrupt parameter shift, yet both models predict this massive jump accurately from the first step onward. Consistent with earlier long-trajectory tests, the pretrained model exhibits instability over extended horizons, with predictions deviating substantially after $t \sim 36\omega_\text{ci}^{-1}$; while the finetuned model maintains stability throughout the entire period, yielding a physically valid final prediction. The predicted zonal field maintains its overall structure with a slightly decreased amplitude, in good agreement with the DNS result.

\begin{figure}[!htbp]
    \centering
    \includegraphics[width=\linewidth]{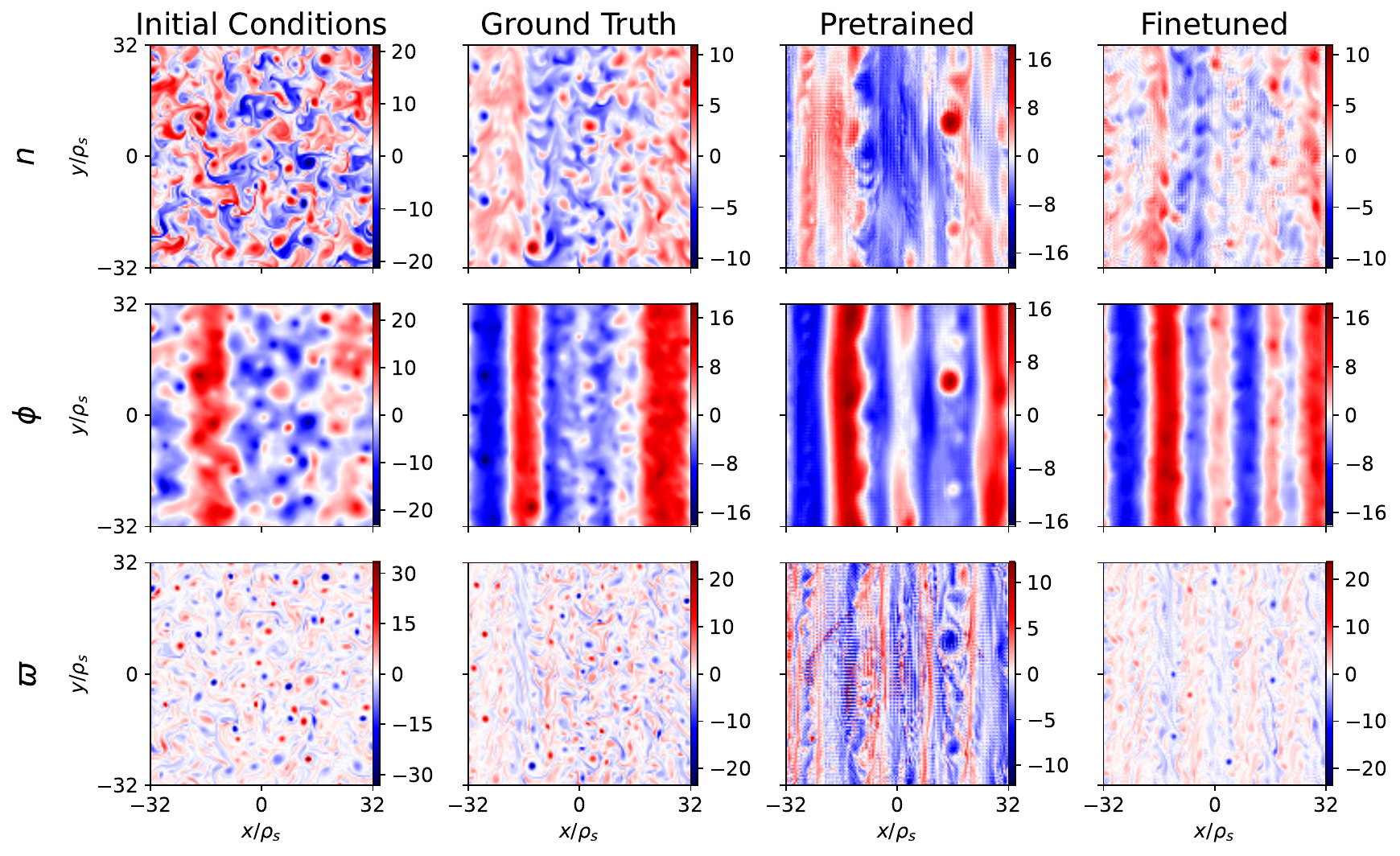}
    \includegraphics[width=\linewidth]{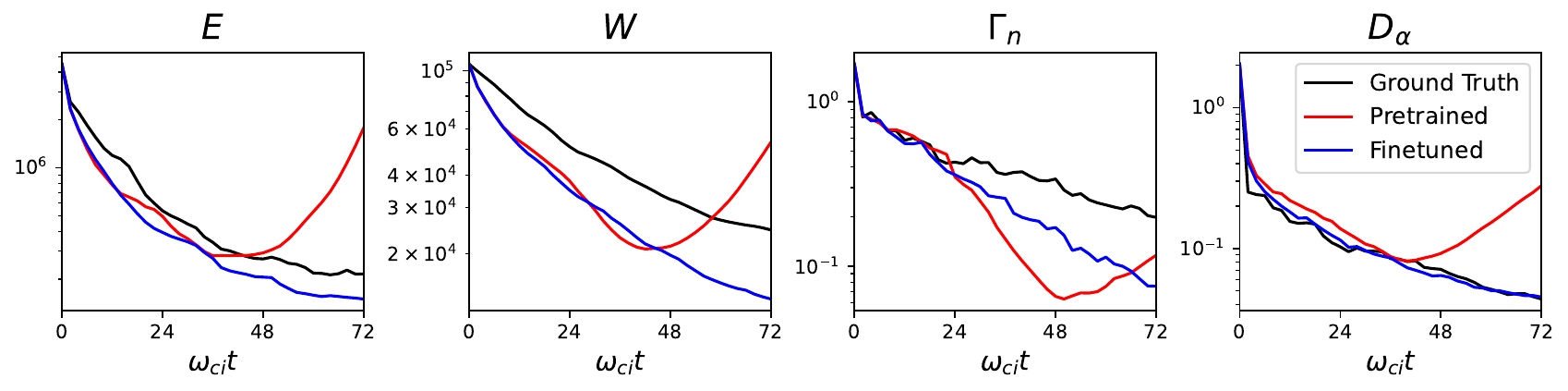}
    \includegraphics[width=\linewidth]{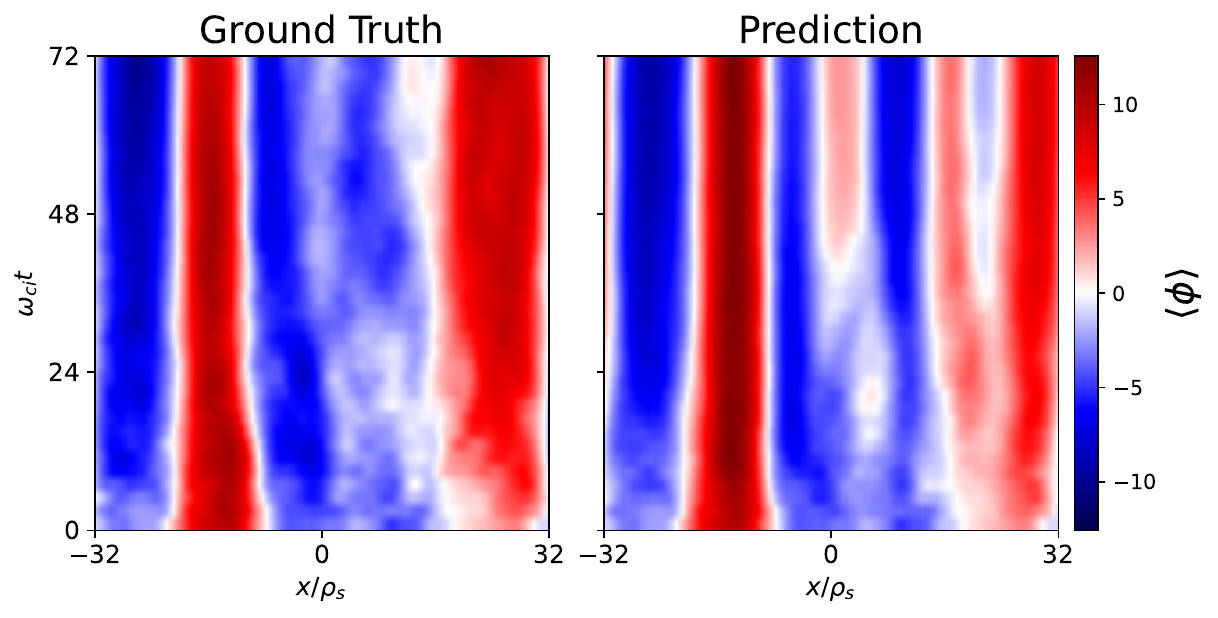}
    \caption{(Top) initial conditions, ground truth, pretrained and finetuned model predictions after $72\omega_\text{ci}^{-1}$ for \textit{dynamical transition} $\alpha=0.1\rightarrow1.0$, $\omega_\text{ci}t_0=0$. (Middle) time evolution of the energy $E$, enstrophy $W$, turbulent flux $\Gamma_n$ and resistive dissipation $D_\alpha$ from the ground truth, pretrained and finetuned models. (Bottom) ground truth and finetuned model trajectories of the zonal component of the potential $\langle\phi\rangle$. }
    \label{fig:Fig3_01_10}
\end{figure}

In Figure \ref{fig:Fig3_01_10}, we present a significantly more challenging generalization test: a transition from a turbulent state to a zonal-flow state via an $\alpha=0.1 \to 1.0$ shift at $t=0$. This scenario serves as an extrapolation task, having been entirely withheld from the training and validation sets.  To succeed, the model must self-consistently suppress turbulent fluctuations and spontaneously generate correct zonal flow structures at the correct rate based solely on the initial condition and the target $\alpha$. Note that this process unfolds on a timescale far exceeding the turbulence autocorrelation time. The finetuned model performs remarkably well, accurately capturing the final zonal flow saturation level. The predicted temporal evolution of the integrated quantities ($E$, $W$, $\Gamma_n$, $D_\alpha$) aligns closely with the ground truth. However, we observe a systematic bias toward lower total energy $E$ and generalized enstrophy $W$ in this scenario, which is the exact opposite of the trend seen in the $\alpha=1.0 \to 0.25$ case. We suspect this is most likely an artifact of the $4\times$ spatial downsampling, which disproportionately hinders the resolution of enstrophy evolution during vortex stretching, splitting and merging. Also, the overall underprediction of total energy relative to the DNS is qualitatively consistent with the turbulent flux (i.e., this system's energy source) which begins to underestimate after $t \sim 24\omega_{\text{ci}}^{-1}$. Moreover, the finetuned model predicts the splitting of the zonal field into finer structures, which is qualitatively confirmed by the DNS. This phenomenon is present in the DNS. Temporally, the splitting is quite consistent: initial zonal field seeds merge right after the rapid transition, and a meaning flow at $x=0$ appears near $t \sim 30\omega_\text{ci}^{-1}$. Spatially and amplitude-wise, however, there are notable deviations: the predicted zonal structure near $x=0$ is stronger and spatially shifted compared to the DNS, and an additional splitting event predicted near $x=16$ does not occur in the DNS (perhaps in a much later time). These exact point-wise discrepancies are physically expected. Given the inherently chaotic nature of the system, any small truncation errors or downsampling artifacts in the initial state will inevitably trigger exponential divergence in the exact trajectories. Therefore, while the point-wise locations differ, the model still captures the correct physical mechanisms and temporal scales of the transition.

Finally, we push the finetuned model to the most extreme extrapolation within our dataset, testing shifts of $\alpha=1.0 \to 0.08$ and $\alpha=0.1 \to 1.1$ (Figures \ref{fig:Fig3_10_008_extra} and \ref{fig:Fig3_01_11_extra} in Appendix B, respectively). Predictive accuracy remains similarly robust. The physical timing of macroscopic zonal structural changes, such as merging during transitions to turbulence-dominated states, and splitting in the opposite scenario, is well captured. We observe the same systematic under- and over-estimations of $E$ and $W$ depending on the direction of the transition, confirming that these biases are scenario-dependent, systematic features of the architecture rather than isolated failures. Ultimately, these extreme tests demonstrate that the finetuned architecture generalizes exceptionally well to highly out-of-distribution dynamical transitions.

\section{Conclusions}\label{sec:con}

We explored the capability of transformer-based neural operators to emulate the complex, multiscale dynamics of magnetized plasmas. Using the Modified Hasegawa-Wakatani (MHW) model as a testbed, we demonstrated that a single, unified finetuned model can successfully capture the nonlinear co-evolution of drift-wave turbulence and zonal flows across a wide parameter space, maintaining stability over time horizons significantly longer than the local turbulence correlation time. Our results show that while standard pretraining yields excellent short-term predictive accuracy, a targeted finetuning procedure is essential for maintaining physical fidelity over extended autoregressive rollouts. The finetuned model effectively prevents the unphysical divergence often seen in data-driven surrogates, maintaining accurate macroscopic statistical properties, total energy conservation, and stable internal representations of resistive dissipation and turbulent flux over long horizons.

Most crucially, the finetuned strategy proved highly capable of predicting a wide range of transient dynamics. For instance, the model successfully reproduced complex, transient physical sequences, such as the nonlinear saturation, the spontaneous collapse of turbulence, and the emergence of macroscopic zonal flow structures. Capturing these stiff temporal dynamics, despite their extreme sparsity in the training data, represents a significant step forward for AI-based plasma modeling.

Speed-wise, our current un-optimized setup achieves a prediction within approximately 5--10 ms per physical simulation time unit (depending on the chosen timestep for the autoregressive rollout), representing a roughly 300--600$\times$ speedup over the baseline DNS. We note that this is an approximate and inherently \textit{unfair} comparison: the DNS is computed on CPUs using double-precision arithmetic at a higher spatial resolution, whereas the neural operator leverages GPU acceleration using single precision at a $4\times$ coarser resolution. Nevertheless, a fundamental portion of this acceleration is solid as it is from the neural operator bypassing both the strict CFL time-stepping and spatial resolution constraints required by traditional numerical solvers. As neural operator architectures and inference engines are actively evolving, there remains substantial potential for even greater computational efficiency.

We acknowledge that the MHW equations represent a simplified, local 2D fluid model. However, the methodology presented here is directly applicable to more realistic, full-scale 3D (perhaps even 5D) global turbulence simulation or experimental data. Certain performance deficiencies in this study, such as the under-resolving of high-$k$ structures, are largely due to the 4$\times$ spatial downsampling required to fit the current training workflow onto a single NVIDIA A100 (80 GB) GPU, compounded by the inherent bias of standard point-wise loss functions toward high-amplitude, low-$k$ features. With expanded computational resources and scale-aware loss formulations, these constraints can be alleviated. Ultimately, this work establishes a robust methodological foundation for extending neural operator surrogates to full 3D fluid and 5D gyrokinetic simulations in complex geometries, paving the way towards real-time scenario optimization and control design in future fusion devices.

\begin{acknowledgments}
Prepared by LLNL under Contract DE-AC52-07NA27344. This work was supported by the U.S. DOE Fusion Energy Sciences MFE-FAST project (SCW01924). This research used resources of the National Energy Research Scientific Computing Center (NERSC), a Department of Energy User Facility using NERSC award FES-ERCAP 0036750 and 0037021. Release no. LLNL-JRNL-2016394.
\end{acknowledgments}

\section*{Author Declarations}
\subsection*{Conflict of Interest}
The authors have no conflicts to disclose.

\section*{Data Availability}

The data that support the findings of this study are available from the corresponding author upon reasonable request.

\section*{Appendix A: Statistical Properties of the MHW Dataset}

\subsection*{A.1 Energy Partition and Turbulence Flux}
To highlight the varying solution state and transitions within the MHW model across the adiabaticity parameter range $\alpha \in \{0.1, 1.0\}$, we analyze the system's energy partition, turbulent flux and resistive dissipation. The total energy is partitioned into four components: zonal internal energy density $\epsilon_{iz}$, perturbed internal energy density $\epsilon_{ip}$, zonal kinetic energy density $\epsilon_{kz}$, and perturbed kinetic energy density $\epsilon_{kp}$. These domain-averaged quantities ($\langle \dots \rangle_{x,y}$) are computed as follows,
\begin{align}
    \epsilon_{iz} = \left\langle \frac{1}{2} \bar{n}^2 \right\rangle_{x,y}, \quad \epsilon_{ip} &= \left\langle \frac{1}{2} n^2 \right\rangle_{x,y} - \epsilon_{iz}, \\ \nonumber
    \epsilon_{kz} = \left\langle \frac{1}{2} (\partial_x \bar{\phi})^2 \right\rangle_{x,y}, \quad \epsilon_{kp} &= \left\langle \frac{1}{2} |\nabla_\perp \phi|^2 \right\rangle_{x,y} - \epsilon_{kz}.
\end{align}

\begin{figure}[!htbp]
    \centering
    \includegraphics[width=\linewidth]{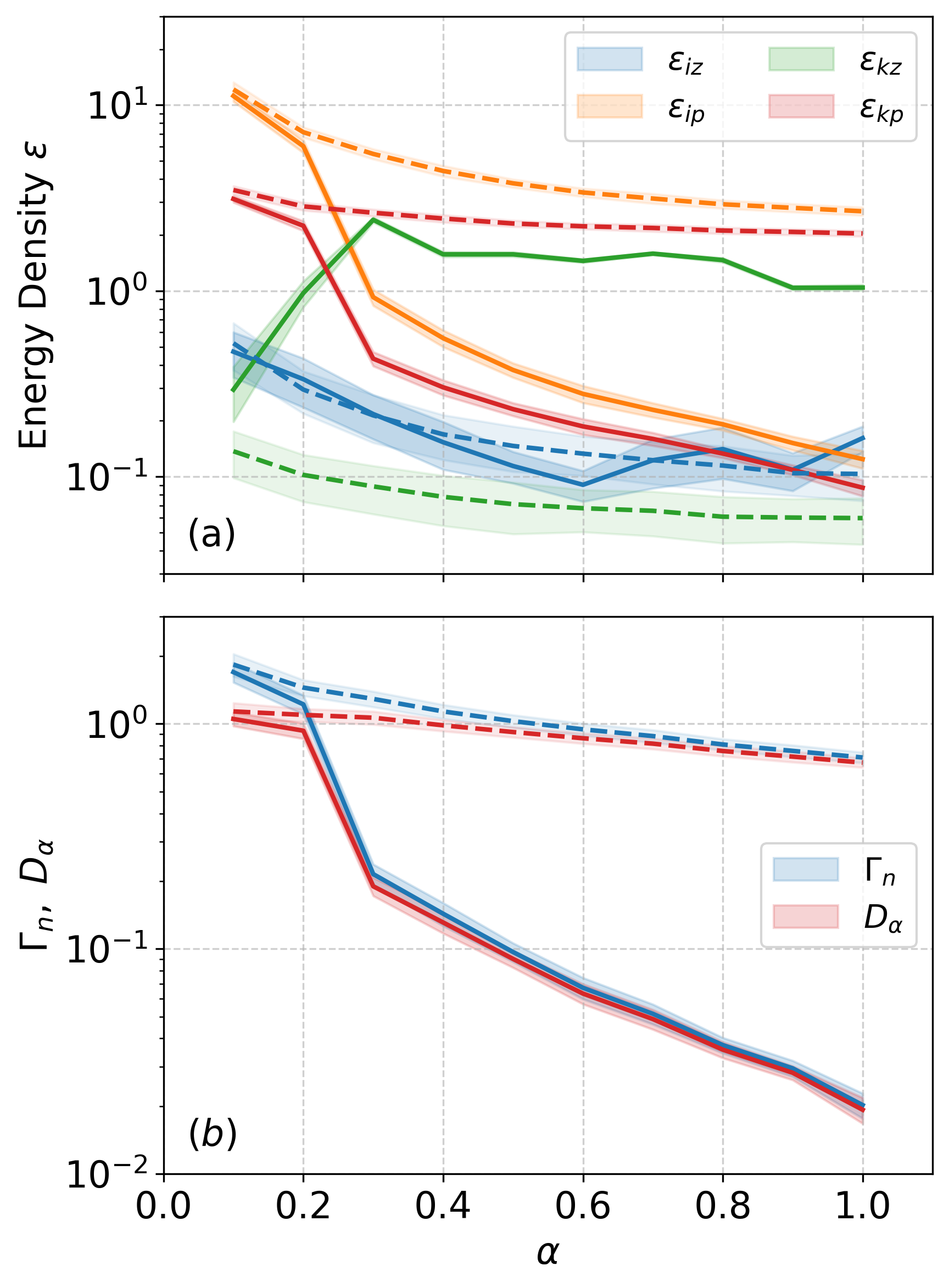}
    \caption{(a) Energy partition and (b) turbulent and dissipative fluxes versus the adiabaticity parameter $\alpha$ for the MHW model (solid lines) and the original HW model (dashed lines). Shaded regions represent the temporal standard deviation.}
    \label{fig:energy_flux}
\end{figure}

The temporal mean and standard deviation of these metrics are evaluated over the simulation time window $t \in [1000, 2000]$ from the \textit{steady-state} simulation dataset, ensuring the data is sampled well within the saturated, statistically steady-state turbulence regime.
These statistical properties are summarized in Figure~\ref{fig:energy_flux}. To illustrate how the MHW model deviates from the original HW model, particularly in the high-adiabaticity regime where self-generated zonal flows play a dominant role, results from the original HW system are plotted for comparison.

\subsection*{A.2 Autocorrelation Scales}

To mitigate statistical noise, the characteristic spatial and temporal scales of the turbulence simulation data are evaluated using ensemble-averaged autocorrelation functions. That is, rather than computing a characteristic scale for each time step (for spatial lengths) or spatial location (for temporal scales) and subsequently averaging these scalars, the correlation functions are first averaged; the characteristic scales are then derived from these ensemble-averaged functions.

\begin{figure}[!htbp]
    \centering
    \includegraphics[width=\linewidth]{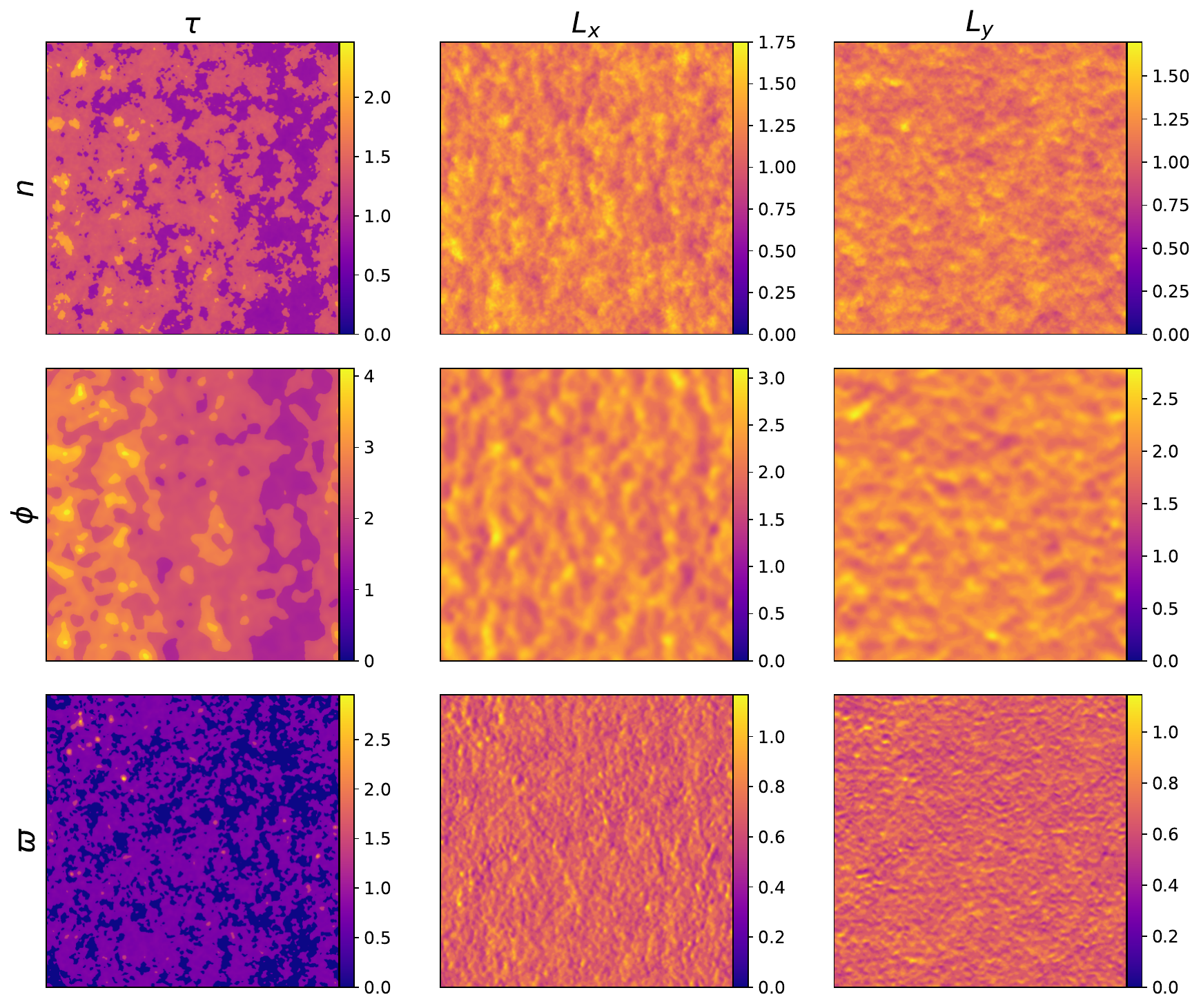}
    \caption{Spatial distribution of autocorrelation times and characteristic length scales of the perturbed density, electrostatic potential, and vorticity for the MHW model with $(\alpha,\kappa)=(0.1,1.0)$.}
    \label{fig:scale_mhw_alpha01}
\end{figure}

\begin{figure}[!htbp]
    \centering
    \includegraphics[width=\linewidth]{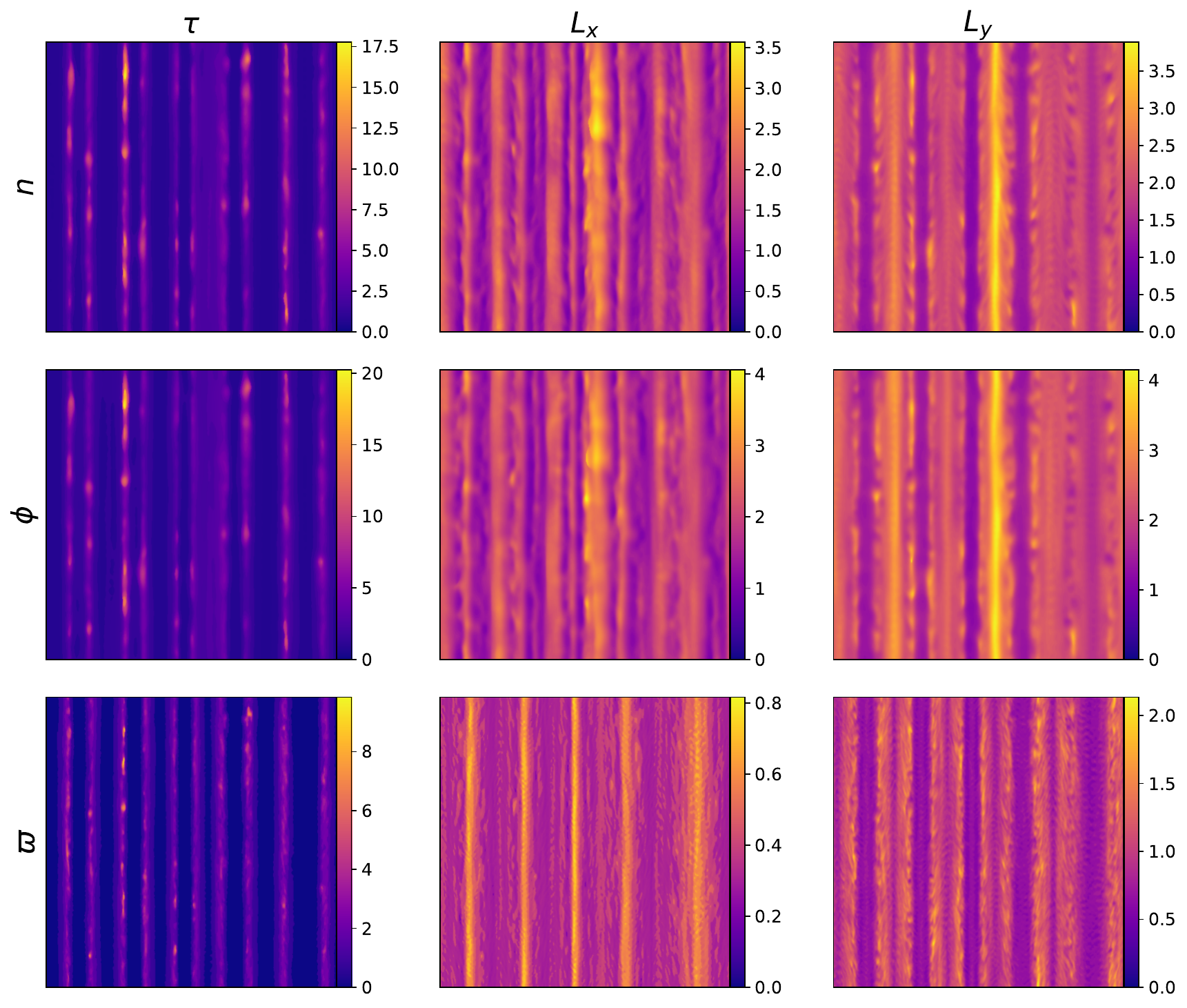}
    \caption{Spatial distribution of autocorrelation times and characteristic length scales of the perturbed density, electrostatic potential, and vorticity for the MHW model with $(\alpha,\kappa)=(1.0,1.0)$.}
    \label{fig:scale_mhw_alpha10}
\end{figure}

\begin{figure}[!htbp]
    \centering
    \includegraphics[width=\linewidth]{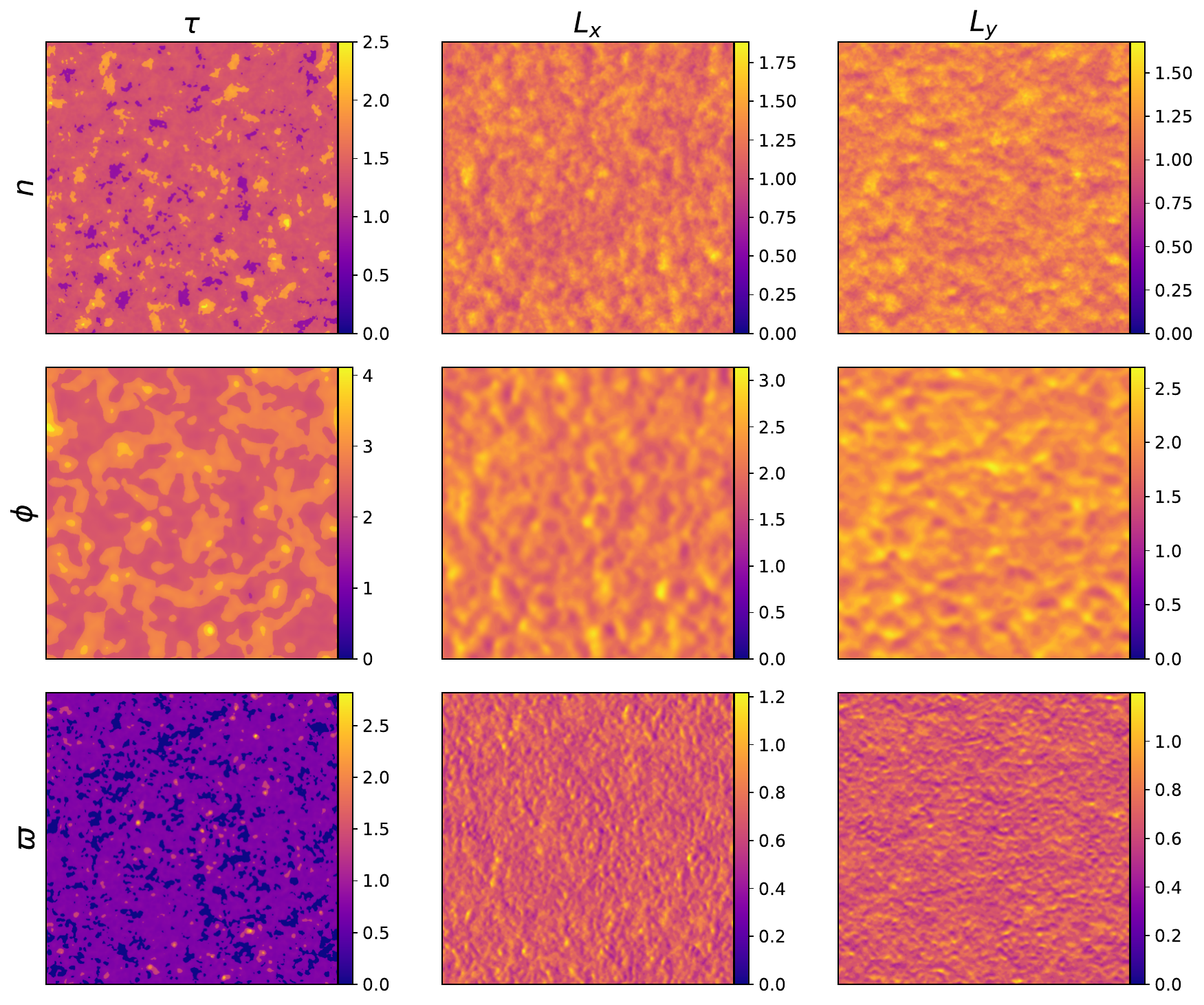}
    \caption{Spatial distribution of autocorrelation times and characteristic length scales of the perturbed density, electrostatic potential, and vorticity for the OHW model with $(\alpha,\kappa)=(0.1,1.0)$.}
    \label{fig:scale_ohw_alpha01}
\end{figure}

\begin{figure}[!htbp]
    \centering
    \includegraphics[width=\linewidth]{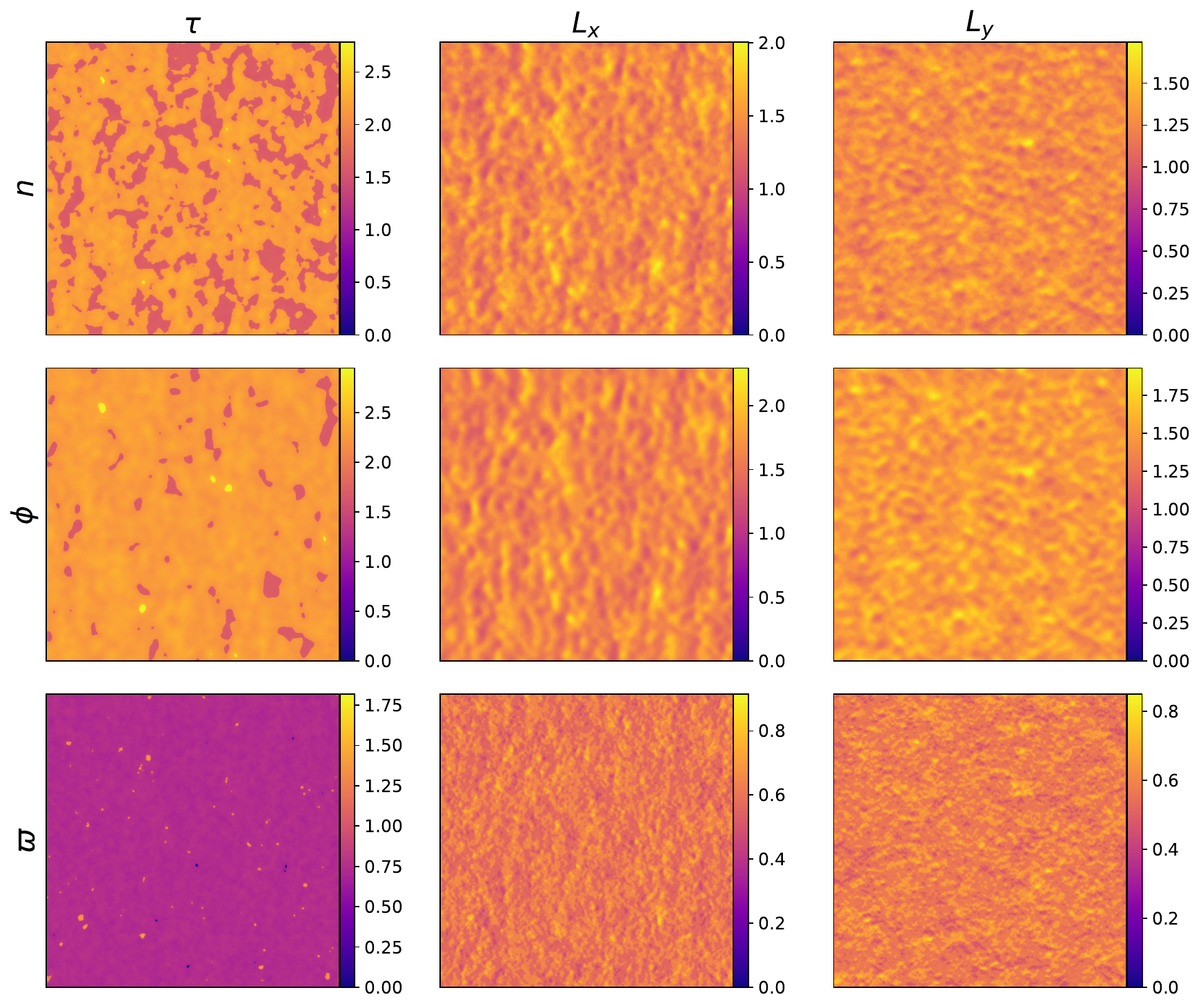}
    \caption{Spatial distribution of autocorrelation times and characteristic length scales of the perturbed density, electrostatic potential, and vorticity for the OHW model with $(\alpha,\kappa)=(1.0,1.0)$.}
    \label{fig:scale_ohw_alpha10}
\end{figure}

The quantity of interest, $f(x, y, t)$, is first decomposed into a macroscopic, instantaneous zonal component $\bar{f}(x,t)=\langle f(x, y, t) \rangle_y$ and a non-zonal contribution $\tilde{f}(x, y, t)=f(x, y, t)-\bar{f}(x,t)$.
The residual fluctuation field $\tilde{f}$ thus represents the turbulent fluctuations, and the corresponding covariance functions, $C$ at a specific spatial location $(x_i, y_j)$ for a separation lag $\delta$ are defined as:
\begin{equation}
    C_x(x_i, y_j, \delta_x) = \frac{1}{N_t} \sum_{t=0}^{N_t} \tilde{f}(x_i, y_j, t) \cdot \tilde{f}(x_i + \delta_x, y_j, t)
\end{equation}
\begin{equation}
    C_y(x_i, y_j, \delta_y) = \frac{1}{N_t} \sum_{t=0}^{N_t} \tilde{f}(x_i, y_j, t) \cdot \tilde{f}(x_i, y_j + \delta_y, t)
\end{equation}
\begin{equation}
    C_\tau(x_i, y_j, \delta_t) = \frac{1}{N_{pairs}} \sum_{t} \tilde{f}(x_i, y_j, t) \cdot \tilde{f}(x_i, y_j, t + \delta_t)
\end{equation}
The normalized autocorrelation function $\rho$ is then obtained by scaling the covariance by the local variance $\sigma^2 = C(0)$:
\begin{equation}
    \rho(\delta) = \frac{C(\delta)}{C(0)}.
\end{equation}
The characteristic integral scale ($L$ or $\tau$) is defined as the integral of the normalized autocorrelation function $\rho(\delta)$, from zero lag up to a decorrelation threshold (e.g., $1/e$ folding time/length) to ensure robustness against tail noise:
\begin{equation}
    L = \int_0^{\delta_\text{cutoff}} \rho(\delta) \, \text{d}\delta \quad \text{where} \quad \rho(\delta_\text{cutoff}) \le e^{-1} \approx 0.3679.
\end{equation}
In this study, the characteristic scales (such as Figures \ref{fig:scale_mhw_alpha01} and \ref{fig:scale_mhw_alpha10}) were evaluated using steady-state simulations over the interval $t=1000$ to $2000$. To highlight the structural differences and to contextualize the increased complexity of the MHW system, the same analysis was applied to the original HW (OHW) model, as shown in Figures \ref{fig:scale_ohw_alpha01} and \ref{fig:scale_ohw_alpha10}, which exhibits homogeneous turbulence in both low and high adiabaticities.

\section*{Appendix B: Further Visualizations}

This appendix provides supplementary figures that support the analysis and conclusions presented in the main manuscript.

\begin{figure}[!htbp]
    \centering
    \includegraphics[width=\linewidth]{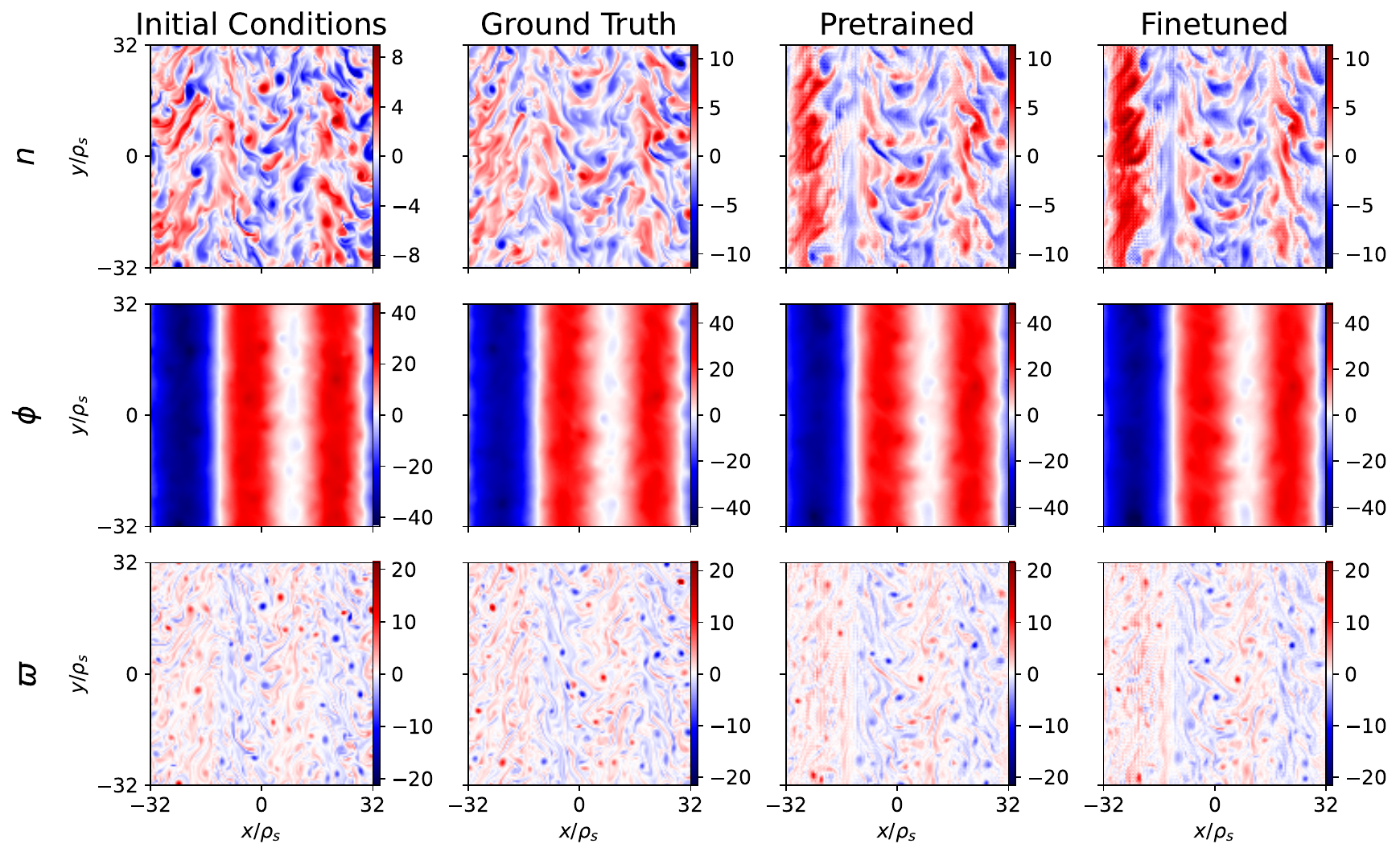}
    \caption{Initial conditions, ground truth, pretrained and finetuned model predictions after $12\omega_\text{ci}^{-1}$ for a held-out $\alpha=0.25$, $\omega_\text{ci}t_0=1938$ \textit{steady-state} trajectory.}
    \label{fig:Fig2_025long}
\end{figure}

\begin{figure}[!htbp]
    \centering
    \includegraphics[width=\linewidth]{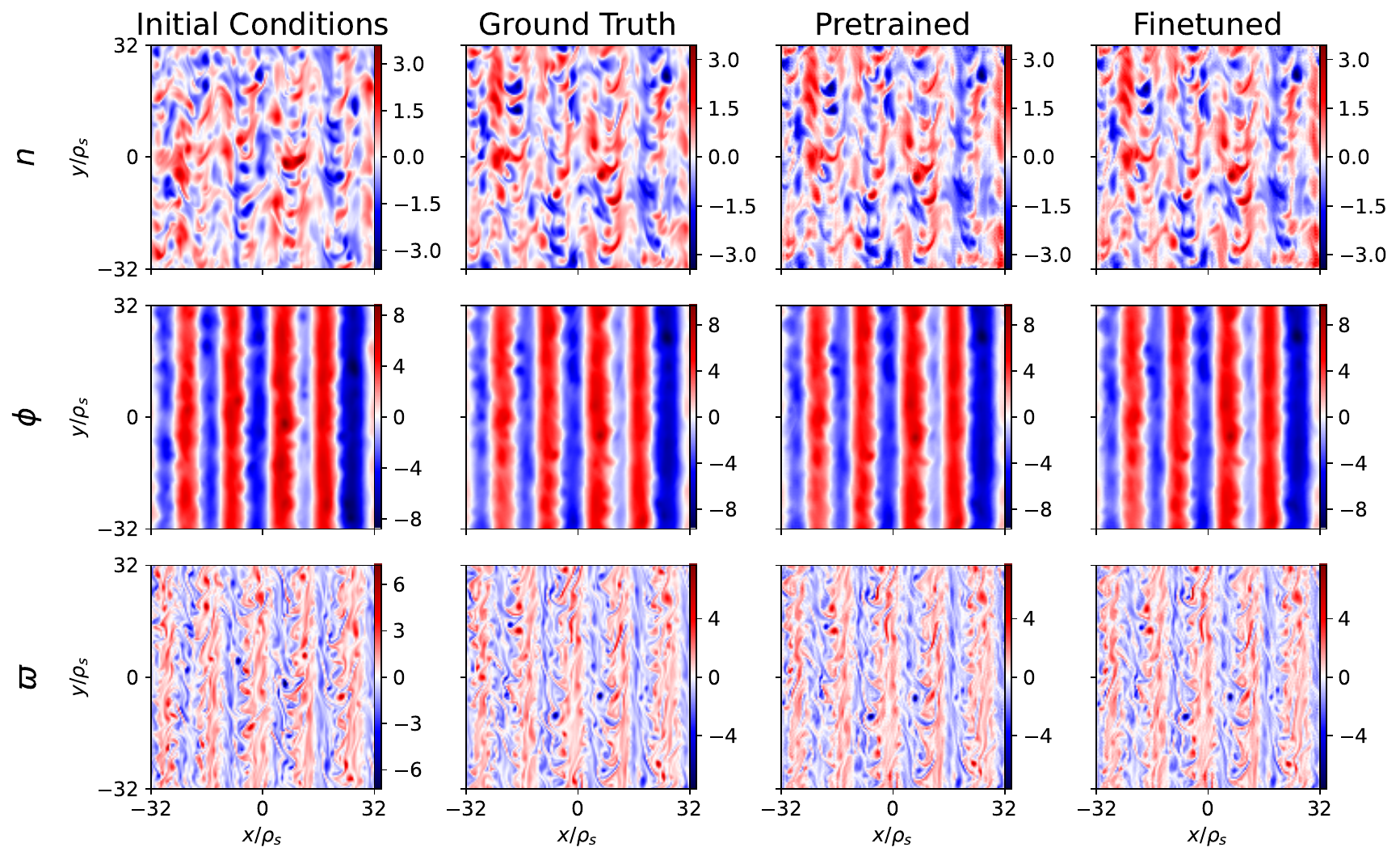}
    \caption{Initial conditions, ground truth, pretrained and finetuned model predictions after $12\omega_\text{ci}^{-1}$ for a held-out $\alpha=0.75$, $\omega_\text{ci}t_0=3268$ \textit{steady-state} trajectory.}
    \label{fig:Fig2_075long}
\end{figure}

\begin{figure}[!htbp]
    \centering
    \includegraphics[width=\linewidth]{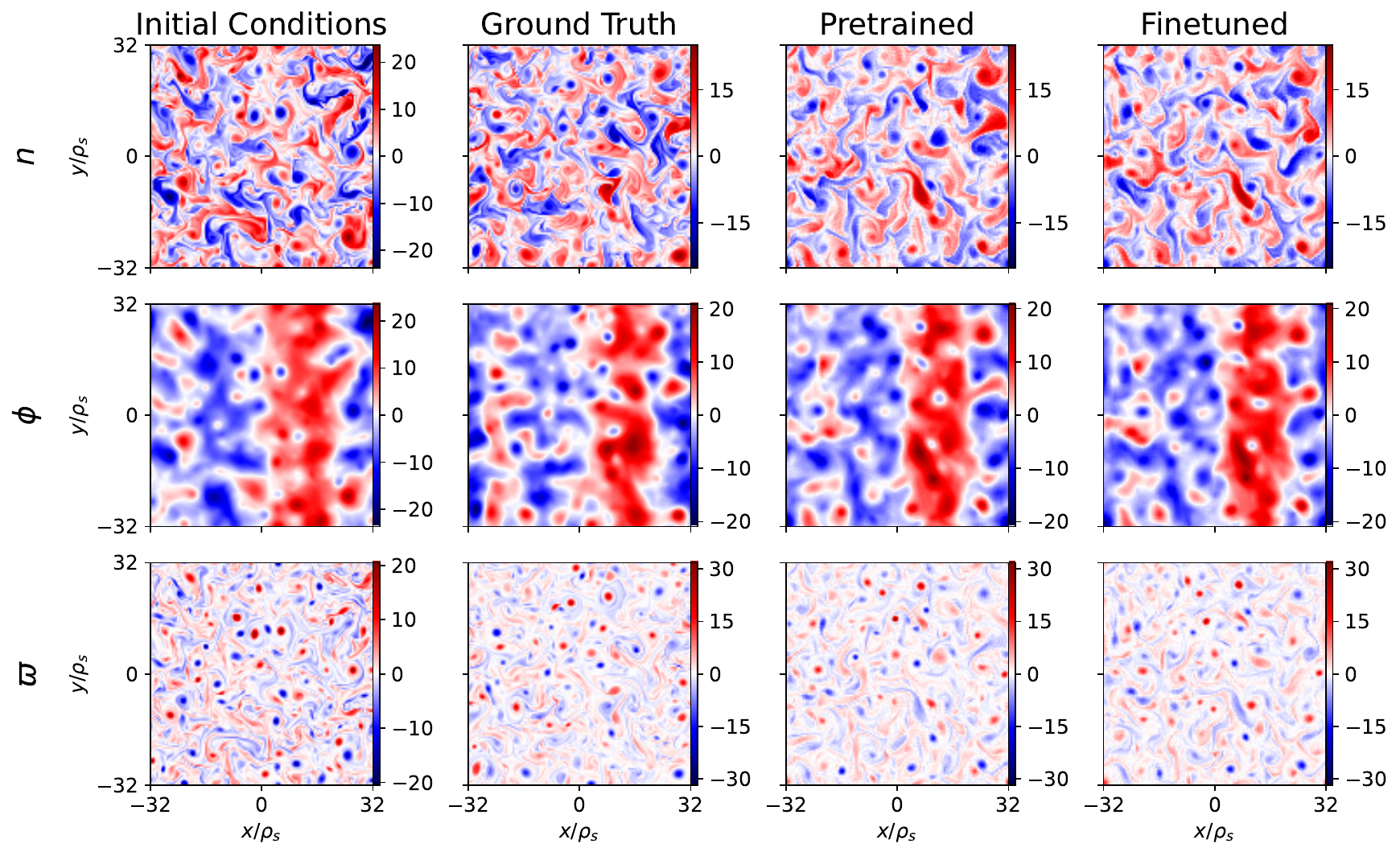}
    \caption{Initial conditions, ground truth, pretrained and finetuned model predictions after $12\omega_\text{ci}^{-1}$ for held-out $\alpha=0.08$.}
    \label{fig:Fig2_008}
\end{figure}

\begin{figure}[!htbp]
    \centering
    \includegraphics[width=\linewidth]{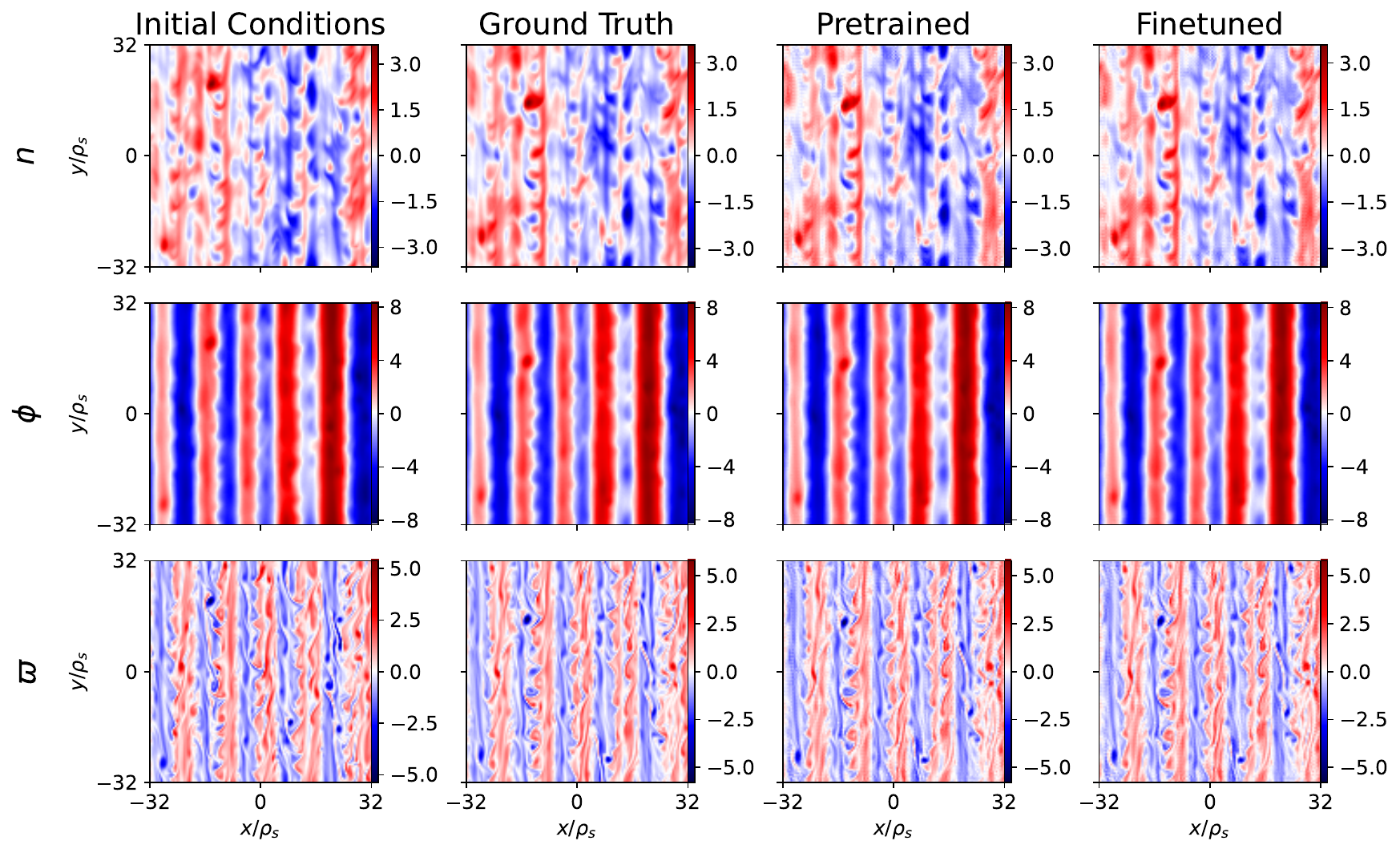}
    \caption{Initial conditions, ground truth, pretrained and finetuned model predictions after $12\omega_\text{ci}^{-1}$ for held-out $\alpha=1.1$.}
    \label{fig:Fig2_11}
\end{figure}

\begin{figure}[!htbp]
    \centering
    \includegraphics[width=\linewidth]{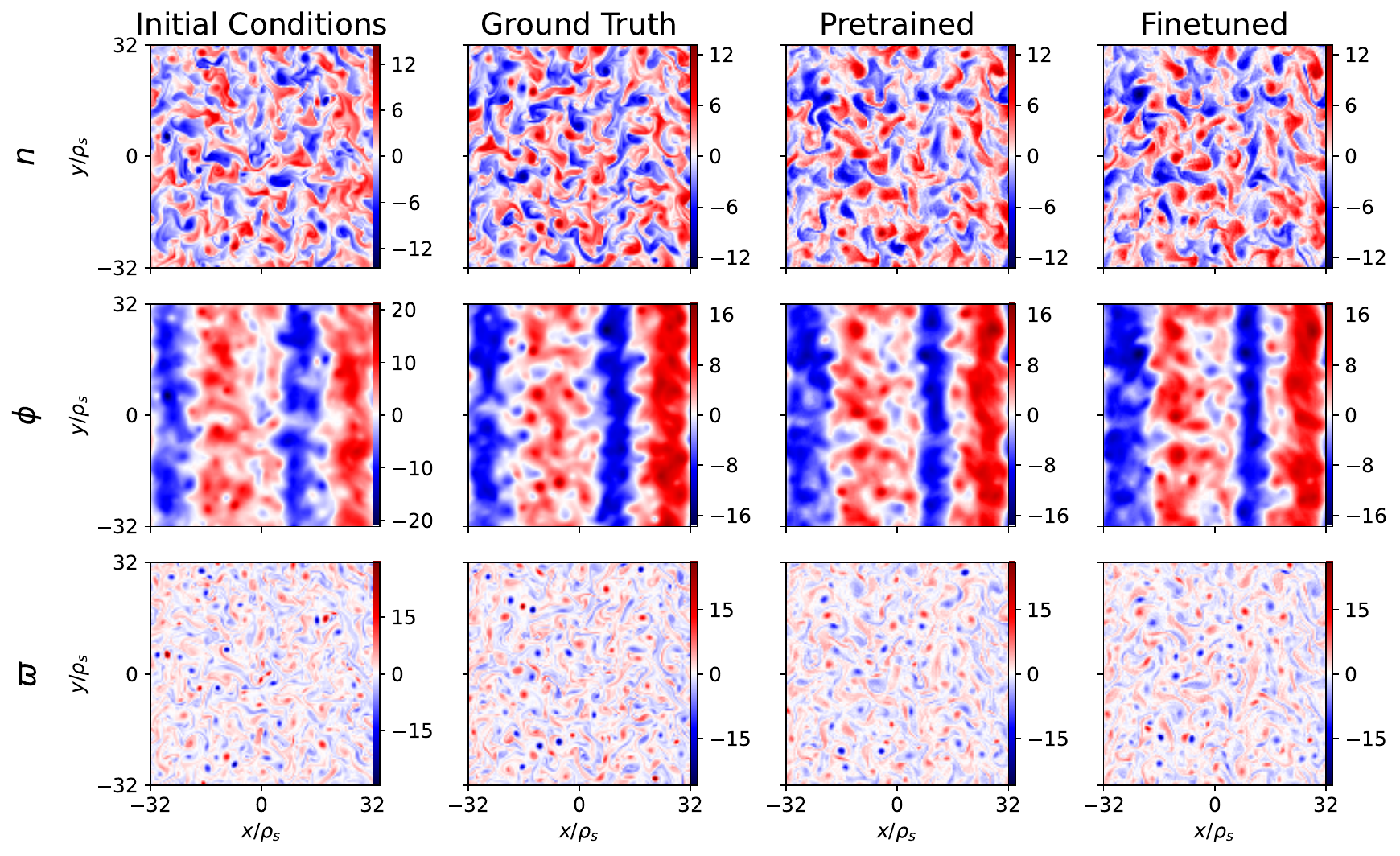}
    \caption{Initial conditions, ground truth, pretrained and finetuned model predictions after $12\omega_\text{ci}^{-1}$ for a held-out $\alpha=0.1\rightarrow 0.25$, $\omega_\text{ci}t_0=266$ \textit{dynamic transition} trajectory.}
    \label{fig:Fig2_01_025}
\end{figure}

\begin{figure}[!htbp]
    \centering
    \includegraphics[width=\linewidth]{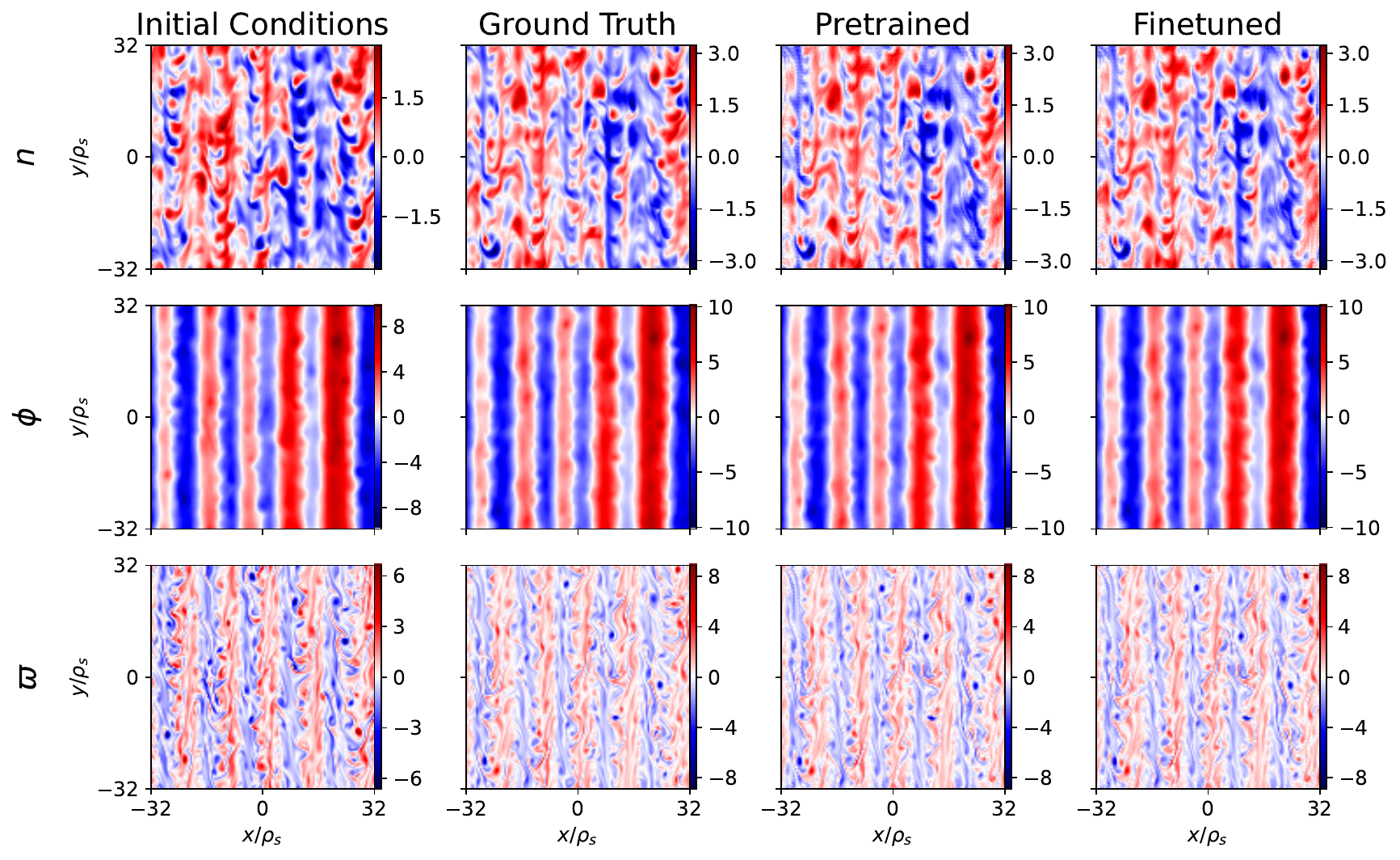}
    \caption{Initial conditions, ground truth, pretrained and finetuned model predictions after $12\omega_\text{ci}^{-1}$ for a held-out $\alpha=1.0\rightarrow 0.75$, $\omega_\text{ci}t_0=266$ \textit{dynamic transition} trajectory.}
    \label{fig:Fig2_10_075}
\end{figure}

\begin{figure}[!htbp]
    \centering
    \includegraphics[width=\linewidth]{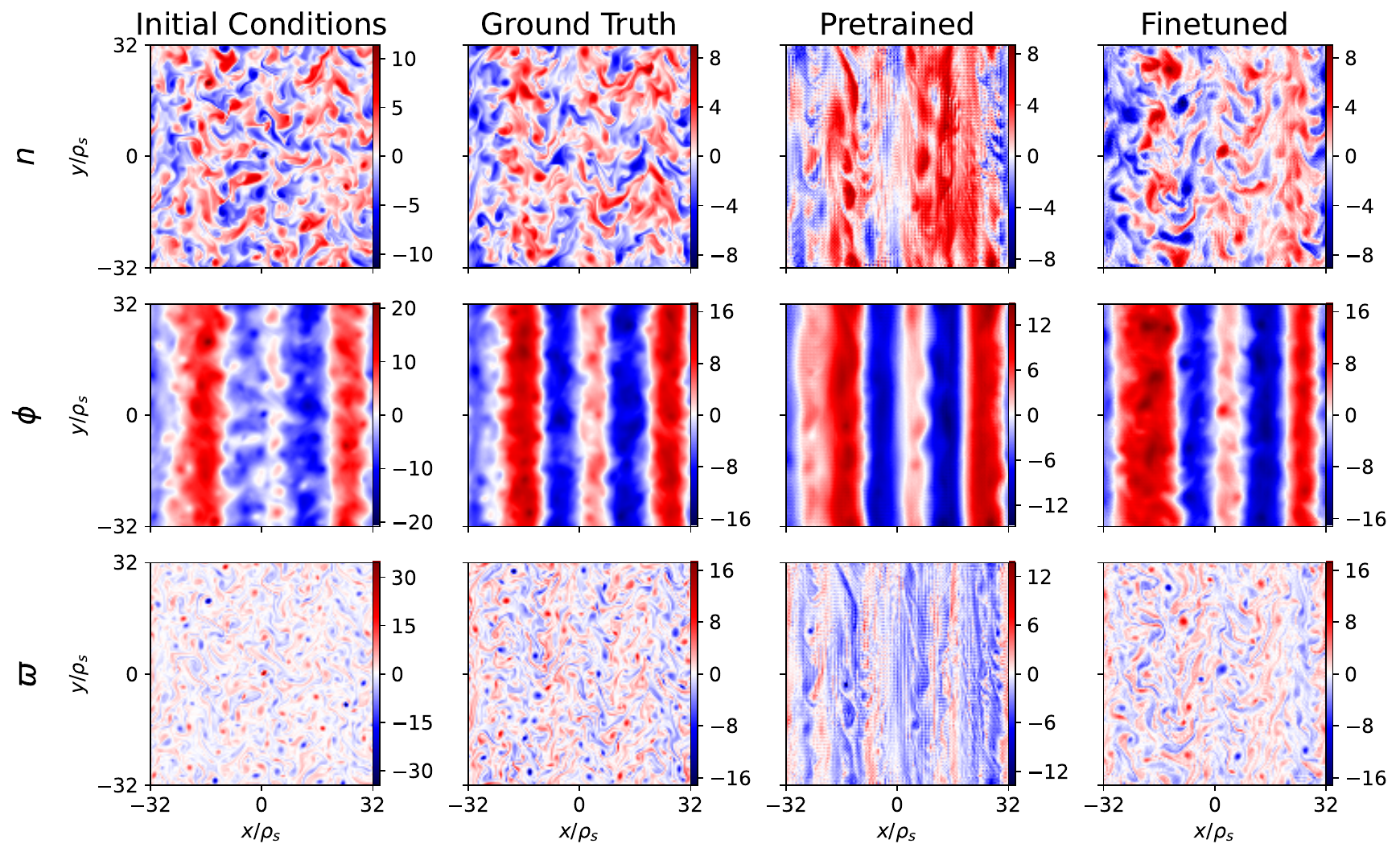}
    \caption{Initial conditions, ground truth, pretrained and finetuned model predictions after $72\omega_\text{ci}^{-1}$ for \textit{dynamical transition} $\alpha=0.1\rightarrow0.4$.}
    \label{fig:Fig3_01_04}
\end{figure}

\begin{figure}[!htbp]
    \centering
    \includegraphics[width=\linewidth]{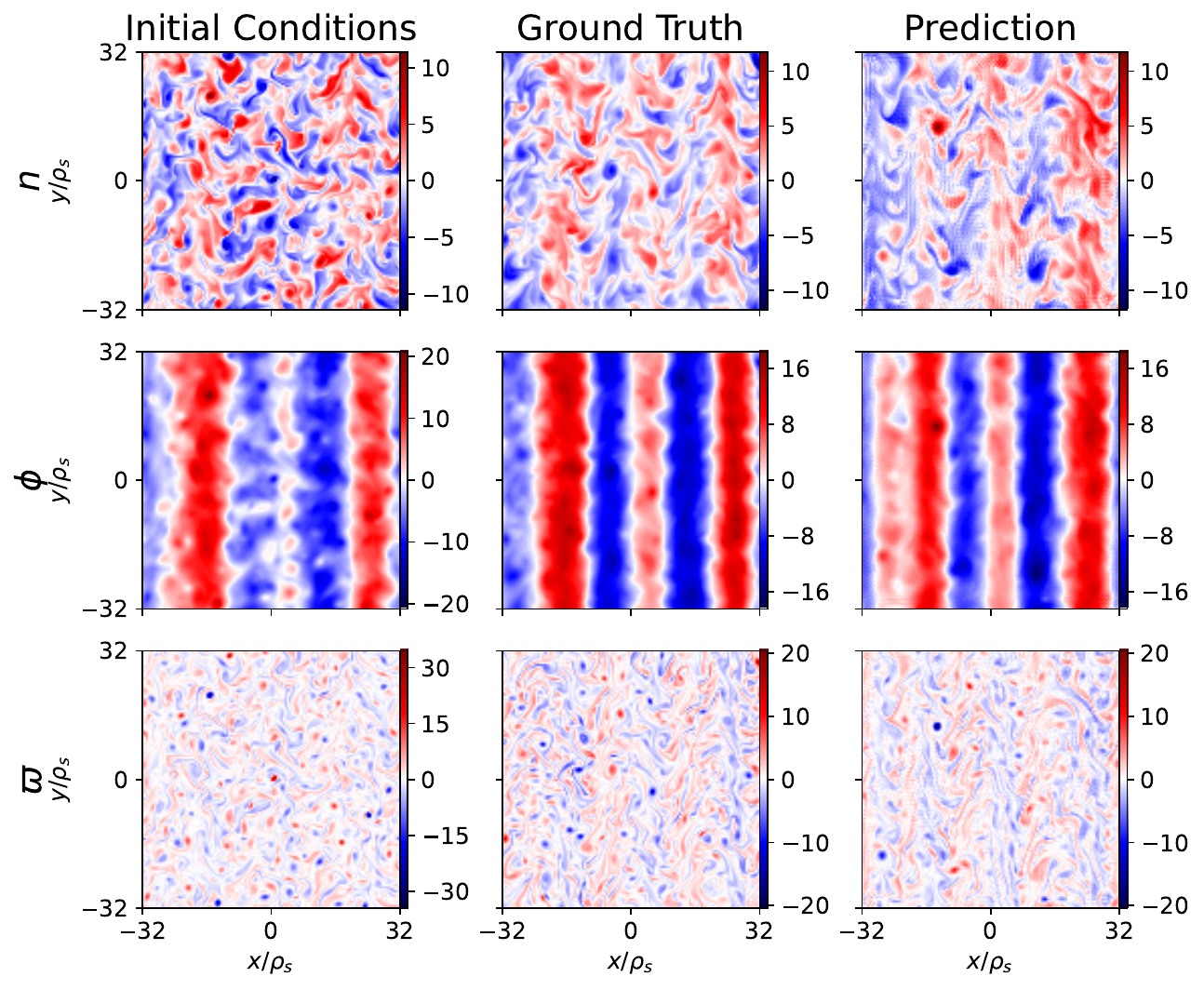}
    \caption{Initial conditions, ground truth and finetuned model prediction after $96\omega_\text{ci}^{-1}$, a $24\omega_\text{ci}^{-1}$ extrapolation beyond the finetuning task, for \textit{dynamical transition} $\alpha=0.1\rightarrow0.4$.}
    \label{fig:Fig3_01_04_extra}
\end{figure}

\begin{figure}[!htbp]
    \centering
    \includegraphics[width=0.49\linewidth]{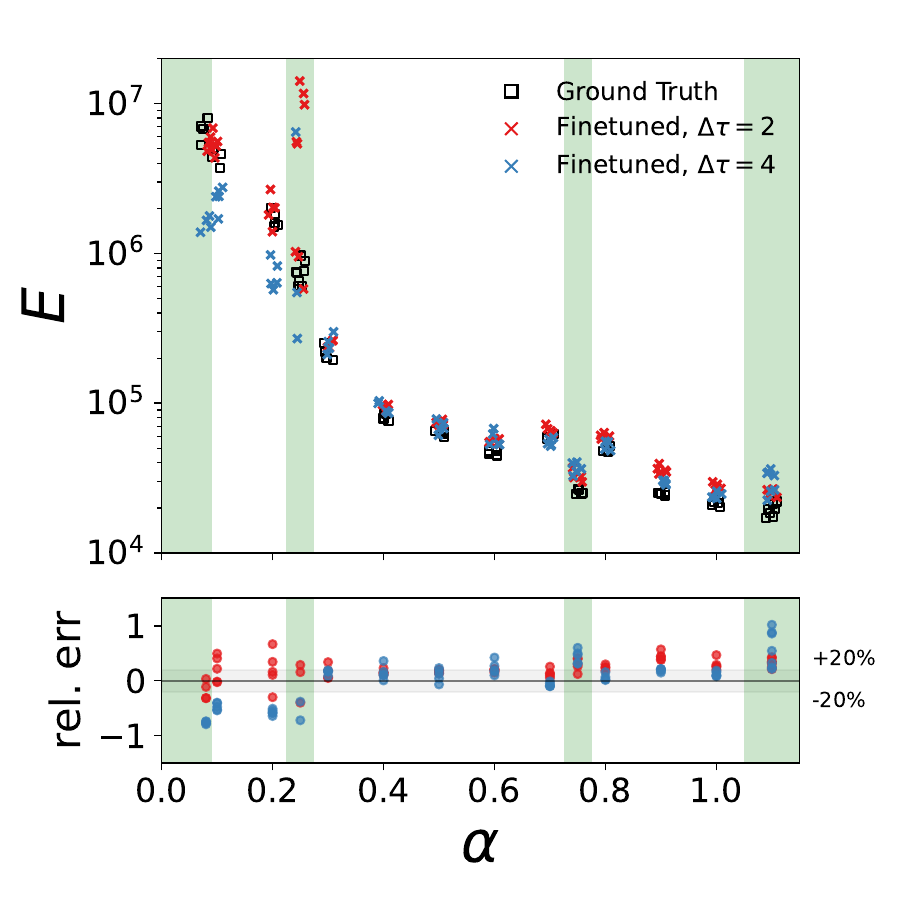}
    \includegraphics[width=0.49\linewidth]{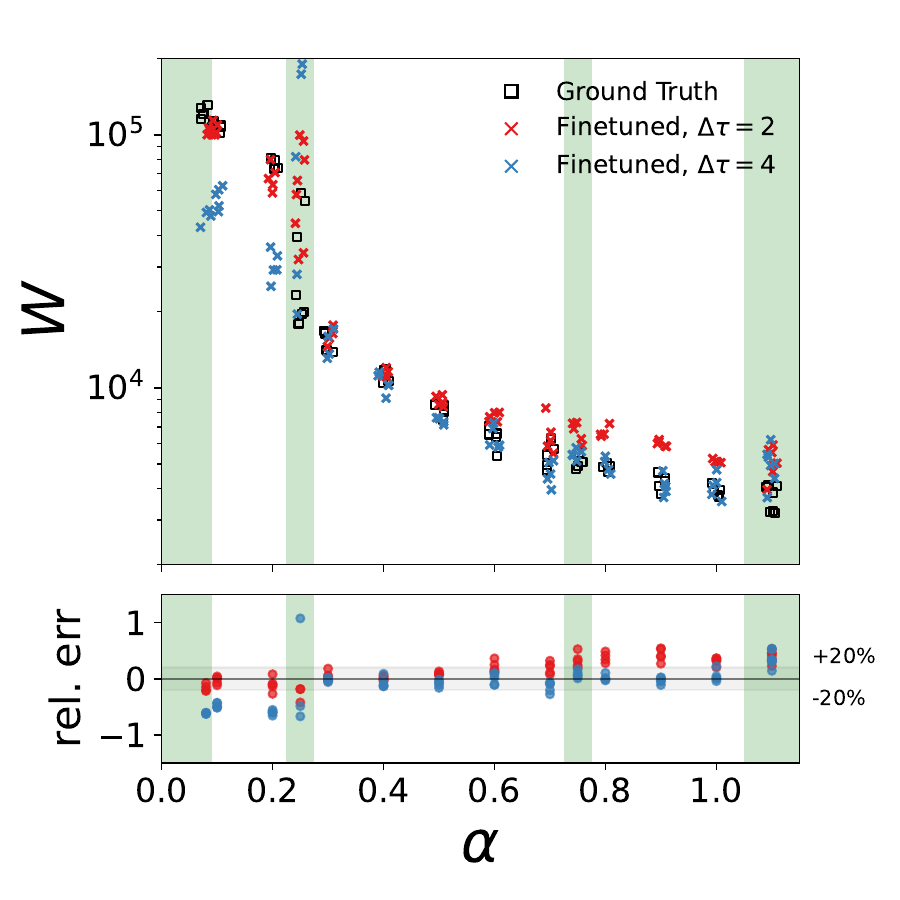}
    \includegraphics[width=0.49\linewidth]{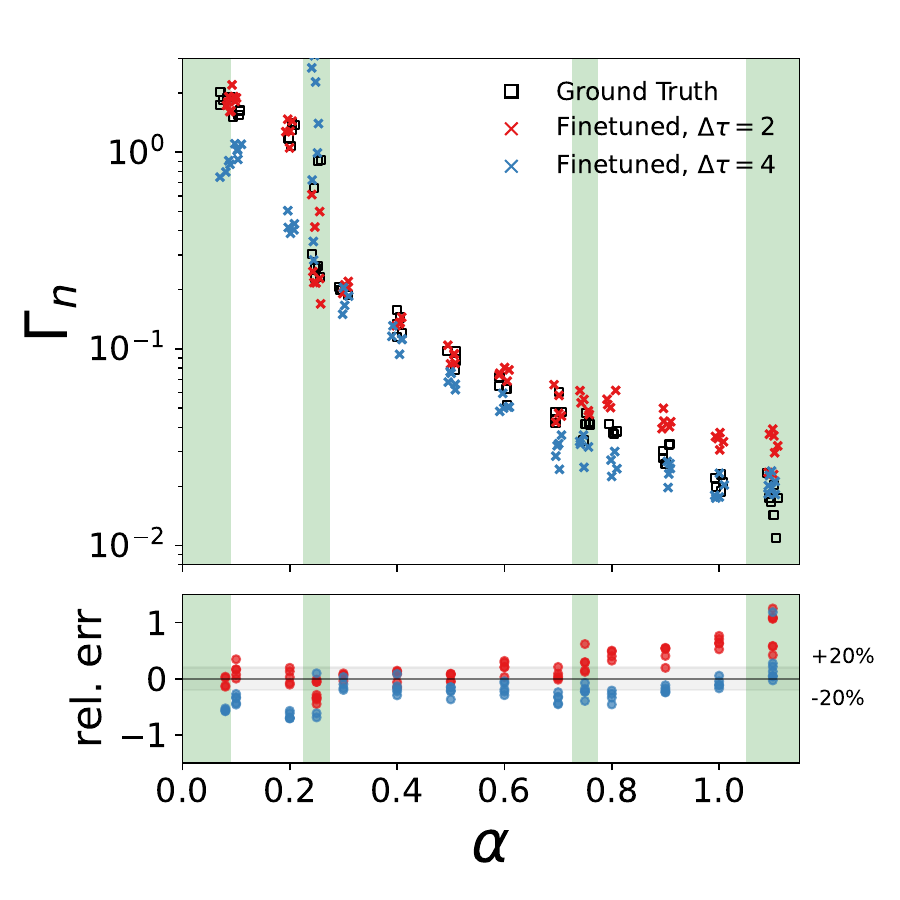}
    \includegraphics[width=0.49\linewidth]{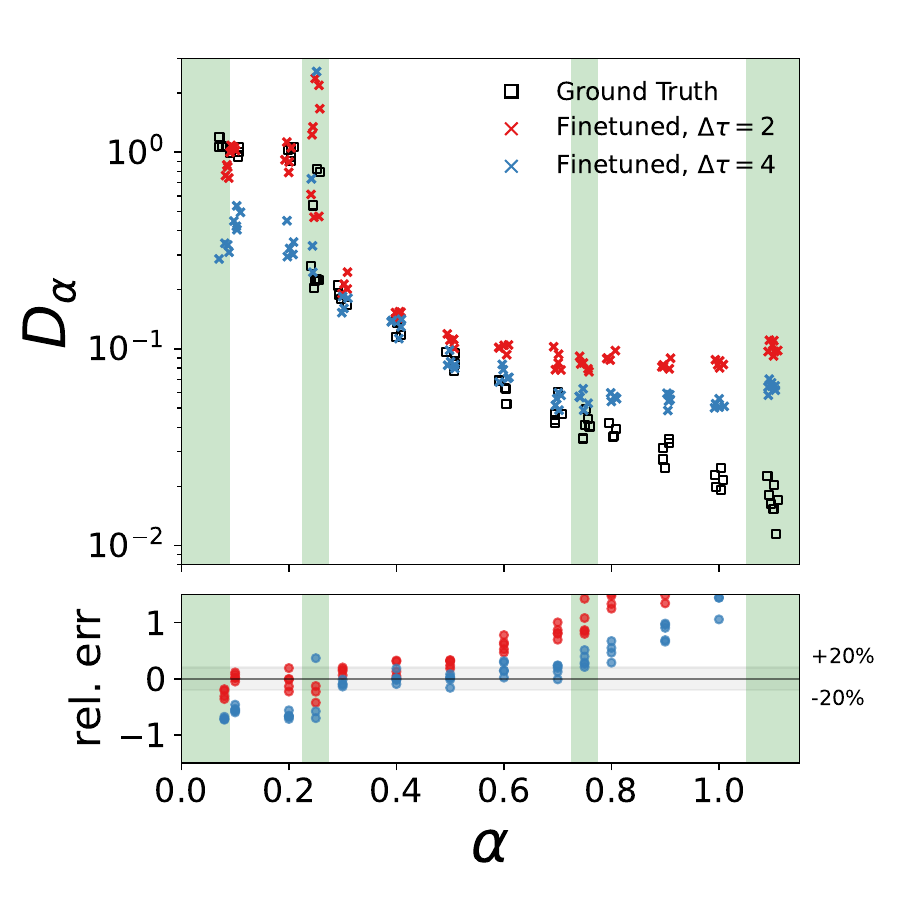}
    \caption{Pairwise comparisons between the ground truth and finetuned model predictions of 72 random held-out \textit{steady-state} test trajectories, including held-out $\alpha\in\{0.08,0.25,0.75,1.1\}$ (green shaded regions), after $72\omega_\text{ci}^{-1}$ of the total energy $E$, enstrophy $W$, turbulent flux $\Gamma_n$ and resistive dissipation $D_\alpha$, using autoregressive rollouts of timesteps $\Delta\tau=2$ (red) and $\Delta\tau=4$ (blue).}
    \label{fig:Fig2_stats_long}
\end{figure}

\begin{figure}[!htbp]
    \centering
    \includegraphics[width=\linewidth]{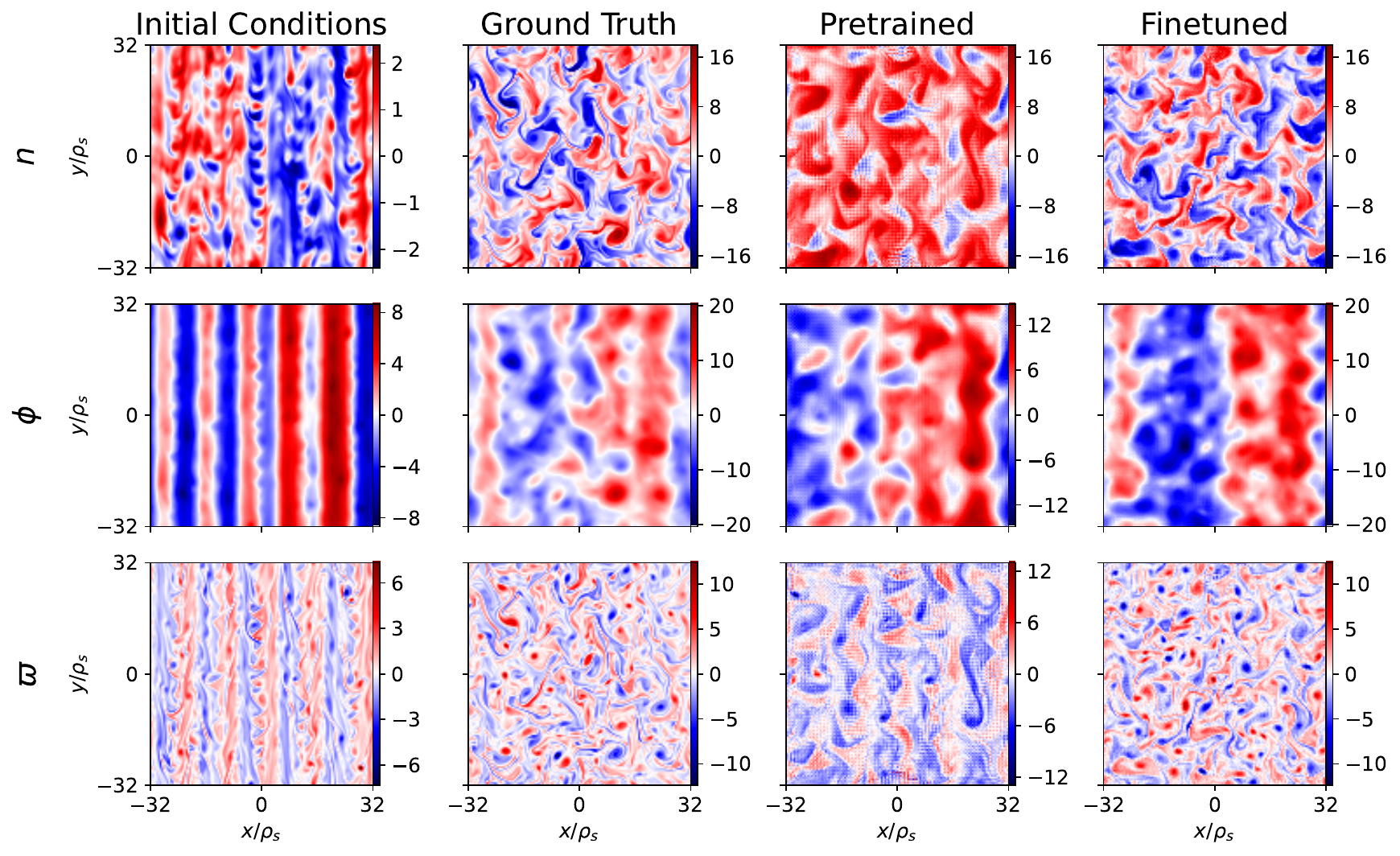}
    \includegraphics[width=\linewidth]{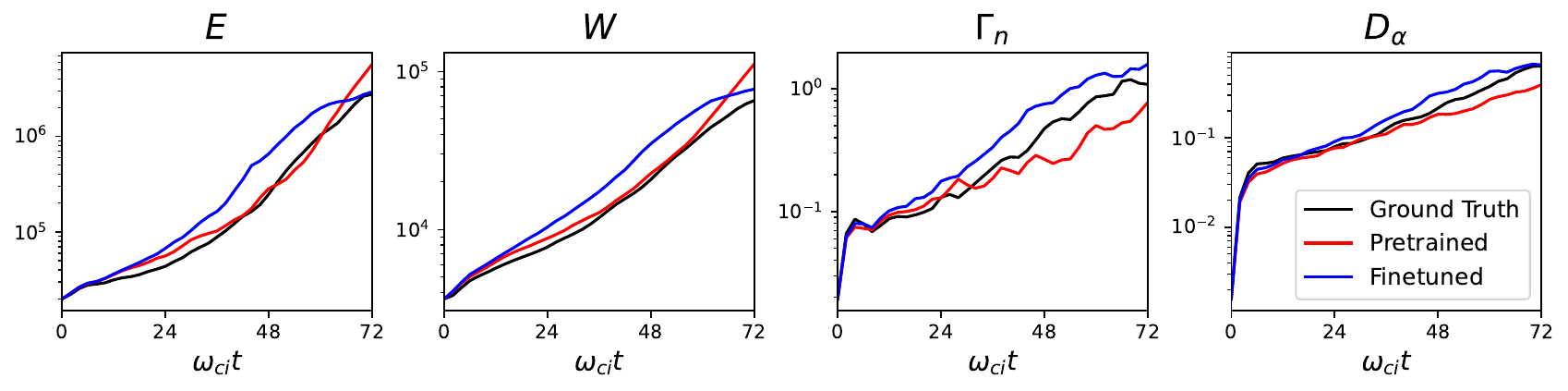}
    \includegraphics[width=\linewidth]{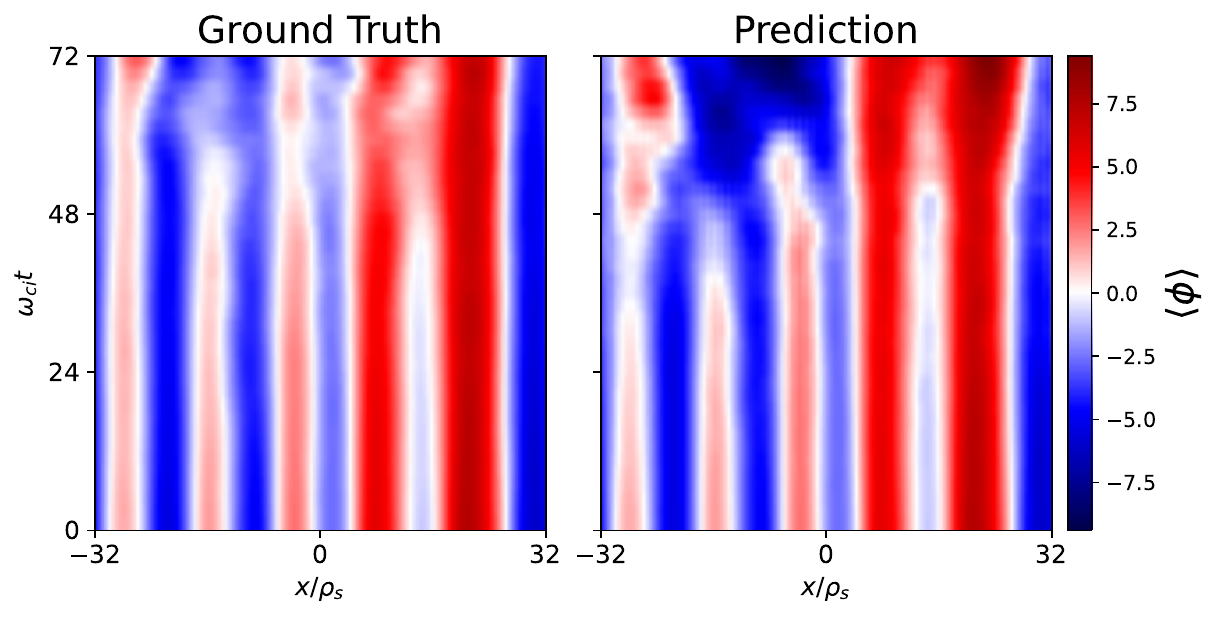}
    \caption{(Top) initial conditions, ground truth, pretrained and finetuned model predictions after $72\omega_\text{ci}^{-1}$ for \textit{dynamical transition} $\alpha=1.0\rightarrow0.08$, $\omega_\text{ci}t_0=0$. (Middle) time evolution of the energy $E$, enstrophy $W$, turbulent flux $\Gamma_n$ and resistive dissipation $D_\alpha$ from the ground truth, pretrained and finetuned models. (Bottom) ground truth and finetuned model trajectories of the zonal component of the potential $\langle\phi\rangle$. }
    \label{fig:Fig3_10_008_extra}
\end{figure}

\begin{figure}[!htbp]
    \centering
    \includegraphics[width=\linewidth]{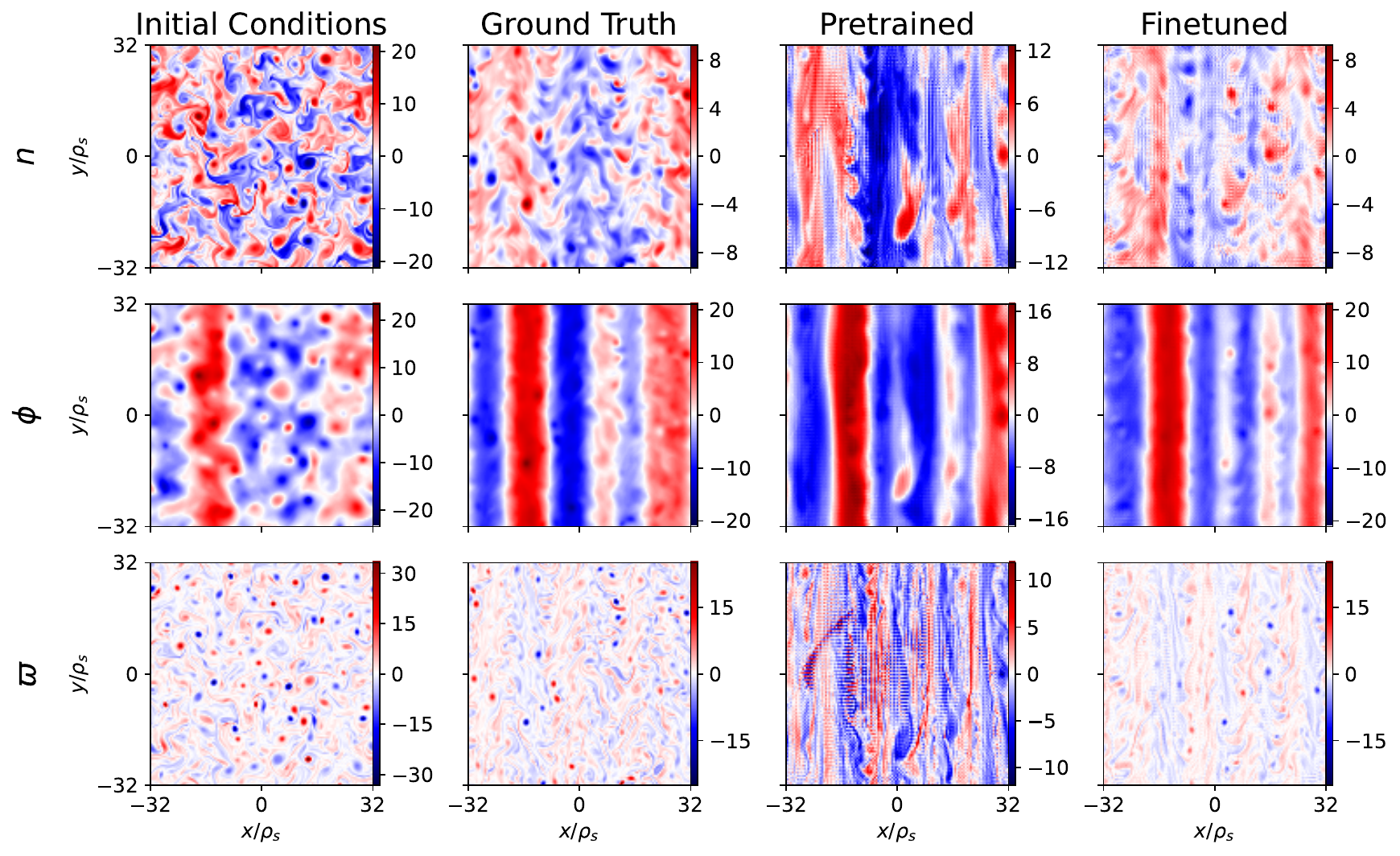}
    \includegraphics[width=\linewidth]{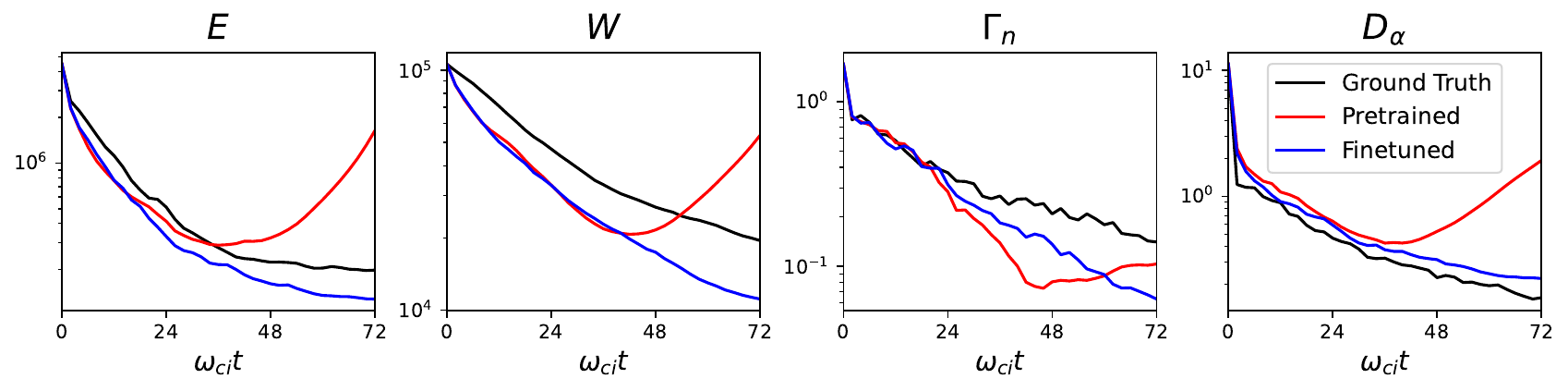}
    \includegraphics[width=\linewidth]{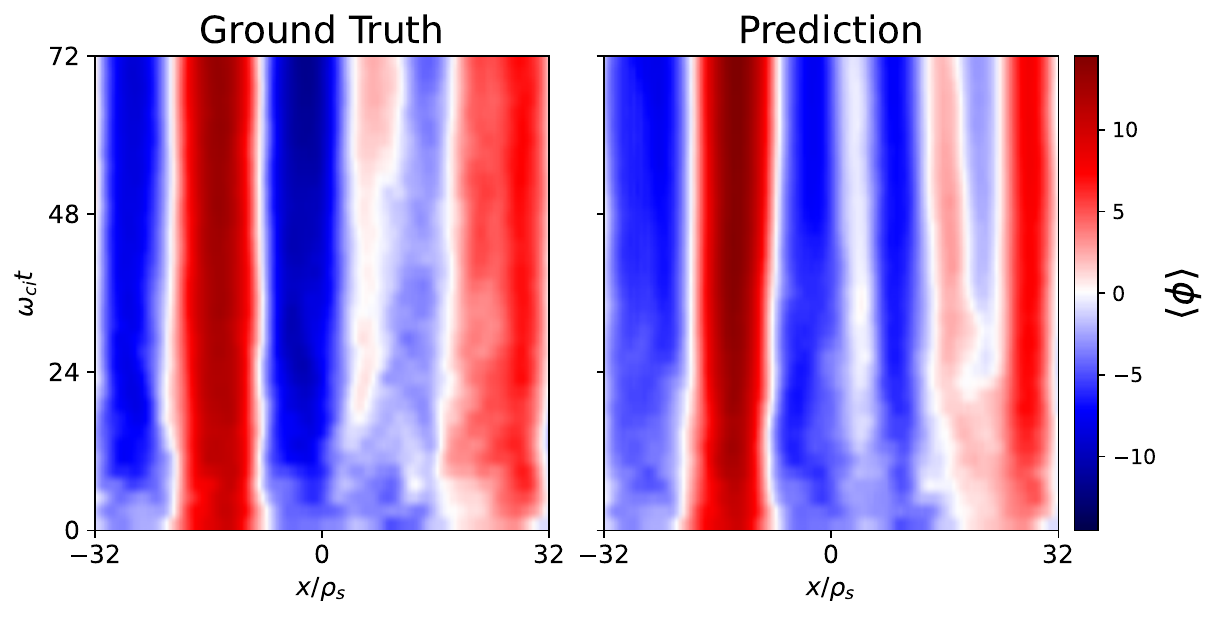}
    \caption{(Top) initial conditions, ground truth, pretrained and finetuned model predictions after $72\omega_\text{ci}^{-1}$ for held-out \textit{dynamical transition} $\alpha=0.1\rightarrow1.1$, $\omega_\text{ci}t_0=0$. (Middle) time evolution of the energy $E$, enstrophy $W$, turbulent flux $\Gamma_n$ and resistive dissipation $D_\alpha$ from the ground truth, pretrained and finetuned models. (Bottom) ground truth and finetuned model trajectories of the zonal component of the potential $\langle\phi\rangle$. }
    \label{fig:Fig3_01_11_extra}
\end{figure}

\bibliography{NOT.bib}

\end{document}